\documentclass[a4paper,11pt]{article}

\usepackage{jheppub}

\usepackage{amsmath, amsthm, amsfonts}
\usepackage[spanish,activeacute,english]{babel}
\usepackage{amsmath}%
\usepackage{amsfonts}%
\usepackage{amssymb}%
\usepackage{graphicx}
\usepackage{cancel}
\usepackage{appendix}
\usepackage{graphicx}
\usepackage{vmargin}
\usepackage{makeidx}
\usepackage{float}
\usepackage{subfig}
\usepackage{braket}
\usepackage{dsfont}
\usepackage{color}
\usepackage{fancybox}
\usepackage{subcaption}

\theoremstyle{definition}

\theoremstyle{remark}

\title{Scalar potential bounds on one-parameter Calabi-Yau threefolds with fluxes    
}
\author{Nana Geraldine Cabo Bizet$^{a}$  \footnote{e-mail address: {\tt nana@fisica.ugto.mx}} , Jos\'e de J. P\'erez Lona  $^{a}$\footnote{e-mail address: {\tt jdj.perezlona@ugto.mx }}, \\
[4mm]
{\small \em \it  $^{a}$Departamento de F\'isica, Divisi\'on de Ciencias e Ingenier\'{\i}as, \\ 
Universidad de Guanajuato, Loma del Bosque 103, \\ C.P. 37150, Le\'on, Guanajuato, M\'exico.}\\
[4mm]

}
\date{{\today}}
\abstract{
	We study the refined de Sitter conjecture (RdSC), the
	Trans-Planckian Censorship Conjecture (TCC) and the species scale
	$\Lambda_s$ in the multifield flux potential of type IIB string theory
	compactified on one-parameter Calabi--Yau threefolds. The scalar
	potential depends on the axio-dilaton and on the single complex-structure
	modulus, and is no-scale along the K\"ahler directions. Restricting to
	$h^{2,1}=1$ allows us to work with the exact periods obtained from the
	Picard--Fuchs equations rather than with their large-complex-structure
	expansion, and therefore to probe the conifold region. For the
	mirror quintic, $(h^{2,1},h^{1,1})=(1,101)$, we scan 200
	flux configurations and find unstable extrema in $96\%$ of the cases,
	with $c'=0.18$. Once the K\"ahler runaway is taken into account, the
	no-scale structure fixes the logarithmic slope of the potential at every
	critical point to the universal value $\sqrt6$, so that the first RdSC
	condition is satisfied throughout. Statistically, the complex structure
	is the heavier sector and the axio-dilaton dominates the unstable
	directions. For the TCC we obtain analytic results valid for the whole
	class of one-parameter models: the slope approaches the universal value
	$2$ at large complex structure, diverges at the conifold, and equals
	exactly $\sqrt2$ in the axio-dilaton directions, where the leading
	potential takes a single $SL(2)$-covariant form, as the multifield
	evolutions confirm numerically. The TCC further translates into a lower
	bound on the compactification volume, $\mathcal{V}>\mathcal{V}_0\sim
	e^{N}$, controlled by the number of e-folds. Finally, the field
	displacements along our trajectories are compatible with the
	species-scale field-range bound, and restoring the volume dependence of
	$\Lambda_s$ shows that $V\lesssim3\Lambda_s^2$ is implied by the TCC
	volume bound, which is thus the genuine constraint on these models.
}

\begin{document}
\maketitle

\section{Introduction}
\label{sec:intro}

For the last forty years a great amount of work has been devoted to
exploring string theory compactifications on Calabi--Yau (CY) varieties
\cite{Candelas:1985en,Candelas:1987kf,Candelas:1989js,Candelas:1990pi,%
Candelas:1990qd,Candelas:1990rm,Grana:2005jc}. A recurring question,
reformulated over time, is whether phenomenologically relevant
four-dimensional theories can be obtained from the dimensional reduction
of strings on these varieties, and how abundant they are
\cite{KKLT,Kachru:2003aw,Grana:2005jc,Douglas:2006es,Honma:2017ghy,%
McAllister:2024cdS}. The spacetime physics is governed by the critical
points of the effective scalar potential $V_{\rm eff}$ of the moduli
fields \cite{KKLT}.

Type IIB flux compactifications have been particularly well studied in
this context, including the consistency conditions coming from the
ten-dimensional equations of motion and the dimensional reduction of the
effective action \cite{GKP,CaboBizet:2016qsa}. An orientifold projection
breaks $\mathcal{N}=2$ supersymmetry down to a four-dimensional
$\mathcal{N}=1$ theory \cite{GKP}. The Ramond--Ramond (R--R) and
Neveu--Schwarz--Neveu--Schwarz (NS--NS) three-form fluxes generate a
Gukov--Vafa--Witten (GVW) superpotential $W$ \cite{Gukov:1999ya} which,
together with the K\"ahler potential $K$, determines $V_{\rm eff}$. The
flux quanta are constrained by the D3-brane tadpole cancellation
condition, which bounds the region of flux space that can be explored.
This potential generically stabilizes the axio-dilaton $\tau$ and the
complex-structure moduli $z_i$, while remaining no-scale along the
K\"ahler directions.

Starting from this configuration, the best-known mechanism to obtain de
Sitter (dS) vacua is the one proposed by Kachru, Kallosh, Linde and
Trivedi (KKLT) \cite{KKLT}. It implements a step-by-step stabilization in
which the axio-dilaton and the complex-structure moduli are fixed first,
at higher energies, while the K\"ahler modulus is stabilized in a
subsequent step; this requires mass hierarchies among the moduli
\cite{Blumenhagen:2022mhq}. Different efforts continue to explore and
scrutinize this setup \cite{Lust:2022xoq,McAllister:2025tasi}, and it has
also been discussed whether dS vacua can be obtained directly from the
effective potential of the axio-dilaton and complex-structure sector,
providing an alternative to KKLT \cite{SS}.

These constructions lead to a vast landscape of solutions
\cite{douglas03,douglas04}. Over the last two decades a complementary
program has emerged, aimed at determining which of these effective
theories are actually consistent with quantum gravity
\cite{Vafa:2005ui,ArkaniHamed:2006dz,Ooguri:2006in,Agmon:2022thq}. This
represents a change of approach: rather than exploring as much of the
apparent landscape as possible, one evaluates the resulting effective
theories against additional consistency criteria. Several of the
resulting conjectures constrain the structure of the scalar potential and
its derivatives, and are the ones we shall test here.

The \emph{refined de Sitter conjecture} (RdSC)
\cite{Obied:2018sgi,Garg:2018reu,Ooguri:2018wrx} states that, in reduced
Planck units, the scalar potential of a consistent effective theory of
quantum gravity must satisfy, wherever $V>0$,
\begin{equation}
\left|\nabla V\right|\;\geq\; c\,V\,,
\qquad\text{or}\qquad
\min\ {\rm eig}\left(\nabla_i\nabla_j V\right)\;\leq\; -\,c'\,V\,,
\label{eq:rdsc}
\end{equation}
where $c,c'>0$ are constants of order one, the gradient and the Hessian
are taken with respect to the field-space metric $G_{ij}$, and the second
condition is invoked only when the first one fails. Throughout this work
we search for $c$ by making vacua statistics.

The \emph{Trans-Planckian Censorship Conjecture} (TCC)
\cite{Bedroya:2019snp} demands that modes whose wavelength is shorter
than the Planck length never exit the Hubble horizon and freeze into
classical perturbations during an expanding phase, a requirement encoded
in the inequality \eqref{eq:tcc-original} of Sec.~\ref{sec:tcc}.
For a positive and asymptotically decreasing scalar potential, this
condition implies a lower bound on the average logarithmic slope of the
potential along an infinite-distance trajectory in moduli space,
\begin{equation}
\left(\frac{|V'|}{V}\right)_{\infty}
\;\geq\;
\frac{2}{\sqrt{(d-1)(d-2)}}
\;\;\overset{d=4}{\longrightarrow}\;\;
\sqrt{\frac{2}{3}}\,,
\label{eq:tcc}
\end{equation}
where the prime denotes the derivative along the trajectory with respect
to the affine parameter and the subscript $\infty$ denotes an asymptotic
average, defined precisely in Sec.~\ref{sec:tcc}. The bound, which is
argued to keep the classical potential reliable
\cite{Bedroya:2019tba}, also applies in the multifield context
\cite{Andriot:2023bsk,Andriot:2025svm}.

Finally, the \emph{species scale} $\Lambda_s$
\cite{Dvali:2007hz,Dvali:2010ws} is the cutoff of an effective theory
containing $N$ light species,
\begin{equation}
\Lambda_s=\frac{M_{{\rm Pl},d}}{N^{1/(d-2)}}
\;\;\overset{d=4}{\longrightarrow}\;\;
\frac{M_{\rm Pl}}{\sqrt{N}}\,,
\label{eq:species-intro}
\end{equation}
and has recently been studied in string theory in connection with the
Distance and Emergent String conjectures
\cite{vdHeisteeg:2022Species,CalderonInfante:2024SSDC}. Requiring the
Hubble scale to stay below the cutoff, and requiring the effective
description to remain valid along a field excursion, gives respectively
\begin{equation}
V\;\lesssim\; 3\,\Lambda_s^2\,,
\qquad
\Delta\phi\;\leq\; -a\log V_0+b\,,
\label{eq:species-bounds-intro}
\end{equation}
in reduced Planck units, with $a>0$ and $b$ of order one
\cite{Heisteeg:2024Bounds,vdHeisteeg:2024SpeciesTCC}. Various setups have
been analyzed for these and related conjectures
\cite{Alvarez-Garcia:2023gdd,Freigang:2023ogu,Castellano:2023aum,%
Blumenhagen:2023tev,Blumenhagen:2023yws}.

In this work we test the three conditions
\eqref{eq:rdsc}, \eqref{eq:tcc} and \eqref{eq:species-bounds-intro} on
one and the same family of explicit type IIB compactifications to four
dimensions, focusing on the axio-dilaton $\tau=C_0+i/g_s\equiv
\tau_1+i\tau_2$ and on the complex-structure sector $z=r\,e^{i\theta}$,
with no-scale behaviour assumed for the K\"ahler moduli. Our approach
differs from previous analyses in three respects.

First, we restrict to geometries with a single complex-structure modulus,
$h^{2,1}=1$. This is not merely a simplifying assumption: for these
varieties the Picard--Fuchs equations can be solved and their solutions
analytically continued, so that the periods, and hence the potential, are
known exactly rather than only through their large-complex-structure
expansion. This is what allows us to explore the neighbourhood of the
conifold point on the same footing as the large-complex-structure (LCS)
region, and the explicit dependence of the potential on the CY periods
turns out to be essential for the results.

Second, we do not evaluate the conjectures only along single-field
asymptotic directions. We construct genuinely multifield cosmological
trajectories and express the potential in terms of the affine parameter
of the field-space evolution, which is the quantity entering the
asymptotic TCC average and the species-scale field-range bound.

Third, evaluating the three conjectures on the same set of vacua allows
us to compare their relative strength, which is a question that is
difficult to address when each conjecture is studied in a different
setup.

Our results can be summarized as follows.
\begin{itemize}
\item      
      The refined dS conjecture is satisfied by $96\%$ of the
      $\sim 10^{2}$ flux configurations analyzed, for $c=0.18$. Once the
      K\"ahler runaway is taken into account, the no-scale structure
      fixes the logarithmic slope of the potential at every critical
      point to the universal value $\sqrt6$, so that no dS vacuum arises
      at this first stabilization step.  This value, familiar from the volume runaway of type IIB
      \cite{CaboBizet:2019sgr,Casas:2024LightTowers} and appearing as
      a lower bound in the asymptotic analysis of
      \cite{CalderonInfante:2022AAE}, is shown here to be saturated at
      every critical point, and to follow from the homogeneity of any
      volume cubic in the two-cycle sizes.
      \cite{CaboBizet:2019sgr,Casas:2024LightTowers,CalderonInfante:2022AAE},
      is shown here to hold as an exact equality for any volume
      homogeneous of degree $3/2$.
      
      \item By analyzing the mass matrix at the critical points, we find a characteristic hierarchy among the moduli: the complex-structure modulus $z$ typically corresponds to the heaviest direction, whereas the $\tau$ modulus predominantly contributes to the lightest, unstable mode. Remarkably, this hierarchy is also reflected in the field dynamics, providing a complementary dynamical confirmation of the stability pattern inferred from the mass spectrum.

      \item The TCC is satisfied along the infinite-distance directions
      associated with the large complex structure and with the dilaton
      $\tau_2$, and also along the large-$\tau_1$ region, since the
      quantity bounded by the conjecture in its multifield form is the
      norm of the gradient, which remains of order one throughout the
      axio-dilaton asymptotic regions. Along the purely axionic
      direction $\tau_1$ the {\it directional} slope instead decays,
      the potential growing rather than decaying there. We check this
      numerically and then confirm it analytically for the whole class. More importantly, imposing the TCC on the multifield evolution
      translates into a lower bound on the compactification volume,
      $\mathcal{V}>\mathcal{V}_0$.
\item The field displacements along the cosmological trajectories are
      compatible with the species-scale bound on $\Delta\phi$. By
      contrast, for a fixed flux configuration the requirement $V\lesssim 3\Lambda_s^2$ is satisfied
      only in restricted regions of the parameter space, and is the most
      stringent of the three conditions for the models analyzed.
\end{itemize}

The paper is organized as follows. In Sec.~\ref{sec:setup} we describe
the four-dimensional effective Lagrangian of type IIB string theory
compactified on one-parameter Calabi--Yau manifolds, and the scalar
potential for the axio-dilaton and the complex-structure modulus
generated by the NS--NS and R--R fluxes. In Sec.~\ref{sec:example} we
specialize to the mirror quintic and explore the critical points of its
potential. Sections~\ref{sec:rdsc}, \ref{sec:tcc} and \ref{sec:species}
are devoted, respectively, to the refined dS conjecture, to the TCC, and
to the bounds arising from the species scale. We present our conclusions
and outlook in Sec.~\ref{sec:conclusions}. Appendix~\ref{app:vacua}
collects the details of the $\sim 10^{2}$ models explored.

\section{Type IIB string theory on Calabi--Yau manifolds with a single parameter}
\label{sec:setup}

In this section we set up the four-dimensional effective theory obtained
by compactifying type IIB string theory on a Calabi--Yau (CY) threefold
with a single complex-structure modulus, and we derive the flux-induced
scalar potential that will be analyzed in the rest of the paper. The
mirror quintic, with Hodge numbers $(h^{2,1},h^{1,1})=(1,101)$, is used
as the reference geometry from Sec.~\ref{sec:example} onwards.

The low-energy limit of type IIB string theory compactified on a CY with
non-trivial three-form fluxes is described by the four-dimensional
$\mathcal{N}=1$ supergravity action
\begin{equation}
    S_4=\int d^4x\, \sqrt{-g}\left[\frac{M_{\rm Pl}^2}{2}R
    - M_{\rm Pl}^2\, K_{a\overline{b}}\,
      \partial_\mu\Phi^a \partial^\mu \overline{\Phi}^{\overline{b}}
    - V(\Phi^l)\right],
    \label{S4}
\end{equation}
where $M_{\rm Pl}^2=V_6/(\kappa_{10}^2 g_s^2)$, $V_6$ is the dimensionful
internal volume and $g_s$ the string coupling. The indices $a,\overline{b}$
run over all the moduli present, namely the complex-structure moduli
$\{z_i\}$, the K\"ahler moduli $\{T_m\}$ and the axio-dilaton $\tau$, and
$K_{a\bar{b}}$ is the moduli-space metric derived from the K\"ahler
potential.

Two conventions follow from \eqref{S4} and will be used throughout the
paper. We work in mostly-plus signature and in reduced Planck units,
$M_{\rm Pl}=1$. Writing the complex fields in terms of real components
$\varphi^i$, the kinetic term of \eqref{S4} becomes
$-\tfrac{1}{2}G_{ij}\partial_\mu\varphi^i\partial^\mu\varphi^j$ with
\begin{equation}
    G_{ij}=2K_{a\overline{b}}\,,
    \label{eq:Gmetric}
\end{equation}
so that the {\it real} field-space metric, the one entering all distances,
gradients and affine parameters in
Secs.~\ref{sec:rdsc}--\ref{sec:species}, carries an extra factor of two
with respect to the K\"ahler metric. In particular, a canonically
normalized field is obtained from
$\partial_\mu\phi_i=\sqrt{2K_{a\bar b}}\,\partial_\mu\varphi^i$.

Let us now briefly describe the scalar fields involved in this setup and
the geometry of their moduli space.

\subsection{Calabi--Yau geometry}
\label{sec:cygeometry}

In order to explore the extrema of the theory it is convenient to
introduce a few expressions from the geometry of moduli space, usually
called special geometry. We concentrate on the complex-structure moduli
space. Let $\left\lbrace \alpha_I,\beta^I\right\rbrace$ be an integral
basis of the cohomology group $H^3(\text{CY},\mathbb{Z})$ and
$\left\lbrace A^I,B_I\right\rbrace$ the dual basis of the homology group
$H_3(\text{CY},\mathbb{Z})$, defined by
\begin{equation}
	\int_{\text{CY}}\alpha_I \wedge \beta^J=\delta_I^J
	=-\int_{\text{CY}}\beta^J \wedge \alpha_I, \label{bc1}
\end{equation}
\begin{equation}
	\int_{\text{CY}}\alpha_I \wedge \alpha_J
	=\int_{\text{CY}}\beta^I \wedge \beta^J=0, \label{bc2}
\end{equation}
\begin{equation}
	\int_{A^I}\alpha_J =\delta^I_J\,,
	\qquad
	\int_{B_I}\beta^J =-\delta_I^J\,, \label{bc3}
\end{equation}
where $I,J=0,\dots,h^{2,1}$. Let $\Omega$ be the holomorphic $(3,0)$ form
of the CY threefold, which encodes the complex-structure moduli
$\left\lbrace z_i \right\rbrace$. Expanding it in the integral basis,
\begin{equation}
	\Omega=\mathcal{X}^I\alpha_I-\mathcal{F}_I \beta^I, \label{3fe}
\end{equation}
and using \eqref{bc3}, the coefficients are identified with the periods
of $\Omega$ over the dual cycles,
\begin{equation}
	\mathcal{X}^I=\int_{A^I} \Omega\,,
	\qquad
	\mathcal{F}_I=\int_{B_I} \Omega\,, \label{coefperiods}
\end{equation}
so that the period vector is
\begin{equation}
    \Pi=\begin{pmatrix}
    \displaystyle\int_{A^I}\Omega \\[4mm]
    \displaystyle\int_{B_I}\Omega
    \end{pmatrix}
    =\begin{pmatrix}
    {\mathcal{X}}^I \\[3mm] {\mathcal{F}}_I
    \end{pmatrix}.
    \label{eq:periodvector}
\end{equation}
Using \eqref{3fe}--\eqref{coefperiods}, the norm of $\Omega$ can be
written in terms of the periods,
\begin{equation}
	\int_{\text{CY}} \Omega \wedge \overline{\Omega}
	=\mathcal{F}_I\overline{\mathcal{X}}^I
	-\mathcal{X}^I\overline{\mathcal{F}}_I
	=\overline{\Pi}^T\Sigma \Pi\,, \label{omomb}
\end{equation}
where $\Sigma$ is the symplectic matrix
\begin{equation}
	\mathrm{\Sigma}=\begin{pmatrix}
		0&\mathbb{I}_{k\times k} \\
		-\mathbb{I}_{k\times k}&0
	\end{pmatrix}, \label{MSIGMA}
\end{equation}
with $k=1+h^{2,1}$; for $h^{2,1}=1$ it is therefore a $4\times4$ matrix.
Note that $\overline{\Pi}^T\Sigma\Pi$ is purely imaginary, since
$\Sigma^T=-\Sigma$.

The complex-structure moduli space of a CY threefold contains singular
points, around which the periods must be defined by analytic
continuation. Transporting the periods around a non-contractible loop in
moduli space produces a monodromy,
\begin{equation}
    \Pi \rightarrow \mathrm{T}\, \Pi,
\end{equation}
with $\mathrm{T}\in \mathrm{Sp}(2h^{2,1}+2,\mathbb{Z})$, so that the
symplectic structure is preserved,
$\mathrm{T}^T\mathrm{\Sigma}\mathrm{T}=\mathrm{\Sigma}$. Since the
periods, or equivalently the cycles, transform under monodromy, so does
the scalar potential, $V\rightarrow V^\prime$. Starting from a given
point of moduli space, for instance a minimum of $V$, the potential
varies continuously as one moves along a monodromy loop, which opens the
way to interesting phenomena such as vacuum tunnelling or the
construction of families of connected vacua; see for example
\cite{Ahlqvist:2010ki,Braun:2011av,Danielsson:2006xw}.

\subsection{Flux compactifications and the scalar potential}
\label{sec:fluxpotential}

Turning on fluxes on a CY orientifold induces the Gukov--Vafa--Witten
(GVW) superpotential $W$, which together with the K\"ahler potential
determines the $\mathcal{N}=1$ scalar potential appearing in \eqref{S4},
\begin{equation}
	V= \frac{1}{\kappa_{10}^2g_s}\, e^{K}
	\left[K^{a\overline{b}}D_aW\overline{D}_{\bar{b}}\overline{W}
	-3\left|W\right|^2\right], \label{scalarpotential}
\end{equation}
where $K^{a\overline{b}}$ is the inverse of the K\"ahler metric
$K_{a\overline{b}}=\partial_{a}\partial_{\overline{b}}K$ and
$D_aW=\partial_aW+(\partial_aK)W$ is the K\"ahler-covariant derivative.
As above, $a,b$ run over all moduli. The GVW superpotential reads
\begin{equation}
    W=\int_{\text{CY}}\Omega \wedge G_{(3)},
\end{equation}
where $G_{(3)}$ is built out of the Ramond--Ramond (R--R) flux $F_{(3)}$,
the Neveu-Schwarz--Neveu-Schwarz (NS--NS) flux $H_{(3)}$ and the
axio-dilaton $\tau=C_{(0)}+ie^{-\phi}\equiv\tau_1+i\tau_2$,
\begin{equation}
    G_{(3)}=F_{(3)}-\tau H_{(3)}.
\end{equation}
The fluxes threading the three-cycles obey Dirac quantization,
\begin{equation}
    \frac{1}{(2\pi)^2\alpha^\prime}\int_{A^I,B_I}F_{(3)}\in \mathbb{Z},
    \hspace{1cm}
    \frac{1}{(2\pi)^2\alpha^\prime}\int_{A^I,B_I}H_{(3)}\in \mathbb{Z}\,,
\end{equation}
and can therefore be expanded in the integral basis
\eqref{bc1}--\eqref{bc3}, so that
\begin{equation}
    W=\int_{\text{CY}}\Omega \wedge
    \left(G_{(3)}^I\alpha_I-{G_{(3)}}_I\beta^I\right)
    =G_{(3)}^I\mathcal{F}_I-{G_{(3)}}_I\mathcal{X}^I
    =\mathrm{G}^T\Sigma\Pi\,,
    \label{eq:WGSigmaPi}
\end{equation}
with the flux quanta collected in the vector
\begin{equation}
    \mathrm{G}=
    \begin{pmatrix}
        G_{(3)}^I \\ {G_{(3)}}_I
    \end{pmatrix}
    =\mathrm{F}-\tau\,\mathrm{H}\,.
\end{equation}
It will be convenient to separate the two flux contributions as
\begin{equation}
    W=\mathrm{F}^T\Sigma\Pi-\tau\,\mathrm{H}^T\Sigma\Pi
    \equiv A+B\tau\,,
    \qquad
    A\equiv \mathrm{F}^T\Sigma\Pi\,,
    \quad
    B\equiv -\,\mathrm{H}^T\Sigma\Pi\,.
    \label{eq:AB}
\end{equation}

The K\"ahler potential is given at tree level by
\begin{equation}
    K=K_{\text{ad}}+K_{\text{cs}}+K_{\text{k}}
    =\underbrace{-\ln\left[ -i(\tau-\overline{\tau})\right]}_{(\text{ad})}
     \underbrace{-\ln\left[ i \int_{\text{CY}} \Omega \wedge
       \overline{\Omega}\right]}_{(\text{cs})}
     \underbrace{-2\ln\left[ \mathcal{V}\right]}_{(\text{k})},
     \label{K}
\end{equation}
where $\mathcal{V}$ is the dimensionless internal volume, a function of
the K\"ahler moduli $\left\lbrace T_m\right\rbrace$, and the labels
{\it ad}, {\it cs} and {\it k} denote the axio-dilaton,
complex-structure and K\"ahler contributions. Using \eqref{omomb}, the
complex-structure piece reads
\begin{equation}
    K_{\text{cs}}=-\ln{\left[i\, \overline{\Pi}^T\Sigma \Pi\right]}\,,
    \label{eq:Kcs}
\end{equation}
the argument of the logarithm being real and positive by the remark below
\eqref{MSIGMA}. Note also that
$-i(\tau-\overline{\tau})=2\tau_2$, so that
$K_{\rm ad}=-\ln(2\tau_2)$ and $K_{\tau\bar\tau}=1/(4\tau_2^2)$.

In what follows we consider no-scale scalar potentials, for which the
K\"ahler-moduli contribution to $|DW|^2$ cancels against the term
proportional to the norm of the superpotential,
\begin{equation}
	K^{m\bar{n}}D_mW\bar{D}_{\bar{n}}\bar{W} -3\left|W\right|^2=0,
	\label{nsc}
\end{equation}
where $m,n$ run over the K\"ahler moduli. Since $W$ does not depend on
them, $D_mW=(\partial_mK)W$ and \eqref{nsc} is equivalent to the no-scale
condition $K^{m\bar n}\partial_mK\,\partial_{\bar n}K=3$. 

In large-volume
compactifications one usually assumes that the volume in the K\"ahler
potential is controlled by a single overall modulus $\rho$,
\begin{equation}
    K_{\text{k}}=-3\ln\left(-i(\rho-\bar{\rho})\right)\,,
    \qquad
    \mathcal{V}=\left(-i(\rho-\bar\rho)\right)^{3/2},
    \label{eq:Kk}
\end{equation}
consistently with \eqref{K}, and in this case \eqref{nsc} is readily
verified. The same holds for several K\"ahler moduli.

It is worth noting that \eqref{nsc} is not an additional requirement to
be checked case by case, but a direct consequence of the homogeneity of
the volume. Working with the variables $u^l=-i(\rho^l-\bar\rho^l)$, on
which $K_{\rm k}$ depends only through the axionic shift symmetry
$\rho^l\to\rho^l+ic^l$, the factors of $i$ cancel and
$K_{l\bar m}=\partial_l\partial_m K_{\rm k}$ becomes a real symmetric
matrix. If $\mathcal{V}$ is homogeneous of degree $n$ in the $u^l$, then
$K_{\rm k}(\lambda u)=K_{\rm k}(u)-2n\log\lambda$, and differentiating
this relation once and twice with respect to $u$ gives
\begin{equation}
    u^l\partial_lK_{\rm k}=-2n\,,
    \qquad
    u^lK_{lm}=-K_m\,,
    \qquad\text{hence}\qquad
    K^{lm}K_m=-u^l\,,
    \label{eq:euler}
\end{equation}
so that contracting with $K_l$ yields
$K^{l\bar m}K_l K_{\bar m}=-u^lK_l=2n$. Since $\mathcal{V}$ is cubic in
the two-cycle volumes and the $u^l$ measure four-cycles, $n=3/2$ and the
no-scale condition \eqref{nsc} follows, whatever the explicit form of
$\mathcal{V}$: swiss-cheese, fibred, anisotropic or generic, and for any
number of K\"ahler moduli. Note that the argument uses only the degree of
homogeneity and the axionic shift symmetry, and that no inversion of the
metric is required; both hypotheses fail once $\alpha'$ or
non-perturbative corrections are included, which is where the no-scale
structure is broken.

 The K\"ahler sector
then contributes to \eqref{scalarpotential} only through the overall
factor $e^{K}$, introducing a runaway dependence that we discuss below.
The no-scale scalar potential therefore takes the form
\begin{equation}
	V= \frac{1}{\kappa_{10}^2g_s}\,e^{K}
	\left[K^{i\overline{j}}D_iW\overline{D}_{\overline{j}}\overline{W}
	\right], \label{nsV}
\end{equation}
with $i,j$ running over the axio-dilaton and the complex-structure
moduli. Being a positive-definite quadratic form, \eqref{nsV} is always
non-negative, and vanishes precisely at the supersymmetric,
imaginary-self-dual configurations.

For a single complex-structure modulus, $h^{2,1}=1$, the two covariant
derivatives in \eqref{nsV} can be evaluated explicitly in terms of the
period vector. Using \eqref{eq:Kcs} one finds
\begin{equation}
    K_{z\bar z}
    =-\,\frac{\widetilde{\Sigma}}
    {\left(i\,\overline{\Pi}^T\Sigma\Pi\right)^{2}}\,,
    \qquad
    D_zW=-\,\frac{\mathrm{G}^T\Sigma\widetilde{\Pi}\Sigma\overline{\Pi}}
    {\overline{\Pi}^T\Sigma\Pi}\,,
    \label{eq:Kzzbar}
\end{equation}
where
\begin{equation}
	\widetilde{\Sigma}(z,\overline{z})
	=\left(\overline{\Pi}^T\Sigma \partial_z\Pi\right)
	 \left( \partial_{\bar{z}}\overline{\Pi}^T\Sigma \Pi\right)
	-\left( \partial_{\bar{z}}\overline{\Pi}^T\Sigma \partial_z\Pi\right)
	 \left( \overline{\Pi}^T\Sigma \Pi\right),
	\label{eq:Sigmatilde}
\end{equation}
\begin{equation}
	\widetilde{\Pi}(z)=\partial_z\Pi \otimes \Pi-\Pi \otimes \partial_z\Pi\,.
	\label{eq:Pitilde}
\end{equation}
One checks directly from \eqref{eq:Sigmatilde} that $\widetilde{\Sigma}$
is real; positivity of the moduli-space metric then requires
$\widetilde{\Sigma}<0$. Likewise, from
$K_{\rm ad}=-\ln(2\tau_2)$ one obtains
$4\tau_2^2\,|D_\tau W|^2=\left|(\mathrm{F}-\bar{\tau}\mathrm{H})^T
\Sigma\Pi\right|^2$. Collecting both contributions, the scalar potential
reads
\begin{equation}
	V=\frac{1}{\kappa_{10}^2g_s}\, e^K\left(
	-\,\frac{\left| \left( \mathrm{F}-\tau \mathrm{H}\right)^T
	\Sigma \widetilde{\Pi}\Sigma \overline{\Pi} \right|^2}
	{\widetilde{\Sigma}}
	+\left|\left(\mathrm{F}-\bar{\tau}\mathrm{H}\right)^T
	\Sigma \Pi\right|^2 \right),
	\label{eq:Vexplicit}
\end{equation}
both terms being manifestly non-negative.

Flux compactifications stabilize the complex-structure moduli and the
axio-dilaton, while non-perturbative corrections are needed to stabilize
the K\"ahler moduli. It is therefore useful to factor out the K\"ahler
dependence and write
\begin{equation}
	V=\frac{1}{\kappa_{10}^2g_s}\, e^{K_{\text{k}}}\,\hat{V}(\tau,z),
	\qquad
    \hat{V}(\tau,z) =e^{K_{\text{ad}}+K_{\text{cs}}}
    \left[K^{i\overline{j}}D_iW\overline{D}_{\overline{j}}\overline{W}
    \right],
    \label{eq:Vhat}
\end{equation}
where $\hat V$ is the slice of the scalar potential along the
axio-dilaton and complex-structure directions. Although the potential may
be extremized along these directions, the K\"ahler moduli enter only
through the overall rescaling $e^{K_{\rm k}}$ and therefore always
provide a runaway direction. In what follows we call an extremum of the
scalar potential a point satisfying
\begin{equation}
	\partial_{i}\hat{V}=\partial_{\overline{i}}\hat{V}=0\,.
	\label{extre}
\end{equation}

Finally, the three-form fluxes contribute to the D3-brane charge,
\begin{equation}
    N_{\text{flux}}=\frac{1}{2\kappa_{10}^2  T_3}
    \int_{\text{CY}}F_{(3)}\wedge H_{(3)}
    =\frac{g_s}{(2\pi)^4{\alpha^\prime}^2}\,
    \mathrm{H}^T\mathrm{\Sigma}\mathrm{F}\,,
    \label{eq:Nflux}
\end{equation}

which must be cancelled against the charge of the localized sources,
giving the tadpole condition \cite{CaboBizet:2016qsa}
\begin{equation}
    N_{D3}-\frac{1}{2}N_{O3}+\frac{\chi(D_{O7})}{12}+N_{\text{flux}}=0\,.
    \label{eq:tadpole}
\end{equation}

This constraint bounds the region of flux space that can be scanned. Throughout our analysis
 we assume  that by incorporating the necessary number
of D3-branes  it
is possible to cancel satisfy it.

Compactifications with constant fluxes fixing the complex structure and
the axio-dilaton through the potential \eqref{nsV} constitute the
standard starting point for type IIB phenomenology in a low-energy
description. From now on we restrict to models with a single parameter,
$h^{2,1}=1$, and adopt the following notation:
\begin{gather}
 \Pi =\begin{pmatrix}
  \Pi_1 \\  \Pi_2\\  \Pi_3 \\  \Pi_4
  \end{pmatrix}, \qquad
  \mathrm{G} =\begin{pmatrix}
  G_1 \\  G_2\\  G_3 \\  G_4
  \end{pmatrix} =\begin{pmatrix}
  F_1\\   F_2\\   F_3\\   F_4
  \end{pmatrix}-\tau\begin{pmatrix}
  H_1\\   H_2\\   H_3\\   H_4
  \end{pmatrix},
  \nonumber\\[2mm]
  W=G_1\Pi_3+G_2\Pi_4-G_3\Pi_1-G_4\Pi_2\,,
  \nonumber\\[2mm]
  N_{\text{flux}}\;\propto\;H_1F_3+H_2F_4-H_3F_1-H_4F_2\,,
  \label{eq:oneparam}
 \end{gather}
where, abusing notation, we omit the subscript indicating the three-form
nature of the fluxes.

\section{The mirror quintic}
\label{sec:example}

We take the Calabi--Yau mirror of the quintic in $\mathbb{CP}^4$ as our
working example \cite{Candelas:1990rm,Candelas:1990qd}. Beyond its historical role as the standard testing
ground for compactifications on CY manifolds, it is well suited to the
present analysis: it has a single complex-structure modulus, its
Picard--Fuchs equations can be solved in closed form, and the resulting
periods can be analytically continued, so that the scalar potential is
under control not only in the large-complex-structure regime but also
near the conifold.

\subsection{Geometry and periods}
\label{sec:quinticgeometry}

Let us briefly recall the topological data of this manifold. Let $W_\psi$
be the one-parameter family of degree-five polynomials in
$\mathbb{CP}^4$,
\begin{equation}
    W_{\psi}=\left\lbrace (x_1,\dots,x_5)\in \mathbb{CP}^4 \;:\;
    P=\sum_{l=1}^{5}a_lx_l^5-5\psi \prod_{l=1}^5 x_l=0\right\rbrace,
    \label{Wfam}
\end{equation}
which is invariant under a $\mathbb{Z}_5^3$ symmetry acting on the
homogeneous coordinates $\left\lbrace x_l\right\rbrace$. The mirror
quintic is obtained by quotienting \eqref{Wfam} by this symmetry,
$W_\psi/\mathbb{Z}_5^3$. It has Hodge numbers $h^{2,1}=1$ and
$h^{1,1}=101$; the single complex-structure modulus $z$ is encoded in the
invariant deformation parameter $\psi$. Its Euler characteristic is
\begin{equation}
    \chi(\mathcal{M})=2\left(h^{1,1}-h^{2,1}\right)=+200\,,
\end{equation}
and its third Betti number is
$b_3=\sum_{j=0}^{3}h^{3-j,j}=4$, which counts the three-cycles and hence
the number of periods.

The family \eqref{Wfam} has three singular points in moduli space: the
conifold, the large complex structure (LCS) point and the orbifold. The
first two arise from $P=0$ together with $dP=0$: the conifold sits at
$x_i=1$, $\psi=1$, while the LCS point is reached as $\psi\to\infty$. The
third one, the orbifold or Gepner point, sits at $\psi=0$, where the
manifold acquires an enhanced symmetry. In this work we restrict to the
conifold and LCS patches. The conifold is of physical interest because it
gives rise to Minkowski vacua with large scale hierarchies, and the LCS
point because it is the only singularity located at infinite distance in
moduli space, which is precisely the regime in which the asymptotic
Swampland conjectures are formulated.

To obtain explicit periods one may either express them in terms of Meijer
$G$-functions, which cover the whole moduli space, or construct power
series solutions around each singular point. Since we are interested in
the neighbourhood of the critical points, we follow the second route and
solve the Picard--Fuchs equations. In terms of the variable
$z=(5^5\psi^5)^{-1}$ these read
\begin{equation}
    \left[ \Theta_{\text{\tiny LCS}}^4
    -5^5\,z_{\text{\tiny LCS}}\prod_{k=1}^4
    \left(\Theta_{\text{\tiny LCS}}+\frac{k}{5} \right) \right]
    \pi_{\text{\tiny LCS}}=0,
    \label{eq:PF}
\end{equation}

where $\Theta_{\text{\tiny LCS}}=z_{\text{\tiny LCS}}
\partial_{z_{\text{\tiny LCS}}}$ and $\pi_{\text{\tiny LCS}}$ is the
period vector in the Frobenius basis,
\begin{equation}
  \pi_{\text{\tiny LCS}}=\begin{pmatrix}
  w^0_{\text{\tiny LCS}} \\
  w^1_{\text{\tiny LCS}}+w^0_{\text{\tiny LCS}}\ln{z_{\text{\tiny LCS}}} \\
  \frac{5}{2}w^2_{\text{\tiny LCS}}
  +\frac{5}{2}w^0_{\text{\tiny LCS}}(\ln{z_{\text{\tiny LCS}}})^2
  +5w^1_{\text{\tiny LCS}}\ln{z_{\text{\tiny LCS}}} \\
  \frac{5}{6}w^3_{\text{\tiny LCS}}
  +\frac{5}{6}w^0_{\text{\tiny LCS}}(\ln{z_{\text{\tiny LCS}}})^3
  +\frac{5}{2}w^1_{\text{\tiny LCS}}(\ln{z_{\text{\tiny LCS}}})^2
  +\frac{5}{2}w^2_{\text{\tiny LCS}}\ln{z_{\text{\tiny LCS}}}
  \end{pmatrix}, \label{eq:PFB}
\end{equation}
with
\begin{align*}
    w^0_{\text{\tiny LCS}}&=1+120\,z_{\text{\tiny LCS}}
    +113400\,z_{\text{\tiny LCS}}^2
    +\mathcal{O}(z_{\text{\tiny LCS}}^3), \\
    w^1_{\text{\tiny LCS}}&=770\,z_{\text{\tiny LCS}}
    +810225\,z_{\text{\tiny LCS}}^2
    +\mathcal{O}(z_{\text{\tiny LCS}}^3),\\
    w^2_{\text{\tiny LCS}}&=2875\,z_{\text{\tiny LCS}}
    +\frac{21040875}{4}z_{\text{\tiny LCS}}^2
    +\mathcal{O}(z_{\text{\tiny LCS}}^3),\\
    w^3_{\text{\tiny LCS}}&=-5750\,z_{\text{\tiny LCS}}
    -\frac{16491875}{4}z_{\text{\tiny LCS}}^2
    +\mathcal{O}(z_{\text{\tiny LCS}}^3).
\end{align*}
The periods in the symplectic basis \eqref{bc1}--\eqref{bc3} are obtained
from the Frobenius basis by a transition matrix,
$\Pi_{\text{\tiny LCS}}=\mathrm{M}_{\text{\tiny LCS}\rightarrow
\text{\tiny S}}\,\pi_{\text{\tiny LCS}}$, with
\begin{equation}
 \mathrm{M}_{\text{\tiny LCS}\rightarrow\text{\tiny S}}
 =\begin{pmatrix}
 -\dfrac{200\,i\,\zeta(3)}{8\pi^{3}}
 &\dfrac{50}{24}\dfrac{1}{2\pi i}&0&\dfrac{1}{(2\pi i)^3} \\[2mm]
 \dfrac{50}{24}&-\dfrac{11}{2}\dfrac{1}{2\pi i}
 &-\dfrac{1}{(2\pi i)^2}&0 \\[2mm]
 1&0&0&0\\[1mm]
 0&\dfrac{1}{2\pi i}&0&0
 \end{pmatrix}.
 \label{eq:transition}
\end{equation}

The transition matrices at the remaining singular points, together with
the corresponding monodromy matrices, were obtained in
\cite{Candelas:1990rm} and recast in the symplectic basis used here in
\cite{CaboBizet:2016qsa}, whose conventions and topological data we adopt. In particular, the conifold
monodromy matrix reads

\begin{equation}
    \mathrm{T}_{\text{\tiny C}}=\begin{pmatrix}
    1 & \hspace{10pt} 0 & \hspace{10pt} 0 & \hspace{10pt} 0\\
    0 & \hspace{10pt} 1 & \hspace{10pt} 0 & \hspace{10pt} 0 \\
    -1 & \hspace{10pt} 0 & \hspace{10pt} 1 & \hspace{10pt} 0\\
    0 & \hspace{10pt} 0 & \hspace{10pt} 0 & \hspace{10pt} 1
    \end{pmatrix},\label{eq:Tconifold}
\end{equation}

from which one reads that $\Pi_1$ is the period of the cycle that shrinks
as $|z_{\text{\tiny C}}|\to 0$, threaded by the fluxes $F_3$ and $H_3$,
while the period of the dual cycle is $\Pi_3$, threaded by $F_1$ and
$H_1$. Under $n$ successive monodromy transformations,
\begin{equation}
    \Pi_{\text{\tiny C}}\to (\mathrm{T}_{\text{\tiny C}})^n\,
    \Pi_{\text{\tiny C}}\,,
    \hspace{.5cm}
    \Pi_3\to \Pi_3^\prime= \Pi_3-n\,\Pi_1\,,
\end{equation}
so that $\Pi_3$ is shifted by $n$ units of $\Pi_1$.

\subsection{Numerical exploration of the extrema}
\label{sec:numerics}

It is convenient to trade the complex moduli for real fields. We use
\begin{equation}
r=|z|\,,\qquad
\theta=\arg z\,,\qquad
\tau_1=\operatorname{Re}\tau\,,\qquad
\tau_2=\operatorname{Im}\tau\,,
\end{equation}
which are also the variables in which the extrema are searched for.

Since the periods are given by power series in $z$, the convergence of
the truncated series has to be controlled. Although the series were
computed up to $600$ terms, solving \eqref{extre} numerically is
demanding, and the extrema were located by a scan with multiple starting
points. The notion of convergence amounts to delimiting the region in
which the Frobenius solutions are reliable: a solution of \eqref{extre}
falling outside the disc of convergence of the expansion used to obtain
it must be discarded. In the following, $z_0$ denotes the value of the
complex-structure modulus at the extremum, and $r_0=|z_0|$.

As a first step, of order $10^3$ flux configurations were scanned using
only $20$ terms in the period series, out of which of order $2*10^2$
convergent extrema were retained; these are the models quoted throughout
the paper and collected in Appendix~\ref{app:vacua}. In what follows, the
number of terms always refers to the Frobenius basis \eqref{eq:PFB}.

\begin{figure}[H]
	\centering
	\includegraphics[width=.8\linewidth]{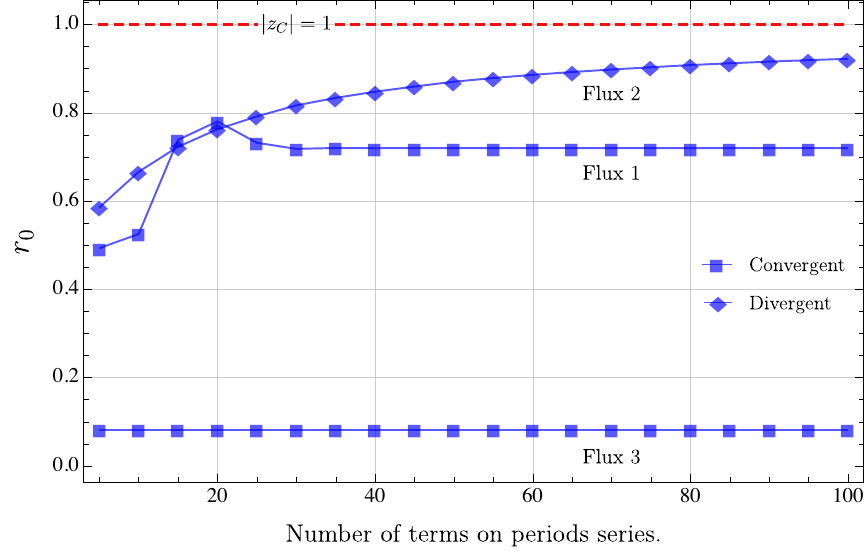}
	\caption{Behaviour of the value of the field $r$ on the vacuum ($r_0$), as a function of the number of terms kept
	in the period series, in the Frobenius basis. The flux
	configurations shown are
	Flux 1: $(F_{(3)},H_{(3)})=(-2,14,4,-1,9,-3,-17,-12)$,
	Flux 2: $(F_{(3)},H_{(3)})=(1,1,-2,-3,-17,0,1,0)$,
	Flux 3: $(F_{(3)},H_{(3)})=(10,20,-9,-5,-1,-11,17,-10)$.}
	\label{fig:r0b}
\end{figure}

\begin{figure}[H]
	\centering
	\includegraphics[width=.8\linewidth]{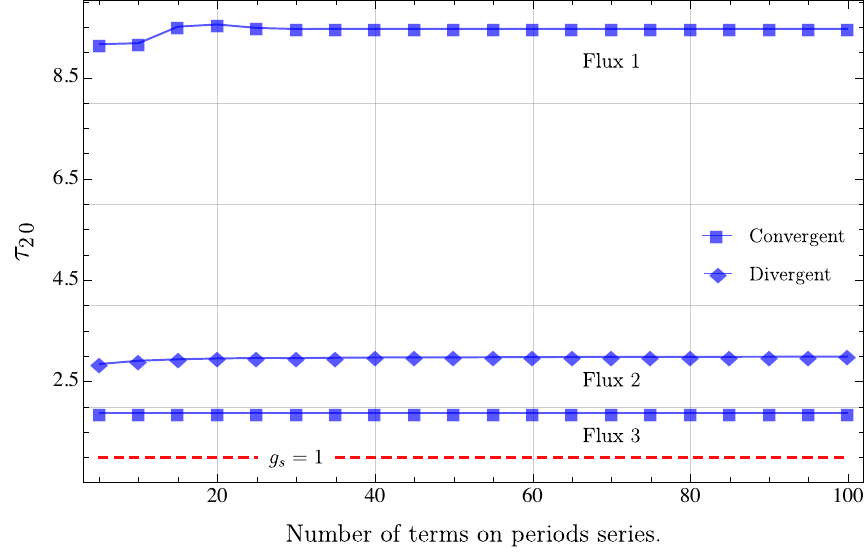}
	\caption{Behaviour of the value of the field $\tau_2$ on the vacuum ($\tau_{2,0}$), as a function of the number of
	terms kept in the period series, for the same three flux
	configurations as in Figure~\ref{fig:r0b}.}
	\label{fig:t20b}
\end{figure}

Figures~\ref{fig:r0b} and \ref{fig:t20b} compare genuine solutions with
spurious ones for the vevs $z_0$ and $\tau_{2,0}$ in the conifold patch.
More than $20$ terms are needed for the solutions to settle. The second
step was to study their sensitivity: as the number of terms grows,
spurious vacua can be identified because $z_0$ exhibits a monotonic
growth associated with non-convergence, whereas $\tau_{2,0}$ typically
oscillates around a constant value; the moduli $\theta$ and $\tau_1$
behave like $z_0$ and $\tau_{2,0}$ respectively. The transition to the
asymptotic regime takes place between $10$ and $40$ terms, and we find
that about $50$ terms suffice for convergence without an excessive
computational cost. Finally, the closer to $|z|=0$, the fewer terms are
needed. This is an extension of what was found inq \cite{CaboBizet:2016qsa}.

\subsection{Results of the vacua exploration}
\label{sec:vacua}

Scanning flux configurations subject to the tadpole condition
\eqref{eq:tadpole} with $N_{\text{flux}}<10^3$, we found a large number
of extrema of the scalar potential $\hat V(\phi^l)$, all of them checked
for convergence. The solutions in the conifold patch are summarized in
Table~\ref{extremaex}.

\begin{table}[H]
	\resizebox{15cm}{!}{
	\centering
	\begin{tabular}{|c|c|c|c|c|c|c|c|c|c|c|c|c|c|c|}
        \hline
        F.C.&$\hat{V}_0$&$r_0$&$\theta_0$&${\tau_1}_0$&${\tau_2}_0$&$N_{\text{Flux}}$&$F_{1}$&$F_{2}$&$F_{3}$&$F_{4}$&$H_{1}$&$H_{2}$&$H_{3}$&$H_{4}$\\
		\hline
        \multicolumn{15}{|c|}{Saddle points} \\
        \hline
        1&6.31277& 0.2355& 6.42538& 6.27781& 7.51705& 50& 17& -1& 0& -1& -1& -1& 3& 0\\
        2& 237.237& 0.154808& 4.99586& -2.01316& 1.06073&
        4& -18& 7& -3& 8& -1& -17& 6& -3\\
        3& 18.3933& 0.415927& 0.755909& -1.83846& 2.25914& 37& 9& 6& -10& 1& 2& 4& 3& -1\\
        4&15.168& 0.447068& 3.04808& 0.929361& 2.12557& 7& 9& -5& -5& 3& 2& 7& 2& 0\\
        5&3.08835& 0.437325& 6.42051& 0.0906062& 4.31071&
        125& 8& -9& 5& -1& -7& 9& 9& -1\\
        \hline
        \multicolumn{15}{|c|}{Minima} \\
        \hline
		6&23.1649& 0.00089899& 0.19875& -0.517303& 1.20469& 39& -1& 2& 10& -2& -5& -9& -5& 1\\
        7&0& 0.0000195051& 9.75971& -0.0337254& 2.64354& 99& 9& -7& -6& 4& -1& -10& 8& 1 \\
        8&631.293& 0.449388& 4.33162& -0.282836& 5.8526& 57&
        16& -2& 3& -20& -17& -7& 9& -1\\
        9& 107.692& 0.00775718& 9.48181& -0.0706286& 1.40464& 87& 7& -2& 2& -3& -10& -19& 18& 1\\
        10&18.21& 0.0000412828& 3.6971& 0.0750008& 1.33977& 47& 5& -10& 4& 0& -3& -9& 9& 1\\
		\hline
	\end{tabular}
	}
	\captionsetup{margin=.5cm}
	\caption{Extrema in the conifold patch. Several hundred flux choices
	were studied numerically; solutions whose complex-structure modulus
	fell outside the convergence disc were discarded, as were those with
	non-perturbative string coupling $g_s>1$, although the latter can be
	relaxed by a flux rescaling, see the discussion below. Entry $7$ is
	a non-supersymmetric Minkowski vacuum, for which $\hat V_0=0$ and the
	fluxes are imaginary self-dual, so that supersymmetry is broken only
	along the K\"ahler directions. Notice that monodromies are relevant and that $\theta_0$ values go outside the branch $(0,2\pi)$.}
	\label{extremaex}
\end{table}

Several of these solutions sit at the edge of the perturbative regime.
The fluxes can, however, be rescaled as $\mathrm{F}\to\sigma  \mathrm{F}$
and $\mathrm{H}\to\xi \mathrm{H}$ with
$\sigma,\xi\in\mathbb{Z}$, under which the vev of the complex-structure
modulus is unchanged while
\begin{equation}
    g_s\to\frac{\xi}{\sigma}\,g_s\,,
    \qquad
    \hat V\to\sigma\xi\,\hat V\,,
    \qquad
    N_{\text{flux}}\to\sigma\xi\,N_{\text{flux}}\,,
    \label{eq:fluxrescaling}
\end{equation}
so that the shape of $\hat V$, and hence its topography, is preserved.
For instance, flux choice $6$ in Table~\ref{extremaex} has
$\tau_{2,0}=1.20469$, i.e. $g_s\simeq0.83$; a rescaling with $\sigma=2$
and $\xi=25$ brings it down to $g_s\simeq0.0664$ while leaving the
complex-structure vev untouched. Note, however, that
$N_{\text{flux}}$ is not invariant: for this example it grows from $39$
to $39\times50=1950$, which exceeds the bound $N_{\text{flux}}<10^3$
imposed by the tadpole condition. Weakening the string coupling by this
mechanism is therefore constrained by the tadpole, and the two
requirements must be imposed simultaneously.

\begin{figure}[H]
	\centering
	\includegraphics[width=1.02\linewidth]{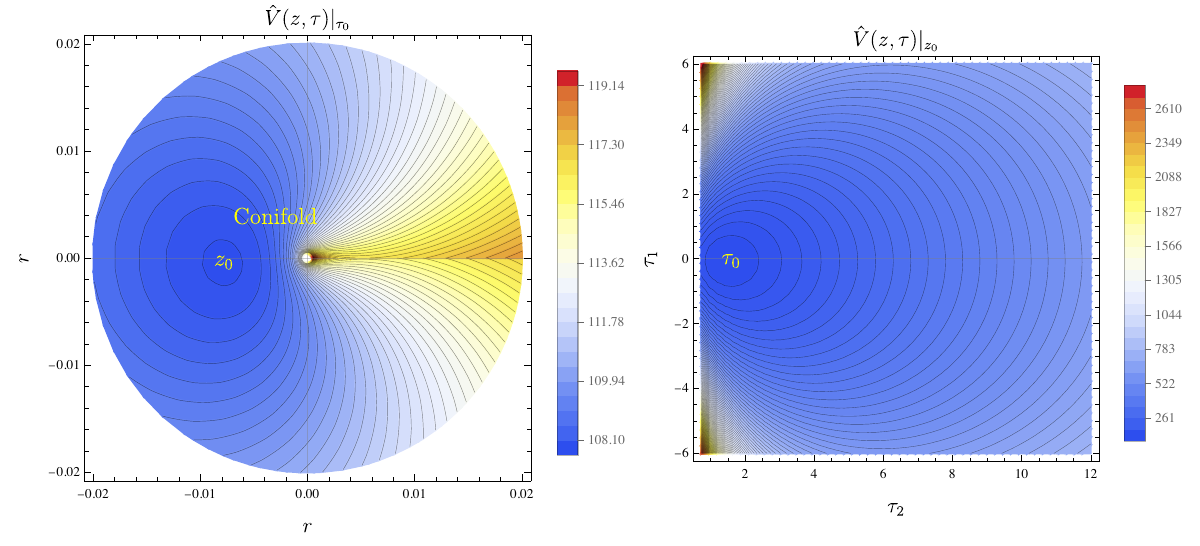}
	\captionsetup{margin=.5cm}
	\caption{The two-dimensional slices of the scalar potential \(\hat V(z,\tau)\) are shown around the minimum \((z_0,\tau_0)\). The flux choice corresponds to entry $9$ of
	Table~\ref{extremaex}. The left panel displays the potential in the complex structure modulus plane, \(z=r e^{i\theta}\), evaluated at the stabilized axio-dilaton value \(\tau=\tau_0\). The contour lines converge toward \(z_0\), indicating a local minimum in the complex-structure direction. The conifold point at \(z=0\) is also indicated, showing that the stabilized value \(z_0\) lies close to, but away from, the conifold singularity. The right panel shows the corresponding potential in the axio-dilaton plane, \(\tau=\tau_2+i\tau_1\), with the complex-structure modulus fixed at \(z=z_0\). The closed contours surrounding \(\tau_0\) demonstrate stabilization along both axio-dilaton directions. Taken together, the two panels illustrate that \((z_0,\tau_0)\) corresponds to a simultaneous local minimum of the potential in the complex-structure and axio-dilaton sectors.
    }
	\label{fig:contourLCS}
\end{figure}
To visualize the stabilization of the complex-structure modulus and axio-dilaton, we evaluate the scalar potential along two complementary two-dimensional slices through the critical point \((z_0,\tau_0)\). Fixing \(\tau=\tau_0\), the potential develops a local minimum at \(z=z_0\), located in the vicinity of the conifold point \(z=0\). Conversely, fixing \(z=z_0\), the potential in the axio-dilaton plane exhibits closed contours centered at \(\tau=\tau_0\). Therefore, these two slices provide a graphical representation of the simultaneous stabilization of the complex-structure and axio-dilaton moduli in \((z_0,\tau_0)\), Figure \ref{fig:contourLCS}. The non-circular and strongly distorted contour structure around \(z_0\) indicates an anisotropic behavior of the scalar potential in the complex structure plane, with the potential varying more rapidly along certain directions than others. This suggests different local curvature scales for fluctuations around the stabilized value. However, because $z$ is generally non-canonically normalized particularly in the vicinity of the conifold the contour structure alone cannot be directly interpreted as a hierarchy of physical masses. Such a hierarchy should instead be established from the eigenvalues of the appropriately normalized Hessian (mass matrix) evaluated at the vacuum \cite{CaboBizet:2024conifold}.

\begin{figure}[H]
	\centering
	\includegraphics[width=.9\linewidth]{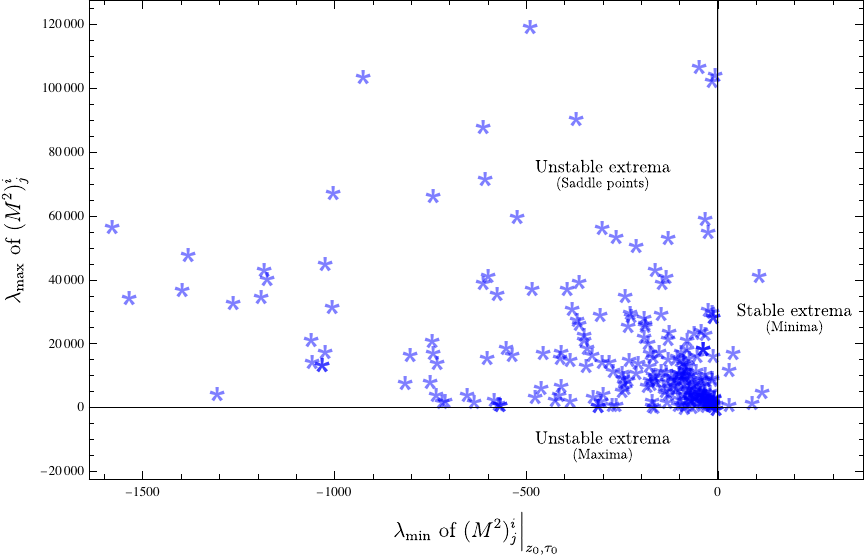}
	\captionsetup{margin=.5cm}
	\caption{Minimum eigenvalue $\lambda_{\text{min}}$ versus maximum
	eigenvalue $\lambda_{\text{max}}$ of the mass matrix evaluated at the
	extrema of $\hat V$. Minima populate the first quadrant
	($\lambda_{\text{min}}>0$, $\lambda_{\text{max}}>0$), unstable
	extrema the second ($\lambda_{\text{min}}<0$,
	$\lambda_{\text{max}}>0$) and maxima the third
	($\lambda_{\text{min}}<0$, $\lambda_{\text{max}}<0$). No maxima are
	found in our scan, which restricts the range of cosmological
	behaviours that can be realized at this stage of the stabilization.}
	\label{fig:vacuaauto}
\end{figure}

In this work the four equations \eqref{extre} were solved simultaneously
for $z$ and $\tau$.  The extrema found in both patches,
conifold and LCS, consist mainly of minima and saddle points, in
agreement with \cite{Johnson:2008vn}. Alternatively, one may first solve
$\partial_\tau \hat V=0$ for $\tau(z)$ and argue that, once $z$ is fixed,
the potential always develops a minimum along the axio-dilaton
directions, which in our observation can not be applied in the majority of the cases, as the axio-dilaton is usually lighter than the complex-structure \cite{CaboBizet:2024conifold}. We present a more detail discussion on Appendix \ref{axiodilatonHeavier}.

To study the stability of the solutions we define the mass matrix at an
extremum as
\begin{equation}
    (M^2)^{i}{}_{j}
    = \left. G^{il}\,\nabla_l\nabla_j \hat V\right|_{z_0,\tau_0}\,,
    \label{mass}
\end{equation}
where $\nabla$ is the covariant derivative in field space; at an extremum
the covariant and ordinary Hessians coincide, since
$\nabla_l\nabla_j\hat V=\partial_l\partial_j\hat V
-\Gamma^k_{lj}\partial_k\hat V$. The field-space metric read off from the
kinetic term of \eqref{S4}, in the conventions of \eqref{eq:Gmetric}, is
\begin{equation}
    G_{ij}= 2\operatorname{diag}
    \left( K_{z\bar{z}}\,,\;\; r^2K_{z\bar{z}}\,,\;\;
    \frac{1}{4\tau_2^{2}}\,,\;\; \frac{1}{4\tau_2^{2}}
    \right).
    \label{eq:Gdiag}
\end{equation}

Diagonalizing \eqref{mass} yields the mass eigenvalues and the
corresponding eigenstates,
\begin{equation}
    \phi'^{\,j}= \sum_i A^j{}_i\,\phi^i\,.
\end{equation}

\begin{figure}[H]
	\centering
	\includegraphics[width=1.\linewidth]{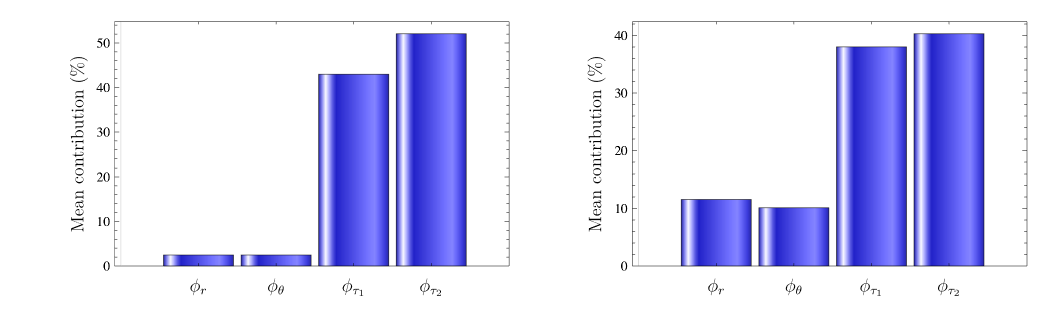}
	\caption{Percentage contribution of each field to the eigenvectors of
the mass matrix. The left panel shows the most unstable direction, that
is, the one with the smallest eigenvalue, and the right panel the two
most unstable directions. Over the $\sim 2*10^{2}$ convergent extrema of
Sec.~\ref{sec:vacua}, the axio-dilaton components $\tau_1$ and $\tau_2$
together dominate the most unstable direction.}
\label{fig:unstable}
\end{figure}

\begin{figure}[H]
	\centering
	\includegraphics[width=1\linewidth]{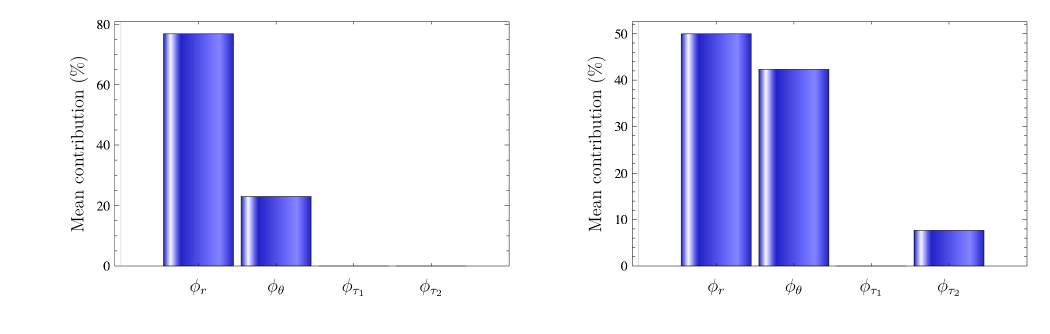}
	\caption{Percentage contribution of each field to the eigenvectors of
the mass matrix. The left panel shows the heaviest direction, that is,
the one with the largest eigenvalue, and the right panel the two
heaviest directions. The complex-structure components $r$ and $\theta$
dominate them, which complements Figure ~\ref{fig:unstable}: statistically,
the axio-dilaton accounts for the lightest modes and the complex
structure for the heaviest.}
\label{fig:heaviest}
\end{figure}

The results of the vacua exploration of $\hat{V}(\tau,z)$ are collected in
Appendix~\ref{app:vacua}, and their stability is illustrated in
Figure~\ref{fig:vacuaauto}. Almost all of the vacua found are unstable:
only  eight minima appear in the full sample. As for the directions along
which the instabilities develop, the leading unstable direction is
dominated by the dilaton $\tau_2$, while the two dominant unstable
directions also involve the axion $\tau_1$, so that the instability
resides in the axio-dilaton sector, as shown in
Figure~\ref{fig:unstable}. The complementary pattern holds for the stable
directions: as displayed in Figure~\ref{fig:heaviest}, the heaviest modes
of the mass matrix are dominated by the complex-structure components $r$
and $\theta$. The two sectors therefore separate statistically, with the
complex structure accounting for the heavy, stable directions and the
axio-dilaton for the light, unstable ones. We find no maxima of the
potential in our exploration: the mass matrix always possesses at least
one stable direction. This is in agreement with the observations in \cite{CaboBizet:2024conifold}.

\section{Refined de Sitter Conjecture}
\label{sec:rdsc}

In this section we explore the refined de Sitter conjecture (RdSC) for
type IIB string theory on a one-parameter Calabi--Yau manifold. We begin
by stating the conjecture.

\medskip
{\it A potential $V(\phi)$ for scalar fields $\phi$ in the low-energy
effective theory of any consistent theory of quantum gravity must
satisfy either}
\begin{equation}
	\left|\nabla V\right|\;\geq\; \frac{c}{M_{\rm Pl}}\, V\,,
	\label{dSC1}
\end{equation}
{\it or}
\begin{equation}
	\min\ {\rm eig}\left(\nabla_i \nabla_j V\right)
	\;\leq\;-\,\frac{c^\prime}{M_{\rm Pl}^2}\, V\,,
	\label{dSC2}
\end{equation}
{\it for some universal constants $c,c^\prime>0$ of order one, where the
left-hand side of \eqref{dSC2} is the minimum eigenvalue of the Hessian
in an orthonormal frame.}\footnote{Both the norm of the gradient and the
Hessian are defined with respect to the orthonormal frame read off from
the kinetic term $-\tfrac12 G_{ij}\partial_\mu\phi^i\partial^\mu\phi^j$,
with $G_{ij}=2K_{a\bar b}$ as fixed in Sec.~\ref{sec:setup},
Eq.~\eqref{eq:Gmetric}. In particular
$|\nabla V|^2=2K^{a\bar b}\partial_aV\,\partial_{\bar b}\overline{V}$.
Numerical values quoted below should be divided by $\sqrt2$ if one
instead adopts $G_{ij}=K_{a\bar b}$.}
\medskip

In essence, the two conditions \eqref{dSC1}--\eqref{dSC2} state that a
universe with a positive cosmological constant must be sufficiently
unstable. The RdSC is automatically satisfied for $V\leq 0$, while for
$V>0$ it forbids a positive local minimum, since at such a point the
gradient vanishes and every eigenvalue of the Hessian is positive.

This is precisely the situation encountered in Sec.~\ref{sec:vacua}. For
the mirror quintic our exploration of the extrema, in agreement with
other works that have studied this model \cite{Johnson:2008vn}, shows a
tendency of the scalar potential to develop saddle points and minima. As
established by the RdSC, minima with $V>0$ would be in conflict with the
conjecture. We show below that, at least at tree level, this tension is
resolved by the runaway direction of the K\"ahler moduli, and that the
resolution is in fact universal.

\subsection{The first condition at an extremum}
\label{sec:rdsc-first}

As discussed above, the axio-dilaton and the complex-structure modulus
are stabilized through \eqref{extre}, so that
$\nabla_{\rm cs+ad}\hat V=0$ at the vacua we consider, where
$\nabla_{\rm cs+ad}$ denotes the gradient with respect to those
directions. Since the tree-level K\"ahler potential splits as
$K=K_{\rm ad}(\tau)+K_{\rm cs}(z)+K_{\rm k}(\rho)$, the moduli-space
metric is block diagonal and the two sectors do not mix. The whole
gradient of the full potential therefore comes from the K\"ahler
directions, and the first condition \eqref{dSC1} reduces to
\begin{equation}
	\left|\nabla_{\rm k} V\right|\;\geq\; c\, V\,,
	\label{eq:rdsc-kahler}
\end{equation}
in reduced Planck units, with $\nabla_{\rm k}$ the gradient with respect
to the K\"ahler moduli $\rho^l$.

The contribution of the K\"ahler moduli to the scalar potential enters
only through the dimensionless volume,
\begin{equation}
K_{\rm k}=-2\ln \mathcal{V}(\rho^l)\,,
\label{eq:Kk-def}
\end{equation}
so that the slice of the scalar potential along the $z$ and $\tau$
directions is rescaled by the volume,
\begin{equation}
V= e^{K_{\rm k}}\,\hat{V}(z,\tau)=\frac{\hat{V}(z,\tau)}{\mathcal{V}^2}\,,
\label{eq:Vkahler}
\end{equation}
so that at a critical point of $\hat V$ the prefactor
$\hat V_0\equiv\hat V(z_0,\tau_0)$ is a constant and the only dependence
on $\rho^l$ is through $e^{K_{\rm k}}$. Consequently
\begin{equation}
    \partial_{\rho^l} V = V\,\partial_{\rho^l}K_{\rm k}\,,
\end{equation}
and therefore
\begin{equation}
    \frac{\left|\nabla V\right|^2}{V^2}\bigg|_{z_0,\tau_0}
    =2\,K^{l\bar m}\,
    \partial_{\rho^l}K_{\rm k}\,\partial_{\bar\rho^m}K_{\rm k}\,.
    \label{eq:ratio-noscale}
\end{equation}

The K\"ahler metric of the K\"ahler-moduli sector reads
\begin{equation}
    G_{\rho^l \bar{\rho}^m}
    = \partial_{\rho^l}\partial_{\bar{\rho}^m}K_{\rm k}
    =-2\,\frac{\mathcal{V}\, \partial_{\rho^l}\partial_{\bar{\rho}^m}\mathcal{V}
    - \partial_{\rho^l}\mathcal{V}\, \partial_{\bar{\rho}^m}\mathcal{V}}
    {\mathcal{V}^2}\,,
    \label{eq:Gkahler}
\end{equation}
and positivity of \eqref{eq:Gkahler} requires in particular
\begin{equation}
\partial_{\rho^l}\mathcal{V}\, \partial_{\bar{\rho}^l}\mathcal{V}
>\mathcal{V}\, \partial_{\rho^l}\partial_{\bar{\rho}^l}\mathcal{V}
\qquad (\text{no sum})\,,
\label{gposit}
\end{equation}
which is necessary, though not by itself sufficient, for the metric to be
positive definite. Whenever \eqref{eq:Gkahler} is diagonal, substituting
it into \eqref{eq:ratio-noscale} gives the ratio explicitly in terms of
the volume and its derivatives,
\begin{equation}
	\frac{\left| \nabla_{\rm k} V\right|}{V}
	=\sqrt{\,\sum_{l=1}^{n}
	\frac{-4\,\partial_{\rho^l}\mathcal{V}\,\partial_{\bar\rho^l}\mathcal{V}}
	{\mathcal{V}\, \partial_{\rho^l}\partial_{\bar{\rho}^l}\mathcal{V}
	- \partial_{\rho^l}\mathcal{V}\, \partial_{\bar{\rho}^l}\mathcal{V}}}\,,
	\label{eq:ratio-explicit}
\end{equation}
where the denominator is negative by \eqref{gposit}, so that the
radicand is positive. Note that \eqref{eq:ratio-explicit} assumes a
diagonal metric; the general case requires the full inverse of
\eqref{eq:Gkahler}, and we return to this point in
Sec.~\ref{sec:rdsc-examples}.

The contraction on the right-hand side is nothing but the left-hand side
of the no-scale condition \eqref{nsc}, which for a K\"ahler potential
$K_{\rm k}=-2\ln\mathcal{V}$ with $\mathcal{V}$ homogeneous of degree
$3/2$ in the variables $u^l=-i(\rho^l-\bar\rho^l)$ takes the value
$K^{l\bar m}\partial_lK_{\rm k}\partial_{\bar m}K_{\rm k}=3$. We
therefore obtain the universal result
\begin{equation}
    \boxed{\;
    \frac{\left|\nabla V\right|}{V}\bigg|_{z_0,\tau_0}=\sqrt{6}\;}
    \label{eq:universal}
\end{equation}
independently of the number of K\"ahler moduli, of the detailed form of
the volume, and of the location of the extremum in the
complex-structure moduli space. The first condition \eqref{dSC1} of the
RdSC is thus satisfied at every extremum of $\hat V$, for any
$c\leq\sqrt6$. 

The identity underlying \eqref{eq:universal} is not new. The
contraction $K^{T\bar T}K_TK_{\bar T}=3$ is the defining property of the
no-scale models of \cite{Cremmer:1983bb}, and the corresponding
exponential runaway of the volume, with rate $\sqrt6$ in four
dimensions, is the reason why type IIB flux potentials were early
recognized to be too steep to drive accelerated expansion
\cite{Hellerman:2001yi,Fischler:2001yj}. The value was obtained
explicitly for the isotropic toroidal GKP model in
\cite{CaboBizet:2019sgr}, appears as the asymptotic exponent along the
volume direction in \cite{Casas:2024LightTowers}, and in the asymptotic
analysis of \cite{CalderonInfante:2022AAE}, Sec.~5.1, the same
contraction is shown to fix the K\"ahler contribution to the de Sitter
coefficient, giving $|\nabla V|/V\geq\sqrt{2n}$ with $n=3$ in type IIB
at large volume. Our result differs in three respects. The bound is
saturated: at a critical point of $\hat V$ the complex-structure and
axio-dilaton contributions vanish, so that $|\nabla V|/V=\sqrt6$ holds
as an equality, in the interior of moduli space and with exact periods
rather than asymptotically. The value of $n$ is explained: by
\eqref{eq:euler} it is the degree of homogeneity of $\mathcal{V}$,
equal to $3$ for any volume cubic in the two-cycle sizes, which is also
what removes the residual $-(3-n)|W|^2$ term and makes the potential
no-scale. And the role of $\sqrt6$ is reversed: for
\cite{CalderonInfante:2022AAE} it is an obstruction to asymptotic
acceleration, whereas here it is the reason why the critical points of
$\hat V$ are not de Sitter vacua and the first RdSC condition holds
automatically. We note that \cite{CalderonInfante:2022AAE} identifies
the complex structure being the lighter sector as the requirement for
it to run with the K\"ahler moduli fixed; the mass spectrum of
Sec.~\ref{sec:vacua}, where it is typically the heavier one, provides a
data point on that question for the present class of models.

It is worth emphasizing what does and does not enter this argument. No
assumption is made about where $z_0$ and $\tau_0$ sit: the extremum may
lie near the conifold, near the large-complex-structure point, or
anywhere in between, and \eqref{eq:universal} holds equally. The only
input is that the K\"ahler sector be exactly no-scale, that is, that
$\mathcal{V}$ be homogeneous of degree $3/2$. This is the same
large-volume, tree-level assumption already made in writing
\eqref{eq:Vkahler}, and it is where the result would be modified: once
$\alpha^\prime$ or worldsheet-instanton corrections spoil the
homogeneity of $\mathcal{V}$, the quotient \eqref{eq:universal} acquires
a moduli dependence.

This should be contrasted with the situation in
Sec.~\ref{sec:tcc}. The asymptotic bound imposed by the TCC is a
statement about infinite-distance limits in moduli space, and can only
be tested in the corresponding asymptotic regions, such as the
neighbourhood of the LCS point. The RdSC, by contrast, is a pointwise
statement, and \eqref{eq:universal} shows that it is met at every
critical point of the flux potential, well inside the moduli space.

\subsection{Examples of K\"ahler moduli dependence}
\label{sec:rdsc-examples}

By \eqref{eq:euler}, the value \eqref{eq:universal} follows from the
homogeneity of $\mathcal{V}$ alone, so that no case-by-case verification
is needed. It is nevertheless instructive to see what each of the
standard parametrizations of the volume does, and does not, add.

\paragraph{Single K\"ahler modulus.} Under the large-volume assumption
the familiar one-modulus dependence is that of \eqref{eq:Kk}, with
$\mathcal{V}=u^{3/2}$ and $K_{\rm k}=-3\ln u$. This is the simplest
homogeneous case, the metric is trivially diagonal, and
$K^{\rho\bar\rho}|\partial_\rho K_{\rm k}|^2=3$, so that the first bound
of the RdSC is satisfied provided
\begin{equation}
c \leq \sqrt{6}\,,
\label{eq:cbound}
\end{equation}
in agreement with \eqref{eq:universal}.

\paragraph{Several K\"ahler moduli.} For a generic Calabi--Yau whose
volume is approximated by the swiss-cheese form of the Large Volume
Scenario \cite{Balasubramanian:2005LVS,Conlon:2005LVSspectrum},
\begin{equation}
	\mathcal{V}=\frac{\alpha}{2\sqrt{2}}
	\left(\left(-i(\rho-\overline{\rho})\right)^{3/2}
	- \sum_i \lambda_i
	\left(-i(\rho_i-\overline{\rho}_i)\right)^{3/2} \right),
	\label{eq:Kk-swiss}
\end{equation}
the volume is again homogeneous of degree $3/2$, so that the quotient is
unchanged, for any number of small moduli and any values of the
$\lambda_i$. What this case does add is a warning about the structure of
the metric. Although $\mathcal{V}$ is a sum of decoupled terms, so that
$\partial_{\rho^l}\partial_{\bar\rho^m}\mathcal{V}=0$ for $l\neq m$, the
second term of \eqref{eq:Gkahler} still contributes off the diagonal,
\begin{equation}
    G_{\rho^l\bar\rho^m}\Big|_{l\neq m}
    =\frac{2\,\partial_{\rho^l}\mathcal{V}\,
    \partial_{\bar\rho^m}\mathcal{V}}{\mathcal{V}^2}
    \;\neq\;0\,,
    \label{eq:offdiag}
\end{equation}
so that the K\"ahler metric is {\it not} diagonal and the simplified
expression \eqref{eq:ratio-explicit} does not apply. This is easy to
overlook, and it is worth quantifying the error it produces. For a single
additional modulus $\hat\rho$,
\begin{equation}
	\mathcal{V}=\frac{\alpha}{2\sqrt{2}}
	\left(\left(-i(\rho-\overline{\rho})\right)^{3/2}
	- \lambda\left(-i(\hat{\rho}-\overline{\hat{\rho}})\right)^{3/2}
	\right),
	\label{eq:Kk-two}
\end{equation}
the diagonal expression \eqref{eq:ratio-explicit} would give
\begin{equation}
     \frac{\left| \nabla V\right|}{V}\bigg|_{z_0,\tau_0}
     \overset{\text{diag.}}{=}\sqrt{6}\,
     \sqrt{\frac{ \operatorname{Im}\rho^3
     +4\lambda{\operatorname{Im}\rho}^{3/2}\operatorname{Im}\hat{\rho}^{3/2}
     +\lambda^2\operatorname{Im}\hat{\rho}^3 }
     {2\operatorname{Im}\rho^3
     +5\lambda{\operatorname{Im}\rho}^{3/2}\operatorname{Im}\hat{\rho}^{3/2}
     +2\lambda^2\operatorname{Im}\hat{\rho}^3 }}\,,
     \label{eq:ratio-two}
\end{equation}
which retains a spurious dependence on the K\"ahler moduli. It reduces to
\eqref{eq:universal} in the degenerate limits $\lambda\rightarrow0$ and
$\operatorname{Im}\hat\rho\rightarrow0$, in which the second modulus
decouples and the metric does become diagonal, but away from them it
departs from the universal value, by $16\%$ at
$\operatorname{Im}\rho^{3/2}=\lambda\operatorname{Im}\hat\rho^{3/2}$,
where it gives $2$ instead of $\sqrt{6}$. Inverting the full metric
restores the constant \eqref{eq:universal}, as \eqref{eq:euler}
guarantees it must.

\paragraph{Anisotropic volume.} The most informative case is a volume of
product type, of the kind arising in anisotropic compactifications
\cite{Cicoli:2011anisotropic},
\begin{equation}
	\mathcal{V}=\sqrt{\prod_{l=1}^{n}
	\left(-i(\rho^l-\bar{\rho}^l)\right)^{a_l}}\,,
	\qquad
	K_{\rm k}=-\sum_{l=1}^{n} a_l\ln u^l\,,
	\label{eq:Kk-aniso}
\end{equation}
because here the degree of homogeneity is left free,
$\tfrac{1}{2}\sum_l a_l$, and the identity \eqref{eq:euler} can be seen
at work. The metric is diagonal, $K_{l\bar l}=a_l/(u^l)^2$, so that
\eqref{eq:ratio-explicit} applies directly and gives
\begin{equation}
   \frac{\left|\nabla V\right|}{V}\bigg|_{z_0,\tau_0}
   =\sqrt{2\sum_{l} a_l}\,,
   \label{eq:aniso}
\end{equation}
which is precisely $\sqrt{2\cdot 2n}$ with $n=\tfrac12\sum_l a_l$, in
agreement with \eqref{eq:euler}. Imposing the no-scale condition, that is
$2n=3$, therefore amounts to $\sum_l a_l=3$, and \eqref{eq:aniso} reduces
once more to \eqref{eq:universal}. The anisotropy, encoded in the
individual $a_l$, drops out of the result: only their sum enters, and it
is fixed. This case thus makes explicit that the universal value is a
statement about the degree of homogeneity of the volume and about nothing else.

\subsection{Discussion}
\label{sec:rdsc-discussion}

Let us now discuss the critical points of the partial potential
$\hat{V}(z,\tau)$ presented in Figure~\ref{fig:vacuaauto} and
Appendix~\ref{app:vacua}, from which one can estimate the constant $c'$
of the second condition \eqref{dSC2}. Requiring the bound to hold for
every critical point found, that is taking the most restrictive value
over the sample, gives $c'=0.18$. Alternatively, one may fit
$\min\,\mathrm{eig}(\nabla_i\nabla_jV)$ against $V$ over the sample; this
procedure, however, yields a mean value with a standard deviation two
orders of magnitude larger than the mean itself, and therefore does not
allow $c'$ to be determined. The two estimates are not directly
comparable, since the former is a bound satisfied by the whole sample
while the latter would be an average slope; we quote the former.

The results above show that a no-scale scenario in which $z$ and $\tau$
are stabilized at a minimum of $\hat V$ is not in conflict with the first
condition of the RdSC: although such a configuration is a minimum of
$\hat V$, it is not a critical point of the full potential, which retains
the K\"ahler runaway, and the corresponding logarithmic slope is the
order-one number \eqref{eq:universal}.\footnote{We thank Susha
Parameswaran for the remark that the K\"ahler derivatives should be
computed in covariant form.} In general the conjecture does not fix the
value of $c$; the TCC, however, singles out
$c=2/\sqrt{(d-1)(d-2)}\to\sqrt{2/3}$ in $d=4$ as an asymptotic bound,
see Sec.~\ref{sec:tcc}, and \eqref{eq:universal} comfortably exceeds it.

We close with a caveat. The analysis above concerns the first condition
\eqref{dSC1} only. The second condition \eqref{dSC2} involves the
Hessian of the full potential, including the mixed K\"ahler and
complex-structure directions, and its evaluation requires a numerical
treatment that we do not attempt analytically here. We note, however, that the two
conditions are alternatives: since \eqref{dSC1} is satisfied at every
extremum by \eqref{eq:universal}, the RdSC is met regardless of the
behaviour of the Hessian.

\section{Trans-Planckian Censorship Conjecture}
\label{sec:tcc}

Inflationary cosmology has been remarkably successful in accounting for
the structure of the universe on large scales: it explains the observed
inhomogeneities as the late-time imprint of quantum fluctuations
generated in the early universe; see \cite{Brandenberger:2003vk} for a
review of cosmological perturbations. There is, however, a well-known
difficulty associated with the redshifting of these primordial
fluctuations. Tracing the modes observed today backwards through a
sufficiently long period of exponential expansion, their wavelengths
become sub-Planckian, so that observable predictions would depend on
physics lying beyond the regime of validity of the effective
description. This is the so-called {\it trans-Planckian problem}.

Motivated mainly by this issue, Bedroya and Vafa proposed the
{\it Trans-Planckian Censorship Conjecture} (TCC) \cite{Bedroya:2019snp},
which states that a field theory consistent with a quantum theory of
gravity cannot give rise to a cosmological expansion in which
perturbations with length scales larger than the Hubble radius can be
traced back to trans-Planckian scales at an earlier time. This is
encoded in the inequality
\begin{equation}
	\frac{a_f}{a_i}\,\ell_{p} \;<\; \frac{1}{H_f}\,,
	\label{eq:tcc-original}
\end{equation}
where $a_i$ and $a_f$ are the cosmological scale
factors\footnote{For simplicity we assume a flat, homogeneous and
isotropic background cosmology, described by the four-dimensional metric
$ds^2=-dt^2+a(t)^2d\mathbf{x}^2$, where $\mathbf{x}$ are the spatial
comoving coordinates.} at the initial and final times, $H$ is the Hubble
expansion rate and $\ell_{p}$ the Planck length. An effective field
theory that leads to the trans-Planckian problem is not consistent with
quantum gravity and belongs to the swampland. Implications of the TCC for
inflationary cosmology have been studied in
\cite{Bedroya:2019tba,Brandenberger:2021pzy}, where an energy bound is
obtained for inflation to provide a suitable scenario for cosmic
structure formation.

As was stated in \cite{Bedroya:2019snp}, in the asymptotic regions of
moduli space the TCC places a bound on the scalar potential, quoted in
\eqref{eq:tcc} above. Let us now make that statement precise. Consider a
trajectory in moduli space parametrized by the affine parameter $s$,
along which the scalar fields approach an infinite-distance limit, and
define
\begin{equation}
V'(\phi(s))
\;\equiv\;
\frac{dV(\phi(s))}{ds}
\;=\;
\frac{d\phi^i}{ds}\,\partial_i V\,,
\qquad
G_{ij}\,\frac{d\phi^i}{ds}\frac{d\phi^j}{ds}=1\,,
\label{eq:Vprime}
\end{equation}
the second equation being the condition that fixes $s$ to be an affine
parameter. Thus $-V'/V$ measures the logarithmic decay rate of the
potential in the direction of motion. For a decreasing positive potential
one has $V'<0$ and $V>0$, hence $-V'/V=|V'|/V$.

In practice we integrate the cosmological evolution in cosmic time and
then convert to the affine parameter by means of
\begin{equation}
\frac{dt}{ds}
=
\frac{1}{\sqrt{G_{ij}\,\dot\phi^{i}\dot\phi^{j}}}\,,
\label{eq:dtds}
\end{equation}
which, integrated numerically, gives $t(s)$ and therefore the potential
along the trajectory, $V(t(s))$.

The asymptotic quotient appearing in \eqref{eq:tcc} is then defined as
\begin{equation}
\left(\frac{|V'|}{V}\right)_{\infty}
=
\liminf_{s_i\rightarrow\infty}\;
\liminf_{s_f\rightarrow\infty}\;
\left\langle
-\frac{V'(\phi(s))}{V(\phi(s))}
\right\rangle_{[s_i,s_f]},
\qquad
\left\langle
-\frac{V'}{V}
\right\rangle_{[s_i,s_f]}
=
\frac{1}{s_f-s_i}
\int_{s_i}^{s_f}\!\! ds\,
\left(-\frac{V'}{V}\right).
\label{eq:tcc-liminf}
\end{equation}
The parameters $s_i,s_f$ denote the values of the affine parameter $s$ at the initial and final points of the trajectory. The use of the $\liminf$ allows for local oscillations or
non-uniform behaviour of the potential, while requiring that its
asymptotic average decay rate not fall below the TCC value. The limits
are nested as in \cite{Bedroya:2019snp}: the inner one is taken at fixed
$s_i$, and the outer one then pushes the starting point of the trajectory
towards the asymptotic region.

In the multifield case one must distinguish two quantities. Following
\cite{Bedroya:2019snp}, let $\{\gamma_\alpha\}$ be a family of causal
paths, that is solutions of the equations of motion with sub-Planckian
initial conditions, which start in the interior of moduli space and
collectively cover its asymptotic region; each $\alpha$ is called an
asymptotic direction. One defines the directional quotient
\begin{equation}
    \left(\frac{|\nabla_I V|}{V}\right)_{\alpha}
    :=
    \liminf_{s_{\alpha,i}\rightarrow\infty}\;
    \liminf_{s_{\alpha,f}\rightarrow\infty}\;
    \left\langle
    \frac{\left|\partial_{s_\alpha}V(\gamma(s_\alpha))\right|}
    {V(\gamma(s_\alpha))}
    \right\rangle_{[s_{\alpha,i},s_{\alpha,f}]},
    \label{eq:tcc-directional}
\end{equation}
which measures $|V'|/V$ along the direction $\alpha$, and the quotient
built from the norm of the full gradient,
\begin{equation}
    \left(\frac{|\nabla V|}{V}\right)_{\alpha}
    :=
    \liminf_{s_{\alpha,i}\rightarrow\infty}\;
    \liminf_{s_{\alpha,f}\rightarrow\infty}\;
    \left\langle
    \frac{\left|\nabla V(\gamma(s_\alpha))\right|}
    {V(\gamma(s_\alpha))}
    \right\rangle_{[s_{\alpha,i},s_{\alpha,f}]}\,.
    \label{eq:tcc-gradient}
\end{equation}
Since the directional derivative is a projection of the gradient, these
satisfy
$\left(|\nabla V|/V\right)_{\alpha}\geq
\left(|\nabla_I V|/V\right)_{\alpha}$, and the bound derived in
\cite{Bedroya:2019snp} holds for both,
\begin{equation}
    \left(\frac{|\nabla V|}{V}\right)_{\alpha}
    \;\geq\;
    \left(\frac{|\nabla_I V|}{V}\right)_{\alpha}
    \;\geq\;
    \frac{2}{\sqrt{(d-1)(d-2)}}\,.
    \label{eq:tcc-both}
\end{equation}
It is the quantity \eqref{eq:tcc-gradient}, built from the norm of the
gradient, that we evaluate in the asymptotic regions analyzed below.
Equivalently, the condition can be written in exponential form as
\begin{equation}
V(s)\;<\;A\,
\exp\!\left[-\frac{2}{\sqrt{(d-1)(d-2)}}\;\Delta\phi_{\rm geo}\right],
\label{eq:tcc-exp}
\end{equation}
with $A$ a positive constant and $\Delta\phi_{\rm geo}$ the geodesic
distance in field space between the initial point and $\phi(s)$.

The TCC has interesting implications for low-energy physics, and we are
interested in particular in what it implies for no-scale scalar
potentials. As pointed out in Sec.~\ref{sec:rdsc}, we may study the slice
$\hat{V}(\phi^i)$ of the scalar potential at fixed K\"ahler moduli, under
the assumption that the remaining K\"ahler directions are stabilized by
non-perturbative corrections \cite{KKLT}. 
If instead the runaway dependence on the K\"ahler moduli is retained, the
full potential $V(\phi^l)$ does not fail the first RdSC condition.
Moreover, our numerical results (see Appendix \ref{app:vacua}) for the critical points indicate that, for this
model, the extrema of $\hat{V}(\phi^l)$ are never maxima, but only minima
or saddle points.

We now turn to the explicit analysis. The effective theory under
consideration contains four real scalar fields,
$\phi^i=\{r,\theta,\tau_1,\tau_2\}$. In the following subsections we
examine in turn the asymptotic directions of moduli space parametrized by
each of them separately, and then a genuinely multifield trajectory,
parametrized by the affine parameter $s$, which corresponds to a mixed
direction in the $(\tau_1,\tau_2)$ plane. The latter allows us to clarify the behaviour of the directional slope along the purely axionic direction, which decays even though the norm of the gradient remains bounded below.

\subsection{Large Complex Structure leading terms}
\label{sec:tcc-lcs}

Let us analyze the behavior of the scalar potential in the asymptotic
region of complex-structure moduli space. Among the singular points of
this moduli space, the large complex structure (LCS) point is the only
one located at infinite distance. In the vicinity of the LCS point, the
periods admit the expansion \cite{Candelas:1990rm,Hosono:1993qy,Morrison:1991cd}
\begin{equation}
\Pi_{\text{\tiny LCS}}(z)\;\sim\;
\begin{pmatrix}
i a_1+i a_2\log z+i a_3\log^3 z\\[1mm]
b_1+i b_2\log z+b_3\log^2 z\\[1mm]
c_1\\[1mm]
i d_1\log z
\end{pmatrix}.
\label{eq:PiLCS}
\end{equation}

The coefficients in \eqref{eq:PiLCS} are fixed by the topological data of
the Calabi--Yau manifold and by the choice of period basis. In the
symplectic basis of Sec.~\ref{sec:setup} they are all {\it real}, the
factors of $i$ having been displayed explicitly. Evaluating $\Pi_{\text{\tiny LCS}}
=\mathrm{M}_{\text{\tiny LCS}\rightarrow\text{\tiny S}}\,
\pi_{\text{\tiny LCS}}$ at leading order, where
$w^0_{\text{\tiny LCS}}\to1$ and $w^{1,2,3}_{\text{\tiny LCS}}\to0$, so
that $\pi_{\text{\tiny LCS}}\to
(1,\,\log z,\,\tfrac52\log^2 z,\,\tfrac56\log^3 z)$, one finds for the
coefficients that will be relevant below
\begin{equation}
    c_1=1\,,\qquad
    d_1=-\frac{1}{2\pi}\,,\qquad
    b_3=\frac{\kappa}{8\pi^2}\,,\qquad
    a_3=\frac{\kappa}{48\pi^3}\,,
    \label{eq:LCScoeffs}
\end{equation}
with $\kappa$ the triple intersection number of the mirror
manifold,\footnote{For the quintic $\kappa=5$, and the remaining
coefficients read $a_1=\chi\zeta(3)/(8\pi^3)=-25\zeta(3)/\pi^3$,
$a_2=-25/(24\pi)$, $b_1=25/12$ and $b_2=11/(4\pi)$, all of them real.}
so that in particular
\begin{equation}
    b_3\,d_1=-\frac{\kappa}{16\pi^3}=-3\,a_3c_1\,.
    \label{eq:LCSrelation}
\end{equation}

Using these asymptotic expansions one can compute the superpotential,
\begin{equation}
    W=\mathrm{G}^T\Sigma \Pi \sim -(F_3-\tau H_3)\,\Pi_1\,,
    \label{eq:WLCS}
\end{equation}
and, together with the K\"ahler potential, obtain the leading behavior of
the scalar potential,
\begin{equation}
    V=e^{K_{\text{k}}}\left[
    -\frac{a_3^2(9a_3c_1-5b_3d_1)
    \left(F_3^2-2F_3H_3\tau_1+H_3^2\tau_1^2+H_3^2\tau_2^2\right)}
    {2(a_3c_1-b_3d_1)(9a_3c_1-b_3d_1)\tau_2}
    \,y^3+\mathcal{O}(y^2)
    \right],
    \label{eq:VLCS}
\end{equation}
where the flux bracket is $\left|F_3-\tau H_3\right|^2$. Including the
next-to-leading contribution, the potential takes the form
\begin{equation}
\begin{aligned}
    V\sim e^{K_{\text{k}}} \left[
    -\frac{a_3^2(9a_3c_1-5b_3d_1)
    \left(F_3^2-2F_3H_3\tau_1+H_3^2\tau_1^2+H_3^2\tau_2^2\right)}
    {2(a_3c_1-b_3d_1)(9a_3c_1-b_3d_1)\tau_2}
    \,y^3
    \right.&\\[1mm]
    \left.
    +\,\frac{a_3 b_3(3a_3c_1+b_3d_1)(F_4H_3-F_3H_4)}
    {(9a_3c_1-b_3d_1)(a_3c_1-b_3d_1)}
    \,y^2+\mathcal{O}(y)
    \right]&,
\end{aligned}
\label{eq:VLCSnlo}
\end{equation}
where we used the identity
$9a_3^2c_1^2-10a_3b_3c_1d_1+b_3^2d_1^2
=(9a_3c_1-b_3d_1)(a_3c_1-b_3d_1)$.
The corresponding norm of the gradient behaves as
\begin{equation}
\begin{aligned}
    \left|\nabla V\right|^2\sim e^{2K_{\text{k}}} \left[
    \frac{a_3^4(9a_3c_1-5b_3d_1)^3
    \left(F_3^2-2F_3H_3\tau_1+H_3^2\tau_1^2+H_3^2\tau_2^2\right)^2}
    {2(a_3c_1-b_3d_1)^2(9a_3c_1-b_3d_1)^3\,{\tau_2}^2}
    \,y^6
    +\mathcal{O}(y^5)
    \right]&.
\end{aligned}
\label{eq:gradVLCS}
\end{equation}

Combining \eqref{eq:VLCS} and \eqref{eq:gradVLCS}, all the flux
dependence cancels and one obtains the quotient
\begin{equation}
    \frac{\left|\nabla V\right|}{V}
    =
    \sqrt{2}\,
    \sqrt{\frac{9a_3c_1-5b_3d_1}{9a_3c_1-b_3d_1}}\;
    \frac{|y|^3}{-\,y^3}\,,
    \label{eq:ratioLCS}
\end{equation}
where we have defined $y=\log r$. Approaching the LCS region corresponds
to $r\to0$, and therefore $y\to-\infty$, so that $|y|^3/(-y^3)=1$.
Substituting the relation \eqref{eq:LCSrelation} and writing
$A\equiv a_3c_1$, the ratio inside the square root becomes
\begin{equation}
    \frac{9a_3c_1-5b_3d_1}{9a_3c_1-b_3d_1}
    =\frac{9A+15A}{9A+3A}=2\,,
\end{equation}
in which the triple intersection number cancels. The gradient-to-potential
ratio therefore approaches a fixed value of order one,
\begin{equation}
    \frac{\left|\nabla V\right|}{V}\bigg|_{\text{\tiny LCS}}=2\,,
    \label{eq:ratioLCSvalue}
\end{equation}
and, since \eqref{eq:LCScoeffs} depends on the geometry only through
$\kappa$, which drops out, this value is the same for every
one-parameter model.

Although this result was also checked numerically, the asymptotic
expression shows analytically that the TCC condition is satisfied for the
flux potentials considered in one-parameter Calabi--Yau
compactifications.

\subsection{Conifold leading terms}
\label{sec:tcc-conifold}

In contrast with the large complex structure point, the conifold is a
finite-distance singularity in moduli space. The asymptotic form of the
Trans-Planckian Censorship bound therefore does not apply there, and one
does not expect the quotient $|\nabla V|/V$ to approach a universal
bounded value analogous to the one obtained in the LCS limit. This can be
seen explicitly from the leading behavior of the periods near the
conifold point.

Let $z_{\text{\tiny C}}$ denote a local coordinate that vanishes at the
conifold singularity, and write
\begin{equation}
    r=\left|z_{\text{\tiny C}}\right|\,,
    \qquad
    y=\log r\,,
\end{equation}
so that the conifold is approached as $r\to0$, that is $y\to-\infty$. The
leading terms of the period vector are
\begin{equation}
  \Pi_{\text{\tiny C}}
  \sim
  \begin{pmatrix}
  i a_1 z_{\text{\tiny C}} \\
  b_1+i b_2+(b_3+i b_4)z_{\text{\tiny C}} \\
  c_1+c_2 z_{\text{\tiny C}}\log z_{\text{\tiny C}} \\
  i d_1+i d_2 z_{\text{\tiny C}}
  \end{pmatrix}.
  \label{eq:PiC}
\end{equation}
The logarithmic term is associated with the vanishing cycle and
characterizes the local monodromy around the conifold point. The
coefficients $a_i$, $b_i$, $c_i$ and $d_i$ depend on the geometry and on
the choice of period basis, but they are not all independent. Indeed,
transporting \eqref{eq:PiC} around the conifold,
$z_{\text{\tiny C}}\to e^{2\pi i}z_{\text{\tiny C}}$, shifts
$\log z_{\text{\tiny C}}$ by $2\pi i$ and therefore sends
$\Pi_3\to\Pi_3+2\pi i\,c_2 z_{\text{\tiny C}}$, whereas the monodromy
matrix $\mathrm{T}_{\text{\tiny C}}$ of Eq.~\eqref{eq:Tconifold} acts by
shifting $\Pi_3$ by one unit of $\Pi_1$. Comparing the two expressions
fixes the modulus of the ratio, and the sign is then determined by
requiring a positive moduli-space metric,
\begin{equation}
    \left|\frac{c_2}{a_1}\right|=\frac{1}{2\pi}\,,
    \qquad\text{and positivity of }K_{z\bar z}\text{ fixes}\qquad
    \frac{c_2}{a_1}=+\frac{1}{2\pi}\,,
    \label{eq:conifold-relation}
\end{equation}
which is the conifold counterpart of the relation
\eqref{eq:LCSrelation} used at large complex structure.

Using these leading terms, the scalar potential behaves as
\begin{equation}
    V
    \sim
    e^{K_{\text{k}}}\left[
    -\,\frac{c_2\left(F_1^2-2F_1H_1\tau_1
    +H_1^2(\tau_1^2+\tau_2^2)\right)}
    {4a_1\tau_2}\,y
    +\mathcal{O}(y^0)\right],
    \label{eq:VC}
\end{equation}
where the flux bracket is $\left|F_1-\tau H_1\right|^2$, the fluxes
$F_1$ and $H_1$ being the ones that thread the cycle dual to the
vanishing one, as discussed in Sec.~\ref{sec:quinticgeometry}. Inserting
\eqref{eq:conifold-relation}, the leading term becomes completely
explicit,
\begin{equation}
    V
    \sim
    -\,e^{K_{\text{k}}}\,
    \frac{\left|F_1-\tau H_1\right|^2}{8\pi\,\tau_2}\,y\,,
    \label{eq:VCexplicit}
\end{equation}
with no free coefficients left. Since $y<0$ in the region of interest,
this is manifestly positive, as it must be for a no-scale potential, and
it grows logarithmically as the conifold is approached.

The norm of the gradient has the leading behavior
\begin{equation}
    \left|\nabla V\right|
    \sim
    e^{K_{\text{k}}}
    \sqrt{
    \frac{
    b_1c_2d_1
    \left(
    F_1^2-2F_1H_1\tau_1
    +H_1^2(\tau_1^2+\tau_2^2)
    \right)^2
    }
    {32a_1^3r^2\tau_2^2y}
    +\mathcal{O}(y^{-2})
    }\,,
    \label{eq:gradVC}
\end{equation}
where positivity of the radicand requires $b_1c_2d_1/a_1^3<0$, consistent
with $b_1d_1<0$, which is in turn what makes
$e^{K_{\text{cs}}}=(-2b_1d_1)^{-1}+\mathcal{O}(z_{\text{\tiny C}})$
positive.

Combining \eqref{eq:VC} and \eqref{eq:gradVC}, the quotient behaves as
\begin{equation}
    \frac{\left|\nabla V\right|}{V}
    \;\sim\;
    \frac{\text{const}}{r\,\left|y\right|^{3/2}}
    \;\xrightarrow[\;r\to0\;]{}\;\infty\,,
    \label{eq:ratioC}
\end{equation}
since $r\to0$ faster than $|y|=|\log r|$ grows. The quotient therefore
depends explicitly on the period coefficients, on the fluxes and on the
direction from which the conifold point is approached, and, unlike the
LCS result, it does not approach a universal value. Note, however, that
its divergence means that no bound of the form $|\nabla V|\geq cV$ can
be violated in this region; combined with the fact that the conifold sits
at finite distance, so that the asymptotic TCC bound does not apply, the
neighbourhood of the conifold places no constraint of the kind studied
here.

We next analyze the TCC in the remaining asymptotic directions of moduli
space,
\begin{equation}
\tau_1\rightarrow\pm\infty,
\qquad
\tau_2\rightarrow+\infty,
\label{eq:asymptotic-directions}
\end{equation}
which are the directions of infinite distance in the axio-dilaton
sector.\footnote{The phase $\theta=\arg z$ is an angle and does not by
itself provide an infinite-distance limit; a trajectory winding around
the conifold does accumulate field-space length, but it also acts on the
periods by the monodromy $\mathrm{T}_{\text{\tiny C}}$, so that the
potential is not single-valued along it. We do not pursue that direction
here.}

For this purpose it is not necessary to expand the periods around a
singular point. Instead, we keep the periods as generic functions of $z$,
together with their derivatives, and extract the asymptotic behavior in
the axio-dilaton directions alone.

\subsection{TCC in the large $\tau_2$ region}
\label{sec:tcc-tau2}

We now analyze another infinite-distance direction in moduli space,
namely the limit in which the imaginary component of the axio-dilaton
becomes large,
\begin{equation}
\tau_2\rightarrow+\infty .
\label{eq:tau2limit}
\end{equation}
Since $\tau=\tau_1+i\tau_2$, this regime corresponds to weak string
coupling, $g_s=1/\tau_2\rightarrow0$. In this limit, the dependence of
the flux superpotential on the axio-dilaton can be made explicit by
writing
\begin{equation}
    W
    =
    \mathrm{F}^T\Sigma\Pi
    -
    \mathrm{H}^T\Sigma\Pi\,\tau
    \equiv
    A+B\tau ,
    \label{eq:Wtau}
\end{equation}
where $A$ and $B$ depend on the complex-structure modulus through the
period vector $\Pi$. In particular,
\begin{equation}
    B=-\,\mathrm{H}^T\Sigma\Pi .
    \label{eq:Bdef}
\end{equation}
Keeping the complex-structure modulus fixed while taking the
large-$\tau_2$ limit, every term of the scalar potential that survives is
quadratic in $B$: the contributions involving $A$ are subleading, being
suppressed by at least one power of $\tau_2$ relative to those retained
below. The leading contribution is linear in $\tau_2$ and takes the form
\begin{equation}
    V
    =
    \frac{1}{2}
    \frac{e^{K_{\rm cs}}}{K_{\rm cs}^{(1,1)}}
    \left[
    K_{\rm cs}^{(1,1)}|B|^2
    +
    \left|
    B^\prime+K_{\rm cs}^{(1,0)}B
    \right|^2
    \right]\tau_2 ,
    \label{eq:Vtau2}
\end{equation}
where a prime denotes differentiation with respect to the
complex-structure modulus and $K_{\rm cs}^{(m,n)}$ denotes $m$
holomorphic and $n$ antiholomorphic derivatives of the complex-structure
K\"ahler potential; explicitly,
\begin{equation}
   K_{\rm cs}^{(1,0)}
   =
   -\,\frac{\overline{\Pi}^T\Sigma\,\partial\Pi}
   {\overline{\Pi}^T\Sigma\Pi}\, .
   \label{eq:K10}
\end{equation}

\subsubsection*{A structural lower bound}

Before evaluating the gradient explicitly, it is worth noting that the
bound follows from the {\it form} of \eqref{eq:Vtau2} alone, without any
knowledge of the periods. Writing the leading potential as
\begin{equation}
    V=\mathfrak{f}(z,\bar z)\,\tau_2\,,
    \qquad
    \mathfrak{f}\equiv
    \frac{1}{2}\frac{e^{K_{\rm cs}}}{K_{\rm cs}^{(1,1)}}\,\mathcal{C}\,,
    \label{eq:Vfactorized}
\end{equation}
with $\mathcal{C}$ defined in \eqref{eq:calC} below, the potential depends
on the axio-dilaton only through the overall factor $\tau_2$. Since
$K_{\rm ad}=-\log(2\tau_2)$ gives $K_{\rm ad}^{\tau\bar\tau}=4\tau_2^2$,
and since $\partial_\tau V=-\tfrac{i}{2}\mathfrak{f}$ for a potential
linear in $\tau_2$, the axio-dilaton contribution to the norm of the
gradient is
\begin{equation}
    K_{\rm ad}^{\tau\bar\tau}\left|\partial_\tau V\right|^2
    =4\tau_2^2\cdot\frac{\mathfrak{f}^2}{4}
    =V^2\,,
    \label{eq:tau-exact}
\end{equation}
{\it exactly}. With the field-space metric $G_{ij}=2K_{a\bar b}$ fixed in
Sec.~\ref{sec:setup}, so that
$|\nabla V|^2=2K^{a\bar b}\partial_aV\,\partial_{\bar b}V$, this gives
\begin{equation}
    \frac{\left|\nabla V\right|^2}{V^2}
    =
    2
    +
    \frac{2\,K_{\rm cs}^{z\bar z}\left|\partial_zV\right|^2}{V^2}
    \;\geq\;
    2\,,
    \label{eq:structural-bound}
\end{equation}
the second term being non-negative because the moduli-space metric is
positive definite, with equality if and only if $\partial_zV=0$. The
bound $|\nabla V|/V\geq\sqrt2$ therefore holds for {\it any} flux
configuration and {\it any} one-parameter geometry, being a consequence
only of the linear dependence of the leading potential on $\tau_2$.

\subsubsection*{The explicit gradient}

To exhibit the complex-structure contribution we introduce the
combinations
\begin{equation}
    \mathcal{C}
    =
    K_{\rm cs}^{(1,1)}|B|^2
    +
    \left|
    B^\prime+K_{\rm cs}^{(1,0)}B
    \right|^2 ,
    \label{eq:calC}
\end{equation}
\begin{equation}
    \mathcal{B}
    =
    K_{\rm cs}^{(1,1)}
    \left(
    K_{\rm cs}^{(1,0)}|B|^2
    +
    2\overline{B}B^\prime
    \right),
    \label{eq:calB}
\end{equation}
\begin{equation}
   \mathcal{E}
   =
   \left(
   \overline{B}^{\,\prime}
   +
   K_{\rm cs}^{(0,1)}\overline{B}
   \right)
   \left(
   B^{\prime\prime}
   +
   K_{\rm cs}^{(1,0)}B^\prime
   +
   K_{\rm cs}^{(2,0)}B
   \right),
   \label{eq:calE}
\end{equation}
\begin{equation}
    \mathcal{F}
    =
    -\frac{1}{K_{\rm cs}^{(1,1)}}
    \left(
    \left|
    B^\prime+K_{\rm cs}^{(1,0)}B
    \right|^2
    K_{\rm cs}^{(2,1)}
    \right).
    \label{eq:calF}
\end{equation}
It is also convenient to record the combination
\begin{equation}
    \mathcal{A}
    =
    \left|B^\prime\right|^2
    +
    2\,\mathrm{Re}
    \left(
    K_{\rm cs}^{(1,0)}B\overline{B}^{\,\prime}
    \right)
    +
    \left(
    \left|K_{\rm cs}^{(1,0)}\right|^2
    +
    K_{\rm cs}^{(1,1)}
    \right)
    |B|^2 ,
    \label{eq:calA}
\end{equation}
which is nothing but $\mathcal{C}$ written out, $\mathcal{A}=\mathcal{C}$,
as follows from expanding the modulus squared in \eqref{eq:calC}. In
terms of these, \eqref{eq:Vtau2} reads compactly
\begin{equation}
    V
    =
    \frac{1}{2}
    \frac{e^{K_{\rm cs}}}{K_{\rm cs}^{(1,1)}}
    \,\mathcal{C}\,\tau_2 ,
    \label{eq:Vcompact}
\end{equation}
while the four combinations above assemble into the logarithmic
derivative of $V$ along the complex-structure direction. Indeed, one
verifies that
\begin{equation}
    \mathcal{B}
    +K_{\rm cs}^{(1,0)}\mathcal{C}
    +\mathcal{E}
    +\mathcal{F}
    \;=\;
    K_{\rm cs}^{(1,0)}\mathcal{C}
    +\mathcal{C}^\prime
    -\mathcal{C}\,
    \frac{K_{\rm cs}^{(2,1)}}{K_{\rm cs}^{(1,1)}}
    \;=\;
    \mathcal{C}\,\frac{\partial_z V}{V}\,,
    \label{eq:identity}
\end{equation}
which is the statement that $\mathcal{B}$, $\mathcal{E}$ and
$\mathcal{F}$ collect, respectively, the derivative of the first term of
$\mathcal{C}$, that of the second, and the contribution of the
$z$-dependence of the metric prefactor. The norm of the gradient
therefore behaves as
\begin{equation}
   \left|\nabla V\right|
   =
   \frac{1}{2}
   \frac{e^{K_{\rm cs}}}{K_{\rm cs}^{(1,1)}}
   \sqrt{
   2\mathcal{C}^2
   +
   2\left|
   \mathcal{B}
   +K_{\rm cs}^{(1,0)}\mathcal{C}
   +\mathcal{E}
   +\mathcal{F}
   \right|^2
   \left(K_{\rm cs}^{(1,1)}\right)^{-1}
   }\,
   \tau_2 ,
   \label{eq:gradVtau2}
\end{equation}
and the ratio between the norm of the gradient and the scalar potential
is
\begin{equation}
   \frac{\left|\nabla V\right|}{V}
   =
   \frac{
   \sqrt{
   2\mathcal{C}^2
   +
   2\left|
   \mathcal{B}
   +K_{\rm cs}^{(1,0)}\mathcal{C}
   +\mathcal{E}
   +\mathcal{F}
   \right|^2
   \left(K_{\rm cs}^{(1,1)}\right)^{-1}
   }
   }{\mathcal{C}}
   \;\geq\;
   \sqrt{2}\,,
   \label{eq:ratiotau2}
\end{equation}
in agreement with the structural bound \eqref{eq:structural-bound}, of
which \eqref{eq:ratiotau2} is the explicit form.

The second contribution under the square root is non-negative, and
therefore the ratio is bounded from below by $\sqrt{2}$, with equality
precisely along the purely dilatonic direction, for which
$\partial_zV=0$. This limiting case admits a transparent interpretation.
Introducing the canonically normalized dilaton, which for the metric
$G_{ij}=2K_{a\bar b}$ of Sec.~\ref{sec:setup} is
$\phi=\tfrac{1}{\sqrt2}\log\tau_2$, the leading potential
\eqref{eq:Vcompact} becomes purely exponential,
\begin{equation}
    V=\mathfrak{f}(z,\bar z)\;e^{\sqrt{2}\,\phi},
    \label{eq:Vexponential}
\end{equation}
so that its logarithmic slope with respect to the affine parameter is the
constant
\begin{equation}
    \frac{\left|V^\prime\right|}{V}=\sqrt{2}\,,
    \label{eq:slope-tau2}
\end{equation}
and the asymptotic average \eqref{eq:tcc-liminf} is trivially equal to
it. This is the quantity the TCC constrains, rather than the norm of the
full gradient, and since $\sqrt{2}>\sqrt{2/3}$ the four-dimensional bound
is met with a wide margin, independently of the flux quanta and of the
geometry, provided that $\mathcal{C}>0$ and the complex-structure metric
remains regular.

Two remarks are in order. First, the exponential rate is fixed to
$\sqrt{2}$ by the linear dependence of the leading potential on $\tau_2$
alone, and coincides with the familiar runaway rate of the type IIB
dilaton; it is therefore insensitive to the details of the
compactification. This rate is precisely the quantity bounded by the
conjecture in its multifield form \cite{Bedroya:2019snp}, and since
$\sqrt{2}>\sqrt{2/3}$ the four-dimensional bound is met with a wide
margin. Second, the potential \eqref{eq:Vtau2} {\it grows} as
$\tau_2\to\infty$, reflecting the fact that at fixed integral NS--NS
flux one has $\left|A+B\tau\right|^2\sim\tau_2^2|B|^2$, so that the
weak-coupling region is energetically disfavoured and the field is not
driven towards it. The opposite region, $\tau_2\to0$, where the
potential does decrease, lies at strong coupling: the expression
\eqref{eq:Vtau2} has been derived in the large-$\tau_2$ regime and
should not be extrapolated there, where the expansion breaks down and
the perturbative description used throughout this work ceases to be
valid.

\subsection{TCC in the large $\tau_1$ region}
\label{sec:tcc-tau1}

We now consider the regime in which the real component of the
axio-dilaton becomes large,
\begin{equation}
\tau_1\rightarrow \pm\infty,
\label{eq:tau1limit}
\end{equation}
while $\tau_2$ and the complex-structure modulus are kept fixed. As in
the large-$\tau_2$ analysis, the dependence on the complex-structure
sector is encoded in the period vector through the function
\begin{equation}
B=-\,\mathrm{H}^T\Sigma\Pi
\label{eq:Bdef2}
\end{equation}
and its derivatives, and the combination $\mathcal{C}$ is the one
defined in \eqref{eq:calC}. In this limit, the leading contribution to
the scalar potential grows quadratically with $\tau_1$ and is given by
\begin{equation}
    V
    =
    \frac{1}{2}
    \frac{e^{K_{\rm cs}}}
    {K_{\rm cs}^{(1,1)}\tau_2}
    \left[
    K_{\rm cs}^{(1,1)}|B|^2
    +
    \left|
    B^\prime+K_{\rm cs}^{(1,0)}B
    \right|^2
    \right]
    \tau_1^2 ,
    \label{eq:Vtau1}
\end{equation}
which, using \eqref{eq:calC}, can be written more compactly as
\begin{equation}
    V
    =
    \frac{1}{2}
    \frac{e^{K_{\rm cs}}}
    {K_{\rm cs}^{(1,1)}\tau_2}
    \,\mathcal{C}\,\tau_1^2 .
    \label{eq:Vtau1compact}
\end{equation}
The corresponding norm of the gradient takes the form
\begin{equation}
   \left|\nabla V\right|
   =
   \frac{1}{2}
   \frac{e^{K_{\rm cs}}}
   {K_{\rm cs}^{(1,1)}\tau_2}
   \sqrt{
   2\mathcal{C}^2
   +
   2\left|
   \mathcal{B}
   +
   K_{\rm cs}^{(1,0)}\mathcal{C}
   +
   \mathcal{E}
   +
   \mathcal{F}
   \right|^2
   \left(K_{\rm cs}^{(1,1)}\right)^{-1}
   }
   \,\tau_1^2 ,
   \label{eq:gradVtau1}
\end{equation}
and the ratio between the norm of the gradient and the potential is
therefore
\begin{equation}
   \frac{\left|\nabla V\right|}{V}
   =
   \frac{
   \sqrt{
   2\mathcal{C}^2
   +
   2\left|
   \mathcal{B}
   +
   K_{\rm cs}^{(1,0)}\mathcal{C}
   +
   \mathcal{E}
   +
   \mathcal{F}
   \right|^2
   \left(K_{\rm cs}^{(1,1)}\right)^{-1}
   }
   }{\mathcal{C}}
   \;\geq\;
   \sqrt{2}.
   \label{eq:ratiotau1}
\end{equation}
Since the second contribution under the square root is non-negative, the
ratio is bounded from below by $\sqrt{2}$. Therefore, in this regime, the
four-dimensional TCC bound is satisfied whenever $\mathcal{C}>0$ and the
complex-structure metric remains regular.

\subsubsection*{A unified description of the two axio-dilaton limits}

The results of this subsection and of Sec.~\ref{sec:tcc-tau2} are two
corners of a single expression. Retaining only the terms quadratic in
$B$, which is what both limits amount to, the flux superpotential
contributes through $\left|B\tau\right|^2=|B|^2|\tau|^2$, and with
$e^{K_{\rm ad}}=(2\tau_2)^{-1}$ the leading potential reads
\begin{equation}
    V
    =
    \frac{1}{2}
    \frac{e^{K_{\rm cs}}}{K_{\rm cs}^{(1,1)}}\,\mathcal{C}\;
    \frac{\left|\tau\right|^2}{\tau_2}
    =
    \frac{1}{2}
    \frac{e^{K_{\rm cs}}}{K_{\rm cs}^{(1,1)}}\,\mathcal{C}\;
    \frac{\tau_1^2+\tau_2^2}{\tau_2}\,,
    \label{eq:Vmaster}
\end{equation}
whose $\tau_2\rightarrow\infty$ limit is \eqref{eq:Vtau2} and whose
$\tau_1\rightarrow\pm\infty$ limit is \eqref{eq:Vtau1}. Writing
$V=\mathfrak{g}(z,\bar z)\left|\tau\right|^2/\tau_2$ and using
$G^{\tau_1\tau_1}=G^{\tau_2\tau_2}=2\tau_2^2$ for the metric
$G_{ij}=2K_{a\bar b}$ of Sec.~\ref{sec:setup}, a direct computation gives
\begin{equation}
    \frac{2\tau_2^2
    \left[\left(\partial_{\tau_1}V\right)^2
    +\left(\partial_{\tau_2}V\right)^2\right]}{V^2}
    =2\,,
    \label{eq:master-exact}
\end{equation}
{\it exactly}, at every point of the $(\tau_1,\tau_2)$ half-plane and not
merely asymptotically. This is the combined effect of the
$SL(2,\mathbb{R})$ covariance of \eqref{eq:Vmaster} and of the hyperbolic
metric on the axio-dilaton moduli space. Consequently
\begin{equation}
    \frac{\left|\nabla V\right|^2}{V^2}
    =2+\frac{2K_{\rm cs}^{z\bar z}\left|\partial_zV\right|^2}{V^2}
    \;\geq\;2\,,
    \label{eq:master-bound}
\end{equation}
which contains both \eqref{eq:ratiotau2} and \eqref{eq:ratiotau1} as
particular cases and shows that in these regions the bound is a
consequence of the $\tau$-dependence alone.

\subsubsection*{Which direction saturates the bound}

It is instructive to see how the value $\sqrt2$ is distributed among the
axio-dilaton directions in the present limit. Evaluating the two
contributions separately for \eqref{eq:Vtau1compact} one finds
\begin{equation}
    \frac{2\tau_2^2\left(\partial_{\tau_1}V\right)^2}{V^2}
    =\frac{8\tau_2^2}{\tau_1^2}
    \;\xrightarrow[\;\tau_1\to\infty\;]{}\;0\,,
    \qquad
    \frac{2\tau_2^2\left(\partial_{\tau_2}V\right)^2}{V^2}=2\,,
    \label{eq:decomposition}
\end{equation}
so that the bound \eqref{eq:ratiotau1} is approached from above, and the
entire asymptotic value comes from the direction transverse to $\tau_1$.
Along the axionic direction itself the situation is qualitatively
different from the dilatonic one. The canonically normalized axion is
$\phi=\tau_1/(\sqrt2\,\tau_2)$, in terms of which
\eqref{eq:Vtau1compact} becomes a power law rather than an exponential,
\begin{equation}
    V=2\,\mathfrak{g}\,\tau_2\,\phi^2\,,
    \qquad
    \frac{\left|V^\prime\right|}{V}=\frac{2}{\phi}
    \;\xrightarrow[\;\phi\to\infty\;]{}\;0\,,
    \label{eq:axion-slope}
\end{equation}
to be compared with $V=\mathfrak{f}\,e^{\sqrt2\phi}$ and
$\left|V^\prime\right|/V=\sqrt2$ in the large-$\tau_2$ region. A
trajectory running along $\tau_1$ at fixed $\tau_2$ and fixed $z$ would
therefore have a vanishing asymptotic logarithmic slope.

This is not in tension with the conjecture. The quantity bounded by the
TCC in its multifield form is the norm of the gradient,
\eqref{eq:tcc-both}, and \eqref{eq:master-exact} shows that it equals
$\sqrt2$ exactly throughout the $\tau$-plane, in particular along the
large-$\tau_1$ region: the bound is satisfied there. The vanishing
directional slope \eqref{eq:axion-slope} concerns a different quantity,
the projection \eqref{eq:tcc-directional} onto the pure $\tau_1$
direction, and merely reflects that this direction is transverse to the
gradient, the whole of the value $2$ in \eqref{eq:decomposition} coming
from $\partial_{\tau_2}V$. Two further observations put this in
perspective. The potential \eqref{eq:Vtau1} {\it grows} as
$\tau_1\rightarrow\pm\infty$, so that no causal trajectory starting in
the interior of moduli space is driven along this direction, and it is
therefore not among the asymptotic directions $\gamma_\alpha$ over
which \eqref{eq:tcc-both} is defined. And $\tau_1$ is an axion lifted
by the fluxes, a point on which the $SL(2,\mathbb{Z})$ duality sheds
some light, as we now discuss.

\subsubsection*{The role of $SL(2,\mathbb{Z})$}

A subtle point concerns the role of the $SL(2,\mathbb{Z})$ duality of
type IIB string theory. In the presence of three-form fluxes, this
duality is preserved only when the axio-dilaton and the NS--NS and R--R
fluxes are transformed simultaneously. Therefore, although the full
theory remains duality invariant, a fixed choice of flux quanta is not,
in general, invariant under the complete modular group. In the present
analysis, the limit $\tau_1\rightarrow\pm\infty$ is considered for fixed
fluxes and should therefore be understood in the covering space of the
axio-dilaton moduli space. Different large values of $\tau_1$ may be
related by an $SL(2,\mathbb{Z})$ transformation together with a
corresponding transformation of the fluxes.

This observation should not be mistaken for a statement about the
trajectory itself. At fixed fluxes the potential \eqref{eq:Vtau1compact}
is not periodic in $\tau_1$, and the field-space distance
$\Delta s=\Delta\tau_1/(\sqrt2\tau_2)$ is unbounded, so that the limit
$\tau_1\rightarrow\pm\infty$ is a genuine infinite-distance limit in the
covering space. What the duality does fix is the origin of the
parabola: the leading potential depends on $\tau_1$ only through
$\tau_1-\tau_1^\ast$, with $\tau_1^\ast=-\mathrm{Re}(A\bar B)/|B|^2$,
and an integer shift of $\tau_1$ maps the configuration to another
representative of the same flux orbit, with the same
$N_{\rm flux}=\mathrm{H}^T\Sigma\mathrm{F}$, since the latter is
$SL(2,\mathbb{Z})$-invariant. The vanishing directional slope
\eqref{eq:axion-slope} therefore reflects that the axion is lifted by
the fluxes, not an artefact of the identification.

\subsubsection*{Simultaneous limits}

We finally comment on simultaneous large-field limits involving $\theta$,
$\tau_1$ and $\tau_2$. In these directions the leading terms of both the
scalar potential and its gradient depend more sensitively on the detailed
structure of the periods and on the particular choice of fluxes, and,
unlike the large-$\tau_1$ and large-$\tau_2$ limits discussed above, no
universal flux-independent bound has been identified . The region
$\tau_2\rightarrow0$ deserves particular caution: it corresponds to
strong string coupling, $g_s=1/\tau_2\rightarrow\infty$, where the
perturbative type IIB description used throughout this work ceases to be
valid, so that no conclusion about the conjecture can be drawn there. A
systematic treatment of the remaining mixed directions, expanding the
potential and its gradient for each flux configuration and evaluating the
associated field displacement, is left for future work; the multifield
trajectories of Sec.~\ref{sec:tcc-multifield} provide a first step in
that direction.
\subsection{Multi-field evolution and the TCC}
\label{sec:tcc-multifield}

In this model the moduli space is parametrized by the axio-dilaton and
the complex-structure modulus, the remaining K\"ahler moduli not being
considered. Taking the real and imaginary parts of the moduli, the action
\eqref{S4} takes the well-known form of $N$ fields minimally coupled to
gravity,
\begin{equation}
	S=\frac{M_{pl}^2}{2}\int d^4x\sqrt{-g}\,
	\left[ \mathcal{R}-\frac{1}{2} G_{ij}\partial_\nu\phi^i
	\partial_\mu\phi^jg^{\mu\nu}-\frac{2}{M_{pl}^4}V(\phi^l) \right],
	\label{eq:Smulti}
\end{equation}
where the $\left\lbrace \phi^i \right\rbrace$ stand for
$\left\lbrace r,\theta,\tau_1,\tau_2\right\rbrace$ and $G_{ij}$ is the
metric in the real field space,
\begin{equation}
    G_{ij}\dot{=} \begin{pmatrix}
        4G_{z\Bar{z}}&0&0&0\\
        0&4r^2G_{z\Bar{z}}&0&0\\
        0&0&4G_{\tau\Bar{\tau}}&0\\
        0&0&0&4G_{\tau\Bar{\tau}}\\
    \end{pmatrix}.
    \label{eq:Gmatrix}
\end{equation}
Since the prefactor $M_{pl}^2/2$ has been kept outside the bracket in
\eqref{eq:Smulti}, the metric appearing in the canonically normalized
kinetic term, and therefore the one that measures distances and enters
the Friedmann equations below, is
\begin{equation}
    \widehat{G}_{ij}=\frac{1}{2}G_{ij}
    =2\,\mathrm{diag}\left(
    G_{z\bar z},\;r^2G_{z\bar z},\;
    G_{\tau\bar\tau},\;G_{\tau\bar\tau}\right)
    =2K_{a\bar b}\,,
    \label{eq:Ghat}
\end{equation}
in agreement with the convention \eqref{eq:Gmetric} fixed in
Sec.~\ref{sec:setup}. In what follows we write $G_{ij}$ for
$\widehat{G}_{ij}$, so that the canonical distance in the induced moduli
space is $ds^2=G_{ij}\,d\phi^id\phi^j$.

The equations of motion for the fields are given by
\begin{equation}
	\Box \phi^m+\Gamma^m_{ji}\partial_\nu \phi^j\partial_\mu \phi^i
	g^{\mu\nu}=G^{mk}\partial_kV(\phi^a),
	\label{eq:eom}
\end{equation}
where the operator $\Box$ is defined as
\begin{equation}
	\Box\phi^m \equiv \frac{1}{\sqrt{-g}}\partial_\nu
	\left( \sqrt{-g}\,g^{\mu\nu} \partial_\mu\phi^m \right),
	\label{eq:box}
\end{equation}
and the Christoffel symbols in field space are
\begin{equation}
	\Gamma^m_{ji}=\frac{1}{2}G^{mk}
	\left[ \partial_jG_{ik}+\partial_iG_{jk}-\partial_kG_{ij}\right].
	\label{eq:christoffel}
\end{equation}
With the goal of obtaining a field evolution, let us assume that, by
isotropy and homogeneity, the field dependence is
$\phi^i(x^\mu)=\phi^i(t)$. Considering a flat FRW spacetime as background
evolution, given by the metric $ds^2=-dt^2+a(t)^2d\vec{x}\,^2$, the
equations of motion for the scalar fields reduce to
\begin{equation}
	\mathcal{D}_t\dot{\phi}^i+3H\dot{\phi}^i+G^{ij}\partial_jV=0,
	\label{eq:eomFRW}
\end{equation}
where the covariant derivative is
\begin{equation}
	\mathcal{D}_tA^i\equiv\dot{A}^i+\Gamma^i_{jk}\dot{\phi}^j\,A^k.
	\label{eq:Dt}
\end{equation}
The Friedmann equations for the $N$-field configuration are
\begin{equation}
	3M_{pl}^2H^2=\frac{1}{2}G_{ij}\dot{\phi}^i\dot{\phi}^j+V
	\text{\hspace{.5cm} and \hspace{.5cm}}
	M_{pl}^2\dot{H}=-\frac{1}{2}G_{ij}\dot{\phi}^i\dot{\phi}^j .
	\label{eq:friedmann}
\end{equation}
Together, the Friedmann and multifield equations of motion describe the
field evolution in such a background. The multifield evolution is
performed by considering as initial conditions $\phi^i(t=0)=\phi^i_0$,
where $\phi^i_0$ is the initial field value, located at any point of the
moduli space or at an extremum of $V$, and initial velocities
$\dot{\phi}^i(t=0)<\mathcal{O}(1)$ in order to preserve the quantum field
description. Once the multifield evolution is obtained, the canonical
distance is measured as
\begin{equation}
    s=\int dt \, \sqrt{G_{ij}\Dot{\phi}^i\Dot{\phi}^j}\,,
    \label{eq:sdist}
\end{equation}
which is the affine parameter of Sec.~\ref{sec:tcc}, so that the
asymptotic average \eqref{eq:tcc-liminf} can be evaluated directly along
the numerical trajectories.

Due to the nature of the period solutions, we can trust our solutions,
and in fact the field variation, at least within the region of
convergence of $\left|z\right|$. For all the flux configurations
analyzed we find that the canonical distance travelled by the fields is
$s\sim\mathcal{O}(1)$, so that we can study the characteristics of the
scalar potential within a small range of field variation.

\begin{figure}[H]
	\centering
	\includegraphics[width=1.01\linewidth]{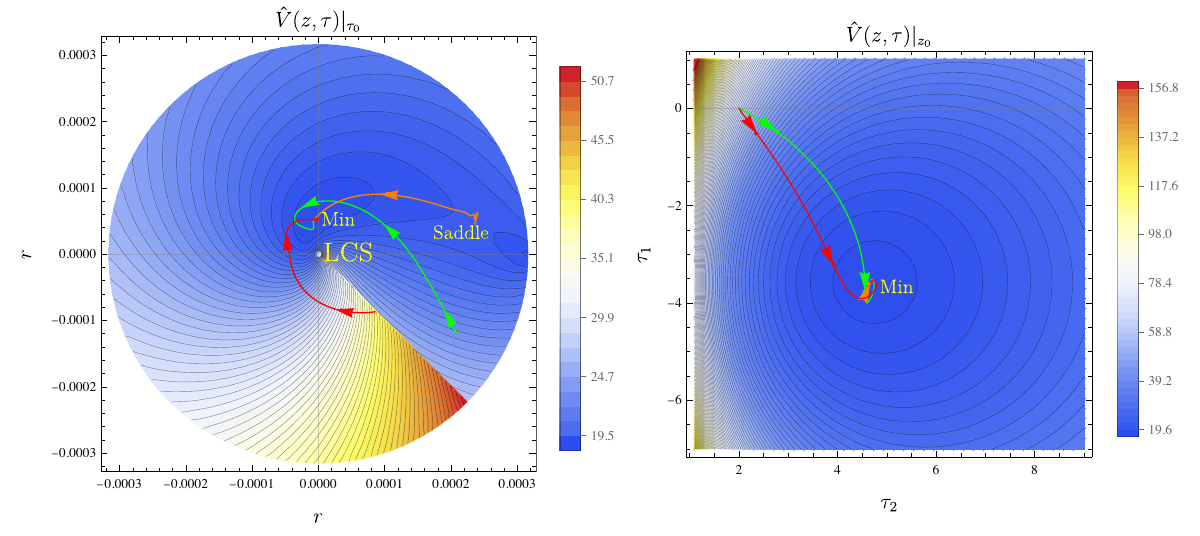}
	\captionsetup{margin=.5cm}
	\caption{Representative trajectories in the complex structure and axio-dilaton moduli spaces for the flux choice \((F_{(3)},H_{(3)})=(-10,1,8,7,1,-9,0,0)\), superimposed on contour plots of the scalar potential. The left panel shows the dynamics of the complex-structure modulus \(z\), while the right panel displays the simultaneous evolution in the axio-dilaton plane. The colored curves represent trajectories starting from different initial conditions. Together, the two panels illustrate the basin of attraction associated with the stabilized solution \((z_0,\tau_0)\).}
	\label{fig:trajectories}
\end{figure}
Despite following substantially different paths through moduli space, the trajectories are attracted toward the same stabilized configuration. In the both planes, the trajectories similarly converge toward the minimum, although their paths exhibit different degrees of curvature depending on their initial conditions.

\subsubsection*{Two classes of trajectories}

To visualize the geodesic evolution, we project the trajectories onto the complex-structure modulus plane. This representation is particularly useful because the distinguished regions of moduli space relevant to our analysis, namely the conifold, LCS, and orbifold regions, are naturally characterized in terms of the complex-structure modulus $z$. The $z$-plane therefore makes it possible to visualize directly whether a trajectory remains within a given patch, crosses the overlap region, or evolves toward a different region of moduli space. This choice should not be interpreted as implying that the axio-dilaton dynamics is irrelevant. The evolution takes place in the full field space, and the initial conditions for $\tau$ can qualitatively modify the projected trajectory in the $z$-plane, as illustrated by the comparison between $\beta_3$ and $\beta_4$. We therefore use the $z$-plane to display the geodesics, while keeping the full coupled dynamics of the complex-structure modulus and the axio-dilaton in their computation.\\

Two qualitatively different behaviours are found. In one class of trajectories the fields move outwards in moduli space, in the direction of the orbifold point; in the other, they remain confined to the neighbourhood of the conifold and LCS patches. Which of the two occurs is controlled by the critical points generated by the given flux choice.

\begin{figure}[H]
	\centering
	\includegraphics[width=.7\linewidth]{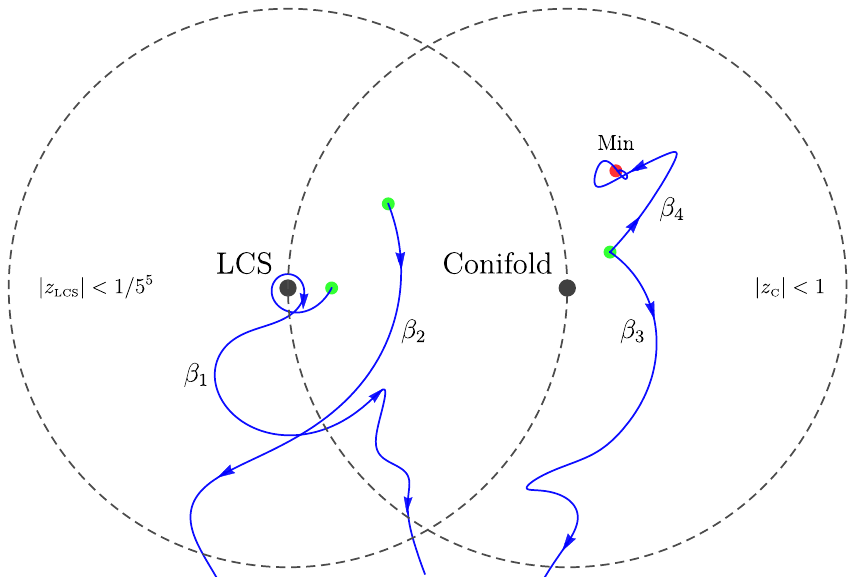}
	\captionsetup{margin=.5cm}
	\caption{For the flux configuration $(F_{(3)},H_{(3)})=(16, -2, 3, -20,-17, -7, 9, -1)$, which generates a local minimum within the conifold patch. Here, the continuous blue lines represent the geodesics, while the green points indicate their initial conditions. Trajectories starting sufficiently close to the minimum remain trapped in its neighborhood,as illustrated by $\beta_4$. By contrast, trajectories starting outside its basin of attraction, such as $\beta_1$, $\beta_2$, and $\beta_3$, escape the conifold/LCS region and move outward in moduli space toward the orbifold point.}
	\label{fig:geo1}
\end{figure}

\begin{figure}[H]
	\centering
	\includegraphics[width=.7\linewidth]{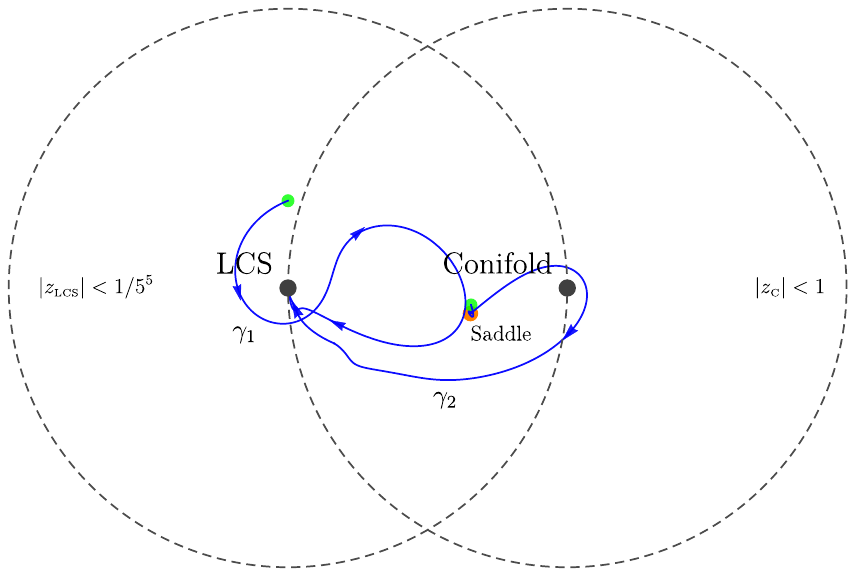}
	\captionsetup{margin=.5cm}
	\caption{Here, the flux choice $(F_{(3)},H_{(3)})=(-10, -3, -5, 2,6, 5, -9, 2)$, which generates a saddle point in the overlap region and a local minimum within the LCS patch. In this case, the LCS minimum attracts trajectories originating in both patches. In particular, trajectories starting in the conifold region can pass through the overlap region and ultimately terminate at the LCS minimum, as illustrated by $\gamma_2$.}
	\label{fig:geo2}
\end{figure}

A local minimum within the conifold patch traps only those trajectories
that start sufficiently close to it, whereas a minimum within the LCS patch can attract trajectories originating in either patch. Importantly,
being attracted to a minimum in the LCS patch does not imply approaching the LCS point itself. Indeed, none of the trajectories reaches the LCS
point asymptotically. Although the TCC bound is satisfied in the formal limit $z \to z_{\mathrm{LCS}}$, as discussed in Sec.~5.1, this limit is not realized by any physical trajectory in moduli space.

A comparison between the trajectories shown in Figs.~8 and~9 also illustrates the sensitivity of the dynamics to the initial conditions.
In Figure~\ref{fig:geo1}, the geodesic $\beta_3$ has the same initial conditions for the fields $r$ and $\theta$ as those considered for $\beta_4$, while the initial conditions for $\tau_1$ and $\tau_2$ are modified. As a
result, the corresponding trajectory no longer evolves toward the local minimum. On the other hand, Figure~\ref{fig:geo2} shows the geodesic $\gamma_2$,
whose initial point corresponds to the saddle point. The sensitivity of the trajectories to the choice of initial conditions will be briefly analyzed in the following section. 

\begin{figure}[H]
	\centering
	\includegraphics[width=0.49\linewidth]{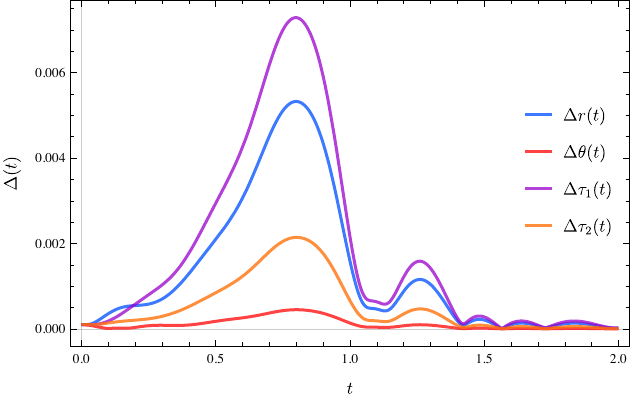}
    \hspace{2 pt }
    \includegraphics[width=0.49\linewidth]{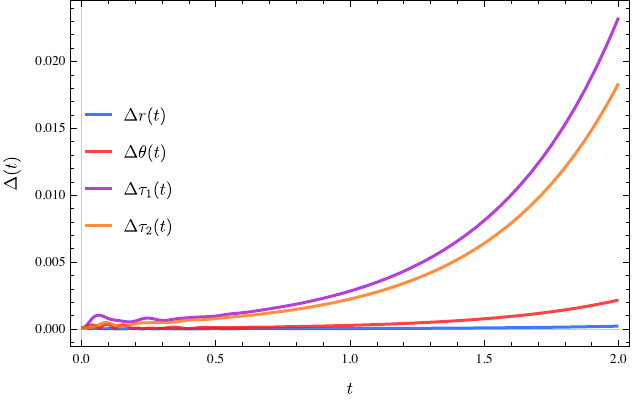}
	\caption{Sensitivity of the geodesic to a perturbation of $\epsilon = 1\times10^{-4}$ in the initial conditions of each field, with the unperturbed initial conditions chosen as the coordinates of a saddle point. Left panel: For the previously illustrated flow configuration $(F_{(3)},H_{(3)})=(-12,12,-13,8,6,-1,-5,-1)$, both a saddle point and a minimum arise within the LCS patch. The deviation $\Delta(t)$ reaches a maximum for each perturbed field and subsequently tends to zero as the geodesic evolves toward the minimum. Among the fields, $r$ and $\tau_1$ induce the largest deviations from the unperturbed geodesic. Right panel: Perturbations of the fields are shown for the flow configuration $(F_{(3)},H_{(3)})=(-10,1,8,7,1,-9,0,0)$. In this case, the chosen fluxes give rise only to a saddle point within the conifold patch. Consequently, the deviation from the unperturbed geodesic becomes increasingly pronounced as the fields evolve, with $\tau_1$ and $\tau_2$ producing the largest deviations.}
	\label{fig:sensitivity}
\end{figure}

\subsection{Sensitivity to initial conditions}

To assess the sensitivity of the field-space trajectory to small variations in the initial conditions, we consider a reference trajectory
\begin{equation}
    \boldsymbol{\phi}(t)
    =
    \left(
    r(t),\,\theta(t),\,\tau_1(t),\,\tau_2(t)
    \right),
\end{equation}
obtained from a given set of initial conditions. We then perturb one initial condition at a time while keeping the remaining ones fixed. For a generic field $\phi_i$, the perturbed initial condition is defined as
\begin{equation}
    \phi_i^{(\epsilon)}(0)
    =
    \phi_i(0)(1+\epsilon),
\end{equation}
where $\epsilon$ is a small dimensionless parameter. Both positive and negative perturbations are considered in order to account for a possible asymmetric response of the trajectory.

The resulting perturbed trajectory is denoted by
\begin{equation}
    \boldsymbol{\phi}_{\epsilon}(t)
    =
    \left(
    r_{\epsilon}(t),\,\theta_{\epsilon}(t),\,
    \tau_{1,\epsilon}(t),\,\tau_{2,\epsilon}(t)
    \right).
\end{equation}
To quantify its deviation from the reference trajectory, we introduce the normalized distance
\begin{equation}
\label{eq:trajectory_distance}
    \Delta(t)
    =
    \left[
    \left(
    \frac{r_{\epsilon}(t)-r(t)}{r_*}
    \right)^2
    +
    \left(
    \frac{\theta_{\epsilon}(t)-\theta(t)}{p_*}
    \right)^2
    +
    \left(
    \frac{\tau_{1,\epsilon}(t)-\tau_1(t)}{\tau_{1,*}}
    \right)^2
    +
    \left(
    \frac{\tau_{2,\epsilon}(t)-\tau_2(t)}{\tau_{2,*}}
    \right)^2
    \right]^{1/2}.
\end{equation}

Here, $r_*$, $\theta_*$, $\tau_{1,*}$, and $\tau_{2,*}$ are characteristic field scales. We choose them to be the corresponding field values at the saddle point of the potential. This normalization is particularly useful because the four fields evolve over different numerical scales; without it, the distance would be artificially dominated by fields whose absolute values are intrinsically larger.

Since the trajectories may eventually converge to the same minimum, their separation at the final integration time does not necessarily provide a meaningful measure of sensitivity. A perturbed trajectory may deviate substantially from the reference trajectory at intermediate times and nevertheless approach the same minimum asymptotically. Then, it is possible characterize the sensitivity using the maximum separation attained during the evolution,
$\Delta_{\mathrm{max}}= \max_{0\leq t\leq t_f}\Delta(t)
$. A dimensionless sensitivity measure can then be defined as
$S_{\mathrm{max}}
    =
    \frac{\Delta_{\mathrm{max}}}{|\epsilon|}$. This quantity measures the maximum deviation of the trajectory generated by a given relative perturbation of the initial condition. By applying perturbations of the same relative magnitude to each field separately, $S_{\mathrm{max}}$ provides a simple way to compare which initial conditions have the strongest influence on the subsequent field-space trajectory.

For each of the four fields, the sensitivity analysis is performed independently
by introducing a small perturbation $\epsilon$ in the corresponding initial
condition. We find that a perturbation of order $\epsilon = 10^{-4}$ is
sufficient to produce an appreciable deviation from the reference geodesic.

Figure~\ref{fig:sensitivity} shows the evolution of $\Delta_i(t)$ for two
representative flux configurations. For the configuration displayed in the
left panel, the scalar potential develops both a minimum and a saddle point
within the conifold patch. At the saddle point, the mass matrix possesses a
negative eigenvalue $\lambda_1$ whose associated eigenvector is
\begin{equation}
    \mathbf{v}_1
    =
    \left(
    0.7405,\,
    -0.6443,\,
    -0.1843,\,
    -0.0481
    \right).
\end{equation}
In contrast, the flux configuration shown in the right panel gives rise only
to a saddle point within the LCS patch. The eigenvector associated with the
negative eigenvalue $\lambda_2$ of the mass matrix evaluated at this critical point is
\begin{equation}
    \mathbf{v}_2
    =
    \left(
    -0.1829,\,
    0.2475,\,
    0.5531,\,
    -0.7741
    \right).
\end{equation}

As shown in Figure~\ref{fig:sensitivity}, the evolution of $\Delta_i(t)$ in
both cases provides a dynamical counterpart to the information encoded in the
mass-matrix eigenvectors, whose components characterize the combination of
fields defining the tachyonic direction. In the left panel, perturbations
along $\tau_1$ and $r$ produce the largest deviations from the reference
geodesic, whereas in the right panel the evolution is predominantly sensitive
to perturbations along $\tau_1$ and $\tau_2$.

These results should be considered together with the previous analysis of the
unstable directions, where a broader set of mass-matrix eigenvectors was
examined. Across those configurations, the components associated with
$\tau_1$ and $\tau_2$ were found to contribute prominently to the unstable
eigenvectors. The sensitivity analysis presented here is consistent with this
general pattern from a dynamical perspective: although the relative
contribution of each field depends on the particular flux configuration, the
canonically normalized fields $\phi_{\tau_1}$ and $\phi_{\tau_2}$ repeatedly
emerge as important contributors to the unstable directions.

A more systematic sensitivity analysis would be required to quantify this
behavior across the full set of flux configurations and lies beyond the main
scope of this work. Nevertheless, the present procedure provides a useful
consistency check of the information extracted from the mass-matrix
eigenvectors and offers a complementary dynamical perspective on the
evolution of the fields.

\subsubsection*{The potential and its logarithmic slope along the evolution}

\begin{figure}[H]
	\centering
	\includegraphics[width=0.78\linewidth]{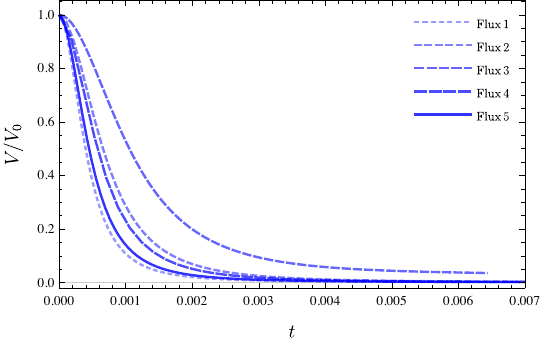}
	\caption{Relative evolution $V/V_0$ of the multifield potential in time, for different flux configurations. All paths begin at $r_0=5\times10^{-7}$, $\theta_0=-5\pi$, $\tau_{1,0}=0$ and $\tau_{2,0}=10$.} 
	\label{fig:P1v}
\end{figure}

The behavior of the scalar potential itself along these trajectories is
shown in Figure ~\ref{fig:P1v}, normalized to its initial
value. In every case the potential decreases monotonically, by up to an
order of magnitude, as the fields roll towards the corresponding critical
point.

\begin{table}[h]
\caption{Table of different flux choices associated with trajectories fo fields shown in Figure 11 and Figure 12.}
\label{tab:gammaA}
\small
\begin{center}
\begin{tabular}{|c|c|}
\hline
Flux choice &
$(F_{(3)},H_{(3)})$
\\ \hline
$1$&
$(-11, 5, 5, 13,-9, -9, 10, 2)$ 
\\ \hline
$2$&
$(-1, 2, 10, -2,-5, -9, -5, 1)$ 
\\ \hline
$3$&
$(-10, 20, 16, 17,-11, 3, -7, -5)$ 
\\ \hline
$4$&
$(-11, 7, 11, -20,-3, 18, 3, -5)$ 
\\ \hline
$5$&
$(18, -5, -17, -12,7, 10, -9, -2)$ 
\\ \hline
\end{tabular}
\end{center}
\end{table}

\begin{figure}[H]
	\centering
	\includegraphics[width=0.78\linewidth]{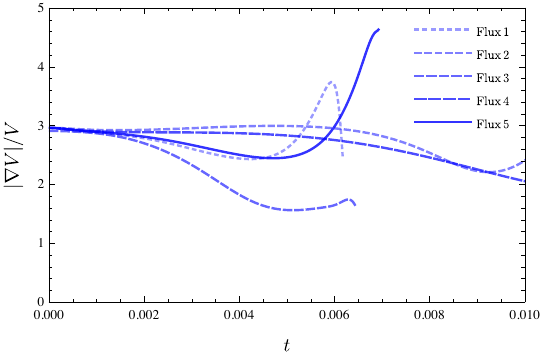}
	\caption{  Evolution of the quotient of the absolute value of the gradient of the scalar field and the scalar field versus the time.
    \label{fig:P1grad}}
\end{figure}

The quantity of direct relevance for the conjecture is the logarithmic
slope of the potential, evaluated along the trajectory rather than along
a fixed asymptotic direction. Its evolution is shown in
Figure~\ref{fig:P1grad}, and provides the numerical
counterpart of the analytic results of
Secs.~\ref{sec:tcc-lcs}--\ref{sec:tcc-tau1}: the quotient remains of
order one throughout the evolution and stays above the four-dimensional
TCC value $\sqrt{2/3}$, as anticipated by the bound
\eqref{eq:master-bound}.

\subsubsection*{Field displacements}

The net field displacements along the trajectories, together with the
canonical distance \eqref{eq:sdist} and the number of e-folds of
accelerated expansion, are collected in Tables~\ref{tab:paths} and
\ref{tab:paths2}. The displacements are sub-Planckian in every component,
while the canonical distance is of order one; the two differ by factors
between $1.3$ and $6.2$, which measures the extent to which the
trajectories are curved in field space rather than radial. This
distinction matters for the field-range bound discussed in
Sec.~\ref{sec:species}, where the relevant quantity is the geodesic
distance between the endpoints and not the length of the path.

\begin{table}[H]
\caption{Field displacements along the multifield evolution. These values are also used to estimate the
species scale in Sec.~\ref{sec:species}. Here, $V_0$ denotes the initial value of the scalar potential,
$\Delta V$ is its variation along the trajectory, and
$\Delta\phi_r$, $\Delta\phi_\theta$, $\Delta\phi_{\tau_1}$, and
$\Delta\phi_{\tau_2}$ denote the corresponding displacements of the
canonically normalized fields. The quantity $s$ is the affine parameter
\eqref{eq:traj-length}, while the last column gives the total number of
$e$-folds.}
\label{tab:paths}
\resizebox{\linewidth}{!}{
\begin{tabular}{|c|c|c|c|c|c|c|c|c|}
\hline
Path&
$V_0$&
$\Delta V$&
$\Delta \phi_r$&
$\Delta \phi_\theta$&
$\Delta \phi_{\tau_1}$&
$\Delta \phi_{\tau_2}$ &
$s$&
$e$-folds
\\ \hline
$\beta_1$ &
$4750.25$&
$-4207.061$&
$0.1325$&
$-0.531$&
$0.1051$&
$-0.1339$&
$2.103$&
$12.0989$
\\ \hline
$\beta_2$ &
$1066.37$&
$-496.412$&
$0.0497$&
$-0.222$&
$0.0699$&
$-0.09183$&
$0.5564$&
$1.5055$
\\ \hline
$\beta_3$ &
$694.639$&
$-399.14$&
$0.06109$&
$-0.0664$&
$-0.01734$&
$0.08468$&
$0.29590$&
$1.6653$
\\ \hline
$\beta_4$ &
$643.88$&
$-12.6658$&
$0.02011$&
$0.01376$&
$-0.00447$&
$0.00604$&
$0.1571$&
$15.083$
\\ \hline
$\gamma_1$ &
$395.957$&
$-369.0955$&
$-0.2927$&
$0.2203$&
$0.6628$&
$-0.4683$&
$3.1209$&
$12.5725$
\\ \hline
$\gamma_{2,1}$ &
$59.652$&
$-21.8457$&
$0.08031$&
$-0.1542$&
$-0.2485$&
$-0.6242$&
$1.5889$&
$28.79$
\\ \hline
$\gamma_{2,2}$ &
$37.7185$&
$-10.862$&
$-0.2835$&
$-0.0498$&
$-0.0561$&
$-0.4955$&
$1.1661$&
$10.8205$
\\ \hline
\end{tabular}
}
\end{table}

The dependence on the initial value of the axionic phase is illustrated
by Tables~\ref{tab:gammaA} and \ref{tab:gammaB}, which repeat the
analysis for $\theta_0=\pi/4$ and $\theta_0=7\pi/4$ with otherwise
identical initial conditions. The displacements change by factors of
order one, and in particular $\Delta\phi_{\tau_1}$ remains negligible in
both cases, consistent with the discussion of the axionic direction in
Sec.~\ref{sec:tcc-tau1}.

\begin{table}[H]
\caption{Field displacements for the path $\gamma_1$, beginning at
$r_0=1\times10^{-5}$, $\theta_0=\pi/4$, $\tau_{1,0}=0$ and
$\tau_{2,0}=1\times10^{4}$.}
\label{tab:paths2}
\small
\begin{center}
\begin{tabular}{|c|c|c|c|c|c|c|}
\hline
$(H_{(3)},F_{(3)})$&
$\Delta \phi_r$&
$\Delta \phi_\theta$&
$\Delta \phi_{\tau_1}$&
$\Delta \phi_{\tau_2}$ &
$s$&
$e$-folds
\\ \hline
$(-1, -10, 8, 1, 9, -7, -6, 4)$ &
$0.3839$&
$-0.0708$&
$-0.00003$&
$-0.2325$&
$0.4549$&
$0.8050$
\\ \hline
$(-5, -9, -5, 1, -1, 2, 10, -2)$ &
$0.3902$&
$-0.0899$&
$-0.00007$&
$-0.238$&
$0.4678$&
$0.8204$
\\ \hline
$(-3, 18, 3, -5, -11, 7, 11, -20)$ &
$0.3469$&
$-0.2129$&
$-0.1676$&
$-5.8364$&
$7.7016$&
$3.1979$
\\ \hline
$(0, -10, -1, 3, -1, 1, 4, 8)$ &
$0.3678$&
$-0.5289$&
$-0.2015$&
$-6.0704$&
$6.6622$&
$2.6938$
\\ \hline
$(0, 7, -1, -3, -1, -4, -2, -9)$ &
$0.0993$&
$0.0260$&
$0.0871$&
$-6.2965$&
$7.4439$&
$4.6570$
\\ \hline
$(-2, -8, 1, 0, 1, 3, -9, 10)$ &
$0.0416$&
$-0.1959$&
$0.0098$&
$-5.3484$&
$5.9770$&
$4.6398$
\\ \hline
\end{tabular}
\end{center}
\end{table}

\begin{table}[H]
\caption{Field displacements for the path $\gamma_2$, beginning at
$r_0=1\times10^{-5}$, $\theta_0=7\pi/4$, $\tau_{1,0}=0$ and
$\tau_{2,0}=1\times10^{4}$, that is, differing from
Table~\ref{tab:gammaA} only in the initial axionic phase.}
\label{tab:gammaB}
\small
\begin{center}
\begin{tabular}{|c|c|c|c|c|c|c|}
\hline
$(H_{(3)},F_{(3)})$&
$\Delta \phi_r$&
$\Delta \phi_\theta$&
$\Delta \phi_{\tau_1}$&
$\Delta \phi_{\tau_2}$ &
$s$&
$e$-folds
\\ \hline
$(-1, -10, 8, 1, 9, -7, -6, 4)$ &
$0.3803$&
$-0.6322$&
$-0.00007$&
$-0.4276$&
$0.8642$&
$0.3639$
\\ \hline
$(-5, -9, -5, 1, -1, 2, 10, -2)$ &
$0.3641$&
$-0.5392$&
$-0.0001$&
$-0.3296$&
$0.7306$&
$0.3098$
\\ \hline
$(-3, 18, 3, -5, -11, 7, 11, -20)$ &
$0.3650$&
$-0.6923$&
$-0.1745$&
$-5.8404$&
$8.1267$&
$3.2427$
\\ \hline
$(0, -10, -1, 3, -1, 1, 4, 8)$ &
$0.3738$&
$-1.0244$&
$-0.2017$&
$-6.0745$&
$7.0686$&
$2.7315$
\\ \hline
$(0, 7, -1, -3, -1, -4, -2, -9)$ &
$0.0853$&
$-0.3976$&
$0.0841$&
$-6.2965$&
$7.6848$&
$4.6741$
\\ \hline
$(-2, -8, 1, 0, 1, 3, -9, 10)$ &
$0.0409$&
$-0.6037$&
$0.0183$&
$-5.3483$&
$6.3880$&
$5.2257$
\\ \hline
\end{tabular}
\end{center}
\end{table}

\subsubsection*{The number of e-folds and a bound on the volume}

A further consequence of the numerical evolution is that the accelerated
expansion is limited in the number of e-folds it can sustain, as the last
column of Tables~\ref{tab:paths}--\ref{tab:gammaB} shows. This is a
direct consequence of the TCC in its original form
\eqref{eq:tcc-original}: writing $N=\log(a_f/a_i)$ and
$\ell_p=M_{pl}^{-1}$, the inequality reads $e^{N}<M_{pl}/H_f$, and using
the Friedmann equation \eqref{eq:friedmann} in the form
$3M_{pl}^2H_f^2\geq V_f$ one obtains
\begin{equation}
    N\;<\;\log\frac{M_{pl}}{H_f}
    \;\leq\;\frac{1}{2}\log\frac{3M_{pl}^4}{V_f}\,,
    \label{eq:efold-bound}
\end{equation}
where $V_f$ denotes the value of the potential at the end of the
expanding phase. The inequality is conservative: the kinetic contribution
to $H_f$ has been dropped, so the true bound on $N$, and hence the bound
on the volume derived below, is tighter than the one quoted.

This bound has a direct implication for the compactification volume. As
discussed in Sec.~\ref{sec:setup}, the K\"ahler moduli enter the
potential only through the overall factor $e^{K_{\rm k}}$, so that, in
reduced Planck units,
\begin{equation}
    V=\frac{\hat{V}(\tau,z)}{\mathcal{V}^2}\,,
    \label{eq:VhatVol}
\end{equation}
with $\hat V$ the dimensionless slice of the potential quoted in our
tables. Substituting \eqref{eq:VhatVol} into \eqref{eq:efold-bound} and
solving for the volume gives
\begin{equation}
    \boxed{\;
    \mathcal{V}\;>\;\mathcal{V}_0\equiv
    e^{N}\sqrt{\frac{\hat{V}_f}{3}}\;}
    \label{eq:volume-bound}
\end{equation}
so that a trajectory sustaining $N$ e-folds is compatible with the TCC
only if the internal volume is large enough. The bound grows
exponentially with the number of e-folds and is therefore controlled by
the longest inflationary stretch found in the scan. Among the
trajectories of Table~\ref{tab:paths}, the strongest case is
$\gamma_{2,1}$, with $N=28.79$ and $\hat V_f=37.81$, which gives
\begin{equation}
    \mathcal{V}\;>\;1.1\times10^{13}\,,
    \qquad
    R\sim\mathcal{V}^{1/6}\sim1.5\times10^{2}
    \label{eq:volume-number}
\end{equation}
in string units. We note in passing that the initial value of the
potential for $\gamma_{2,2}$ coincides, to within the quoted precision,
with the final value for $\gamma_{2,1}$, which suggests that the two are
successive stages of a single evolution. If they are chained, the total
number of e-folds is $N=39.61$ and the bound tightens considerably, to
$\mathcal{V}>4.8\times10^{17}$.

The volume obtained above is large but perfectly admissible, well
inside the regime in which the tree-level K\"ahler potential
\eqref{eq:Kk} and the no-scale structure used throughout this work are
reliable. It is worth stressing that the constraint runs in the opposite
direction to the usual concern about large-volume compactifications:
rather than bounding the volume from above, the TCC requires it to be
{\it bounded from below} once a given amount of accelerated expansion has
taken place. We note finally that \eqref{eq:volume-bound} has been
obtained at fixed K\"ahler moduli, the volume entering only through the
prefactor $e^{K_{\rm k}}$; a complete treatment would require following
the K\"ahler directions along the evolution as well.

Let us collect the results of this section. In the two singular regions
of the complex-structure moduli space the logarithmic slope of the
potential is fixed: it approaches the universal value $2$ at large
complex structure, Eq.~\eqref{eq:ratioLCSvalue}, and it diverges at the
conifold, Eq.~\eqref{eq:ratioC}, which sits at finite distance and where
no bound of this type can be violated. In the axio-dilaton directions the
leading potential takes the single form \eqref{eq:Vmaster}, for which the
axio-dilaton sector contributes exactly $2$ to
$\left|\nabla V\right|^2/V^2$, so that $\left|\nabla V\right|/V\geq
\sqrt2$ throughout, the dilatonic direction saturating the bound while
along the purely axionic one only the directional slope, and not the
norm of the gradient bounded by the conjecture, vanishes, as discussed
in Sec.~\ref{sec:tcc-tau1}. The numerical multifield evolutions confirm
this picture: the quotient stays of order one along every trajectory
analyzed, comfortably above the four-dimensional value $\sqrt{2/3}$.

The most restrictive consequence of the TCC in this setting is therefore
not a constraint on the shape of the potential, which is satisfied
throughout, but the lower bound \eqref{eq:volume-bound} on the
compactification volume implied by the finite number of e-folds that the
evolution can sustain.

\section{Scalar field range bounds from the species scale}
\label{sec:species}

In this section we analyze numerically to what extent the bounds
associated with the species scale are fulfilled in our models. We study
trajectories close to the boundary of the complex-structure moduli space,
that is, in the neighbourhood of the LCS region, which constitutes a limit
of infinite distance.

Two kinds of trajectories occur: some run towards the LCS point, while
others start in the asymptotic region but move away from it. We first
show two representative examples, one in each of the two patches, and
then evaluate the potentials for multiple flux configurations. The field
displacements along the trajectories studied are collected in
Table~\ref{tab:paths}, together with the affine parameter $s$ that
measures the length of each path. The numerical results show that the field displacements remain
bounded and, in fact, sub-Planckian, so that along the multifield
evolution the trajectories never approach the range at which the effective
description would break down at the species scale. As we discuss below,
this compatibility is obtained with a wide margin, and should therefore be
read as a consistency check rather than as a sharp test of the bound. The
comparison of the potential values with the square of the species scale,
on the other hand, is a much more stringent condition, and we return to it
in Sec.~\ref{sec:species-potential}.

\begin{figure}[H]
\centering
\includegraphics[width=.95\linewidth]{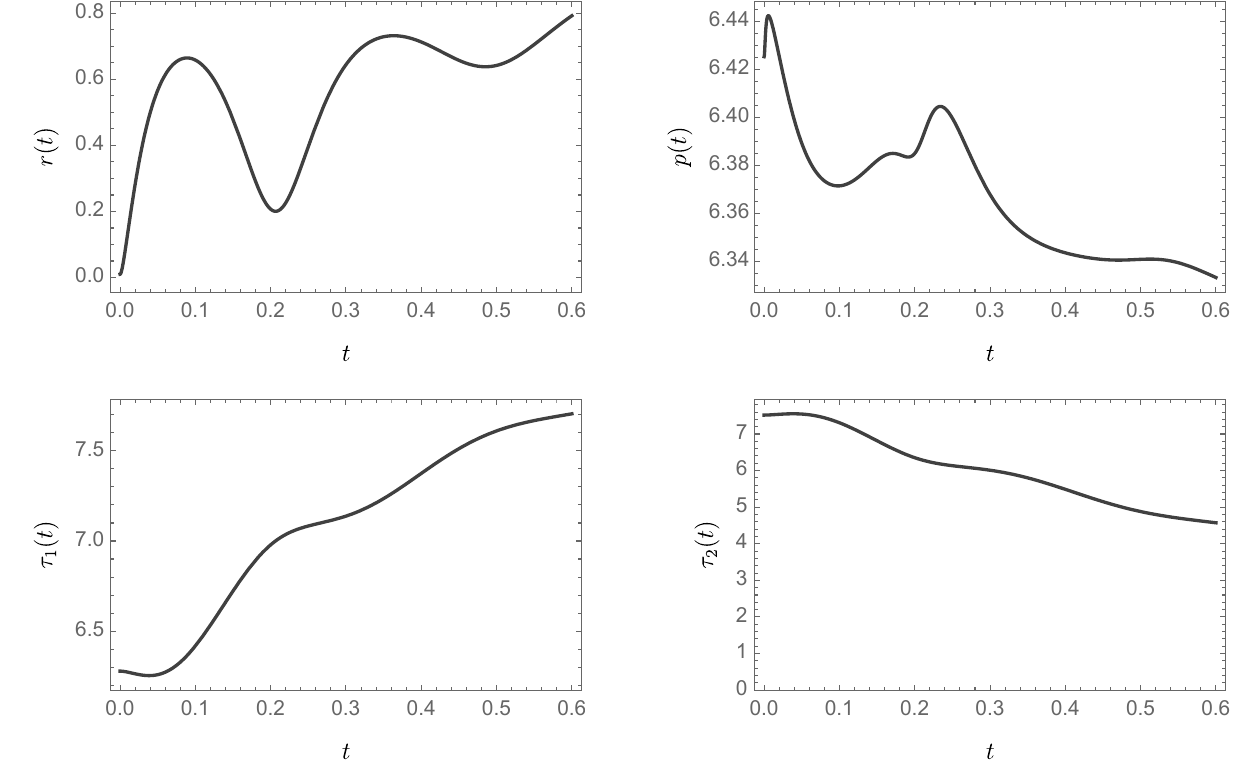}
\caption{Conifold region. Multi-field evolution for the flux
configuration $H_{(3)}=(-1,-1,3,0)$, $F_{(3)}=(17,-1,0,-1)$. Evaluating
the expressions we obtain
$\left|\partial_s V\right|/V\big|_{t=30}=0.07899$, with a dimensionless
distance in moduli space $s=0.8$.}
\label{fig:mfeCON}
\end{figure}

\begin{figure}[H]
\centering
\includegraphics[width=.95\linewidth]{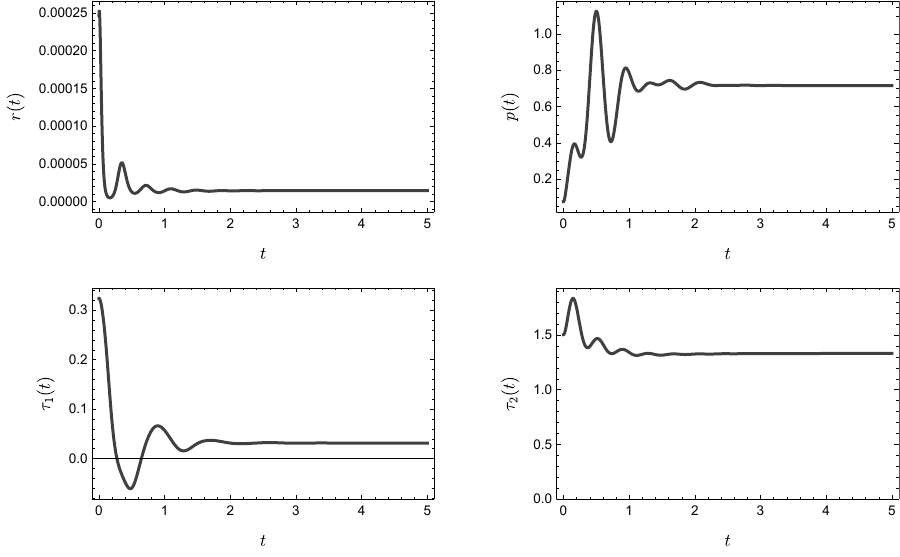}
\caption{LCS region, also known as the point of maximal unipotent
monodromy (MUM). Multi-field evolution for the same flux configuration as
in Figure ~\ref{fig:mfeCON}, $H_{(3)}=(-1,-1,3,0)$ and
$F_{(3)}=(17,-1,0,-1)$. Here
$\left|\partial_s V\right|/V\big|_{t=3}=2.17$, with a dimensionless
distance in moduli space $s=1.33$.}
\label{fig:mfeLCS}
\end{figure}

Figures~\ref{fig:mfeCON} and \ref{fig:mfeLCS} display the multifield
evolution for one and the same flux configuration, in the conifold and in
the LCS patch respectively. The two behave very differently. In the LCS
region the logarithmic slope along the trajectory is
$\left|\partial_sV\right|/V\simeq2.17$, in good agreement with the
universal value $2$ obtained analytically in
Eq.~\eqref{eq:ratioLCSvalue}, and comfortably above the four-dimensional
TCC value $\sqrt{2/3}\simeq0.816$. Near the conifold, instead, the same
quantity is $\left|\partial_sV\right|/V\simeq0.079$, an order of
magnitude below that value. This does not signal a violation of the
conjecture: the conifold sits at {\it finite} distance in moduli space, so
that the asymptotic form of the bound does not apply there, and moreover
the quantity that diverges as the conifold is approached,
Eq.~\eqref{eq:ratioC}, is the norm of the full gradient rather than the
derivative along this particular trajectory. The contrast between the two
figures illustrates how strongly the behaviour of the potential depends on
the region of moduli space, even at fixed fluxes.

\subsection{The species scale and the field-range bound}
\label{sec:species-setup}

The species scale was introduced by Dvali and collaborators
\cite{Dvali:2007hz,Dvali:2010ws}. It states that in a four-dimensional
quantum field theory coupled to gravity and containing $N$ light species,
the effective quantum-gravity cutoff is lowered from the Planck scale to
\begin{equation}
    \Lambda_s
    =
    \frac{M_{\rm Pl}}{\sqrt{N}}\,.
    \label{eq:species-def}
\end{equation}
The original argument relies on the consistency of black-hole evaporation:
a black hole must be able to resolve and decay into the available particle
species without spoiling the semiclassical description.

More recently, the species scale has been studied in string
compactifications in connection with the Distance Conjecture and the
Emergent String Conjecture. In asymptotic regions of moduli space its
logarithmic variation with respect to a canonically normalized scalar
field is expected to be bounded on both sides,
\begin{equation}
    \frac{1}{\sqrt{(d-1)(d-2)}}
    \;\lesssim\;
    M_{\rm Pl}\left|
    \frac{\Lambda_s'(\phi)}{\Lambda_s(\phi)}
    \right|
    \;\lesssim\;
    \frac{1}{\sqrt{d-2}}\,,
    \label{eq:species-slope}
\end{equation}
where the prime denotes a derivative with respect to the canonically
normalized field. The lower bound encodes the exponential decay of
$\Lambda_s$ along infinite-distance limits, while the upper bound is
saturated in emergent-string limits
\cite{CalderonInfante:2024SSDC,vdHeisteeg:2022Species}. In $d=4$ this
reads $1/\sqrt{6}\lesssim M_{\rm Pl}|\Lambda_s'/\Lambda_s|\lesssim
1/\sqrt{2}$.

For a cosmological effective theory one further requires the Hubble scale
to remain below the species scale, $H\lesssim\Lambda_s$. Using the
four-dimensional Friedmann relation $3M_{\rm Pl}^2H^2\simeq V$, this gives
\begin{equation}
    V
    \;\lesssim\;
    3M_{\rm Pl}^2\Lambda_s^2
    \;\;\overset{M_{\rm Pl}=1}{\longrightarrow}\;\;
    V\;\lesssim\;3\Lambda_s^2\,,
    \label{eq:V-vs-Lambda}
\end{equation}
which, up to order-one factors, is commonly written as
$V\lesssim\Lambda_s^2$. Note that this is the same relation that produced
the bound \eqref{eq:volume-bound} on the compactification volume in
Sec.~\ref{sec:tcc-multifield}, with $\Lambda_s$ in place of $M_{\rm Pl}$:
requiring the Hubble scale to stay below the cutoff is a strictly stronger
demand than requiring it to stay below the Planck scale.

For positive and sufficiently slowly varying potentials, the requirement
\eqref{eq:V-vs-Lambda} translates into a logarithmic control of the
allowed field range \cite{Heisteeg:2024Bounds},
\begin{equation}
    \Delta\phi
    \;\leq\;
    -a\log V_0+b
    \;=\;
    a\log\!\left(\frac{1}{V_0}\right)+b\,,
    \label{eq:species-range}
\end{equation}
where $V_0$ is expressed in reduced Planck units and $a>0$ is fixed by the
asymptotic behaviour of the tower of states controlling $\Lambda_s$. The
constant $b$ depends on the normalization of the species scale and on the
location of the desert point. Both constants are expected to be of order
one in the regime where the bound applies
\cite{CalderonInfante:2024SSDC,vdHeisteeg:2024SpeciesTCC}. We stress that
$\Delta\phi$ in \eqref{eq:species-range} is a {\it geodesic} distance in
field space; we return to this point below.

Along a geodesic approaching the large-complex-structure point one has
$F_1\simeq\tfrac{1}{12}c_2\operatorname{Re}t$, where $t$ is the flat
coordinate, mirror to the K\"ahler parameter of the quintic, and
$c_2=50$, so that $\Lambda_s\propto(\operatorname{Re}t)^{-1/2}$
\cite{vdHeisteeg:2022Species}. The trajectories studied in the present
work, however, are multifield and are not, in general, geodesics, so
that no such closed form is available for them. We therefore test the
field-range bound numerically along the corresponding cosmological
evolutions.

\subsubsection*{Canonically normalized variables}

With the conventions fixed in Sec.~\ref{sec:setup}, the kinetic sector of
the axio-dilaton and complex-structure moduli reads
\begin{equation}
    \mathcal{L}_{\rm kin}
    =
    -K_{a\bar b}\,\partial_\mu z^a\partial^\mu\bar z^{\bar b}
    =
    -\frac{1}{2}\,G_{ij}\,\partial_\mu\varphi^i\partial^\mu\varphi^j\,,
    \qquad
    G_{ij}=2K_{a\bar b}\,,
    \label{eq:kinetic}
\end{equation}
so that the {\it locally} canonically normalized variables associated with
$z=re^{i\theta}$ and $\tau=\tau_1+i\tau_2$ are obtained from
$\partial_\mu\phi_i=M_{\rm Pl}\sqrt{2K_{a\bar b}}\,\partial_\mu\varphi^i$,
explicitly
\begin{align}
    \partial_{\mu}\phi_r
    &=
    M_{\rm Pl}\,\sqrt{2K_{z\bar z}}\;\partial_{\mu}r\,,
    &
    \partial_{\mu}\phi_\theta
    &=
    M_{\rm Pl}\,\sqrt{2K_{z\bar z}}\;r\,\partial_{\mu}\theta\,,
    \label{eq:can-z}
    \\[1mm]
    \partial_{\mu}\phi_{\tau_1}
    &=
    M_{\rm Pl}\,\frac{\partial_{\mu}\tau_1}{\sqrt{2}\,\tau_2}\,,
    &
    \partial_{\mu}\phi_{\tau_2}
    &=
    M_{\rm Pl}\,\frac{\partial_{\mu}\tau_2}{\sqrt{2}\,\tau_2}\,,
    \label{eq:can-tau}
\end{align}
where we used $K_{\tau\bar\tau}=1/(4\tau_2^2)$, so that
$\sqrt{2K_{\tau\bar\tau}}=1/(\sqrt{2}\tau_2)$: both sectors follow the
same rule. With \eqref{eq:can-z}--\eqref{eq:can-tau} the kinetic
Lagrangian takes the canonical form
\begin{equation}
    \mathcal{L}_{\rm kin}
    =
    -\frac{1}{2}\sum_i\partial_\mu\phi_i\,\partial^\mu\phi_i,
    \label{eq:Lcanonical}
\end{equation}
locally along the trajectory.

For each component the net displacement between the initial and the final
point is
\begin{equation}
    \Delta\phi_i
    =
    \phi_i(t_f)-\phi_i(t_0)
    =
    \int_{t_0}^{t_f}\dot\phi_i\,dt\,,
    \qquad i=r,\theta,\tau_1,\tau_2\,.
    \label{eq:displacement}
\end{equation}
In a multifield system, however, the invariant quantity is the total
length of the trajectory in field space,
\begin{equation}
    \Delta\phi_{\rm traj}
    =
    \int_{t_0}^{t_f}\!dt\,
    \sqrt{G_{IJ}\bigl(\varphi(t)\bigr)\,
    \dot\varphi^{I}(t)\dot\varphi^{J}(t)}
    =
    \int_{t_0}^{t_f}\!dt\,
    \sqrt{
    \dot\phi_r^{\,2}+\dot\phi_\theta^{\,2}
    +\dot\phi_{\tau_1}^{\,2}+\dot\phi_{\tau_2}^{\,2}}\,,
    \label{eq:traj-length}
\end{equation}
the second equality holding in the locally canonical variables; this is
precisely the affine parameter $s$ of Eq.~\eqref{eq:sdist}. When only the
initial and final values are available we use the endpoint estimate
\begin{equation}
    \Delta\phi_{\rm tot}
    \approx
    \sqrt{
    (\Delta\phi_r)^2+(\Delta\phi_\theta)^2
    +(\Delta\phi_{\tau_1})^2+(\Delta\phi_{\tau_2})^2}\,.
    \label{eq:endpoint}
\end{equation}
It is important to be precise about which of these quantities is being
compared with \eqref{eq:species-range}. The bound refers to the geodesic
distance $\Delta\phi_{\rm geo}$ between the endpoints, and one has
\begin{equation}
    \Delta\phi_{\rm geo}\;\leq\;\Delta\phi_{\rm traj}\,,
    \label{eq:geo-le-traj}
\end{equation}
since the evolution is not a geodesic. The endpoint estimate
\eqref{eq:endpoint} is instead the Euclidean distance in the {\it locally}
canonical coordinates and, the field space being curved, it coincides with
neither of the two; it typically underestimates $\Delta\phi_{\rm traj}$
when the evolution is significantly curved. Consequently, our numerical
statement that the displacements remain bounded should be read as a
consistency check rather than as a sharp test of the conjecture, and we
quote $\Delta\phi_{\rm traj}$ whenever the full evolution is available.

\subsubsection*{Numerical test of the field-range bound}

We consider several types of trajectories in the multifield evolution.
Some of them remain within the LCS region, where the period expansion is
under perturbative control and semi-analytical estimates are possible. The
corresponding displacements are collected in
Table \ref{tab:paths}.

The two measures of distance introduced above differ appreciably. Taking
the ratio $s/\Delta\phi_{\rm tot}$ for the trajectories of
Table~\ref{tab:paths} one finds values between $1.3$ and $6.2$,
so that the endpoint estimate underestimates the length of the path by up
to almost an order of magnitude. This is a measure of how strongly the
trajectories are curved in field space, and it is the reason why the
conclusions below are stated as a consistency check: the quantity plotted
is the smaller of the two, and the bound \eqref{eq:species-range} would be
tested more stringently by $\Delta\phi_{\rm traj}$.

It is worth being explicit about the size of the margin involved. The
largest excursion in Table~\ref{tab:paths} is
$\Delta\phi_{\rm tot}\simeq0.89$ for $\gamma_1$, and even the longest
path length is $s\simeq3.1$, whereas the right-hand side of
\eqref{eq:species-range} for order-one $a$ and $b$ and for the values
$V_0\sim10^{2}$--$10^{3}$ of the table is itself of order one or larger.
The bound is therefore satisfied, but by a factor that is not small, and
this family of trajectories does not probe the regime in which
\eqref{eq:species-range} becomes restrictive. The reason is structural:
the trajectories are trapped by the critical points discussed in
Sec.~\ref{sec:tcc-multifield} and never travel far in field space, so that
the field-range bound cannot be tested sharply within this setup. What the
numerical analysis does establish is the absence of any tension: no
trajectory is driven to a distance at which the tower of states would
invalidate the effective description used to obtain it.

\begin{figure}[H]
\centering
\includegraphics[width=.6\linewidth]{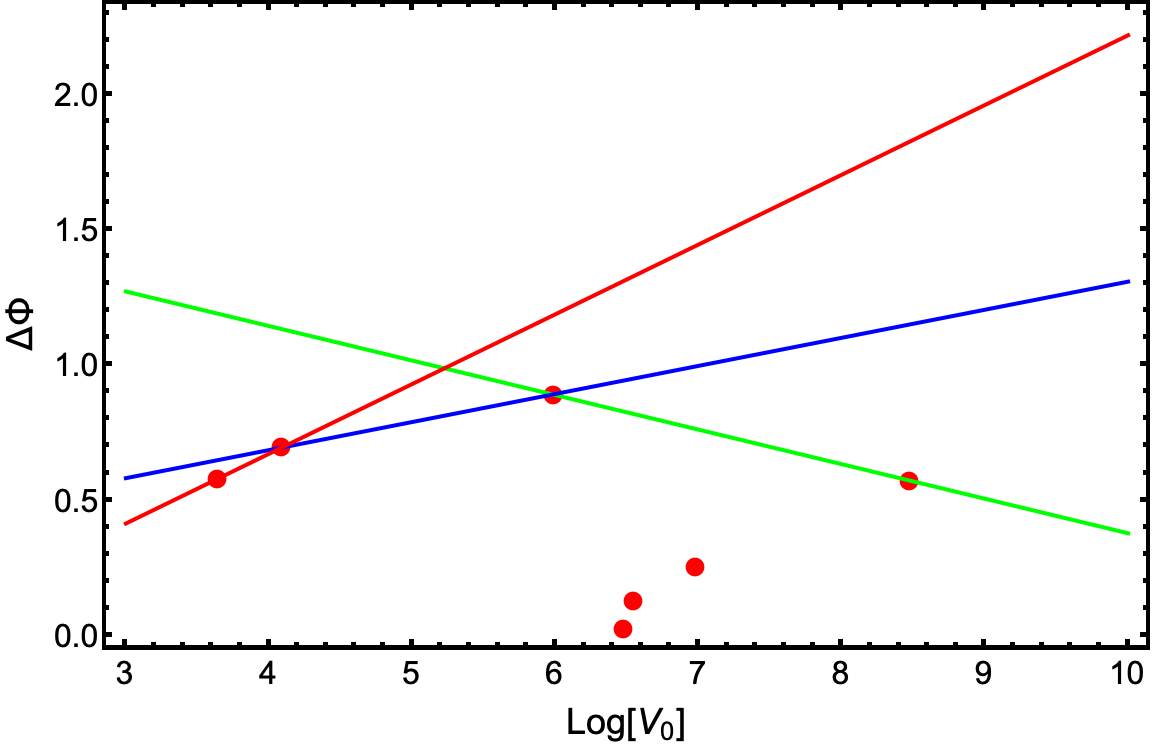}

\caption{
Endpoint estimates of the total field displacement
$\Delta\phi_{\rm tot}$, Eq.~\eqref{eq:endpoint}, as a function of
$\log V_0$. The red and green lines correspond, respectively, to
$(a,b)=(0.257867,-0.360319)$ and $(0.103775,0.269690)$ in
$\Delta\phi=-a\log V_0+b$; the blue line, $(a,b)=(-0.127688,1.65414)$,
has $a<0$ and does not have the form dictated by
Eq.~\eqref{eq:species-range}. The lines are empirical comparisons with the
numerical data and should not be interpreted as independent determinations
of the universal coefficients of the species-scale bound.
}
\label{fig:species-lines}
\end{figure}

As a first exploratory test of \eqref{eq:species-range} we compare the
seven computed endpoint displacements with linear functions of $\log V_0$.
The lines shown in Figure ~\ref{fig:species-lines} correspond to
\begin{equation}
    (a,b)
    =
    (0.257867,\,-0.360319)\,,
    \qquad
    (0.103775,\,0.269690)\,,
    \label{eq:ab-values}
\end{equation}
which provide empirical upper envelopes for different subsets of the
numerical data. A third linear envelope obtained from the fit,
$(a,b)=(-0.127688,\,1.65414)$, has $a<0$ and is therefore {\it not} of the
form required by \eqref{eq:species-range}; we display it in
Figure~\ref{fig:species-lines} only for completeness. We stress that this comparison is purely illustrative. With seven data
points and two free parameters, a linear envelope can always be found, and
its coefficients carry no information about the universal constants of
\eqref{eq:species-range}, which in addition require $a>0$, $V_0$ expressed
in reduced Planck units, and an evolution in the slowly varying,
positive-potential regime. Extracting $a$ and $b$ meaningfully would
require sampling trajectories with a much wider spread in $V_0$ and in
field-space distance than the ones accessible here.

\subsection{Comparing the flux potential with the species scale}
\label{sec:species-potential}

We now perform a more direct test, by comparing the scalar potential with
the square of the moduli-dependent species scale. Following
\cite{vdHeisteeg:2022Species} we use
\begin{equation}
   \Lambda_s
   =
   \frac{M_{\rm Pl}}{\sqrt{F_1+N_{\rm des}}}
   \;<\;
   \frac{M_{\rm Pl}}{\sqrt{F_1}}\,,
   \label{eq:LambdaF1}
\end{equation}
where $F_1$ is the genus-one topological free energy and $N_{\rm des}$
parametrizes the finite contribution associated with the desert point. In
the numerical analysis we follow \cite{vdHeisteeg:2022Species} and take
$N_{\rm des}=1$.

The genus-one free energy is obtained from the integrated
Bershadsky--Cecotti--Ooguri--Vafa (BCOV) relation \cite{BCOV93},
\begin{equation}
F_1
=
\frac{1}{2}
\left(
3+h^{1,1}-\frac{\chi}{12}
\right)
K(z,\bar z)
-\frac{1}{2}\log\det|G_{i\bar j}|
+\log|f(z)|^2\,,
\label{eq:BCOV}
\end{equation}
where $\chi$ is the Euler characteristic, $K$ the K\"ahler potential on the
relevant moduli space, $G_{i\bar j}$ the corresponding metric and $f(z)$
the holomorphic ambiguity, fixed by the boundary conditions at the
singular loci. Expression \eqref{eq:BCOV} is defined up to a
moduli-independent additive constant, which corresponds to a
normalization ambiguity of the inferred species scale; this ambiguity
enters the comparison performed below and we comment on it at the end of
this subsection.

\begin{figure}[H]
\centering
\includegraphics[width=.6\linewidth]{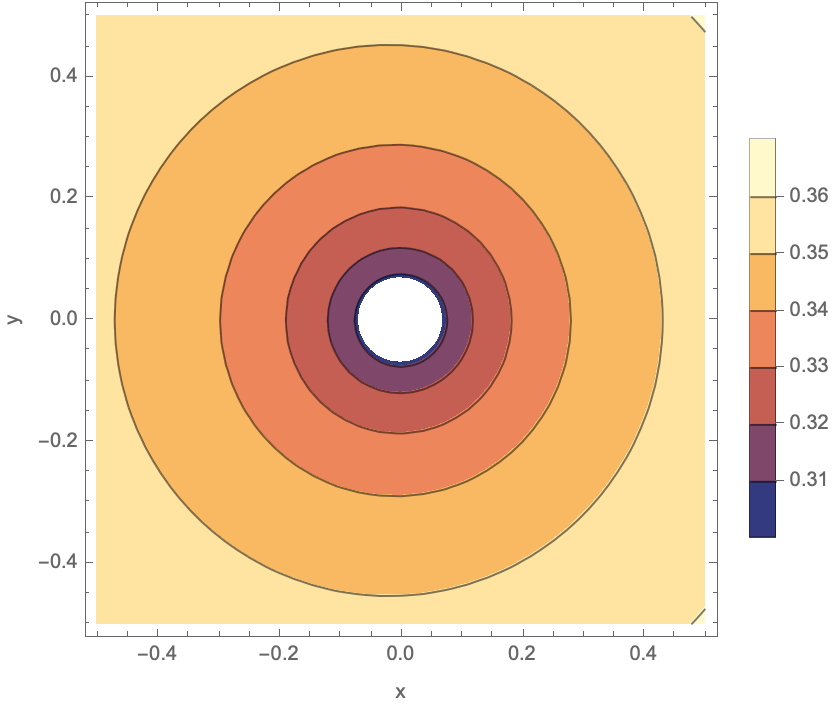}

\caption{
Contour plot of the geometric estimate
$\Lambda_s=M_{\rm Pl}/\sqrt{F_1}$, Eq.~\eqref{eq:LambdaF1},
near the LCS point of the mirror quintic, with $M_{\rm Pl}=1$. The species-scale condition places a strong restriction
on the allowed magnitude of the flux-induced scalar potential at fixed volume. The figure
illustrates a partial exploration for a representative fixed choice of
fluxes; a complete scan over flux configurations is left for future work.
}
\label{fig:F1contour}
\end{figure}

For the mirror quintic in type IIB we employ the genus-one free energy of
the B-model complex-structure moduli space. This quantity underlies the
proposal of a moduli-dependent species scale in
\cite{vdHeisteeg:2022Species}, originally formulated in the underlying
four-dimensional $\mathcal{N}=2$ Calabi--Yau compactification and
subsequently applied to type IIB orientifolds with fluxes in
\cite{vdHeisteeg:2024SpeciesTCC}. Using $h^{1,1}=1$ and $\chi=-200$ for
the quintic in $\mathbb{P}^4$,\footnote{Note that here $\chi$ refers to
the quintic itself and not to its mirror, whose Euler characteristic is
$+200$; see Sec.~\ref{sec:quinticgeometry}.} the prefactor in
\eqref{eq:BCOV} evaluates to $\tfrac{1}{2}(3+1+200/12)=31/3$, and the
explicit expression in the convention of \cite{vdHeisteeg:2022Species}
reads
\begin{equation}
    F_1
    =
    \frac{31}{3}K(z,\bar z)
    -\frac{1}{2}\log|G_{z\bar z}|
    +
    \log\left|
    z^{-31/12}\,(1-5^5z)^{-1/12}
    \right|^2 .
    \label{eq:F1quintic}
\end{equation}
The exponents in the holomorphic ambiguity encode the required behaviour
at the LCS and conifold loci; in particular the second one reproduces the
universal $-\tfrac{1}{12}\log\Delta$ behaviour near the conifold
discriminant $\Delta=1-5^5z$, consistently with the location of the
conifold point discussed in Sec.~\ref{sec:quinticgeometry}.
Equation \eqref{eq:LambdaF1} can then be evaluated along a trajectory
$z(t)$, keeping the K\"ahler moduli fixed in the present approximation.
Figure~\ref{fig:F1contour} shows the resulting species scale near the LCS
point.

In this approximation $F_1$, and hence $\Lambda_s$, depends only on the
complex-structure modulus. By contrast, the flux potential
\eqref{eq:Vexplicit} depends explicitly on the NS--NS and R--R flux quanta
and on the axio-dilaton, so that the condition $V\lesssim3\Lambda_s^2$
does not follow automatically from the geometry: it constrains the allowed
fluxes, the axio-dilaton and the region of complex-structure moduli space.
We therefore explore whether
\begin{equation}
    V\;<\;\frac{3}{F_1+N_{\rm des}}
    \label{eq:species-test}
\end{equation}
holds in finite regions of parameter space for representative flux
configurations.

Before quoting numbers, one caveat must be stated, because it conditions
the interpretation of the whole comparison. The species scale
\eqref{eq:LambdaF1} is computed from the genus-one free energy of the
underlying $\mathcal{N}=2$ Calabi--Yau, and therefore counts the states of
the tower associated with the {\it complex-structure} sector; it depends
on $z$ alone. The flux potential, on the other hand, has a magnitude
governed by the compactification volume and the string coupling through
the prefactor $e^{K_{\rm k}}$ and the flux quanta. The two sides of
\eqref{eq:species-test} thus refer to different sectors of the
compactification, one of which has been frozen. The comparison below
should be read accordingly.

With that proviso, our preliminary numerical exploration indicates that,
{\it at fixed compactification volume}, the region satisfying
\eqref{eq:species-test} is considerably more restricted than the regions
satisfying the RdSC and the TCC conditions. Three comments are in
order.

First, since the flux superpotential is linear in the flux quanta, the
scalar potential scales quadratically under a common rescaling of the
fluxes, whereas the geometric estimate of $F_1$ is unchanged at fixed
complex structure. Reducing the flux magnitudes therefore facilitates
\eqref{eq:species-test}. This is, however, not a free parameter: the flux
quanta are integers, the tadpole contribution $N_{\rm flux}$ is itself
quadratic in them, as recalled in \eqref{eq:fluxrescaling}, and small
fluxes are in tension with the requirement of stabilizing the axio-dilaton
and complex-structure moduli.

Second, and most importantly, the K\"ahler sector has been frozen.
Writing $V=e^{K_{\rm k}}\hat V(\tau,z)\sim\mathcal{V}^{-2}\hat V(\tau,z)$,
the condition \eqref{eq:species-test} becomes
\begin{equation}
    \hat V(\tau,z)\;\lesssim\;3\,\mathcal{V}^{2}\,\Lambda_s^{2}\,,
    \label{eq:species-volume}
\end{equation}
whose right-hand side grows without bound as $\mathcal{V}\to\infty$ if
$\Lambda_s$ is treated as volume independent, as it is in the present
approximation. In that approximation the condition is therefore satisfied
for any flux configuration at sufficiently large volume, and the statement
that it is restrictive holds {\it only} at fixed $\mathcal{V}$. This
should not be read as a bound in itself: in the full orientifold
$\Lambda_s$ does depend on the volume, through the Kaluza--Klein towers of
the compactification, and it is that dependence which determines whether
\eqref{eq:species-volume} constrains $\mathcal{V}$ from above. Should it
do so, the result would combine with the lower bound
\eqref{eq:volume-bound} obtained from the TCC into a genuine window for
the compactification volume, which is arguably the most interesting
question raised by the present analysis. Establishing it requires
identifying the tower that dominates $\Lambda_s$ in the orientifold and
its scaling with $\mathcal{V}$, which we leave for future work. We note
that the two conditions are closely related to begin with, since
\eqref{eq:V-vs-Lambda} is nothing but \eqref{eq:efold-bound} with the
species scale in place of the Planck mass, and is therefore strictly
stronger. The dependence on $g_s$ must likewise be examined using the
complete normalization of the four-dimensional Einstein-frame potential,
and should not be inferred solely from the explicit prefactor in
\eqref{eq:Vexplicit}.

Third, \eqref{eq:BCOV} is defined up to an additive constant and $F_1$ is
not positive definite over the whole moduli space; we restrict the
comparison to the region where $F_1+N_{\rm des}>0$, and we note that the
quantitative strength of the bound inherits the corresponding
normalization ambiguity of $\Lambda_s$.

With these caveats, we find representative regions in which
\eqref{eq:species-test} is satisfied, while a systematic scan over flux
choices, the axio-dilaton and the complex-structure modulus is beyond the
scope of the present work. Summarizing, of the three conditions studied
in this paper the field-range bound \eqref{eq:species-range} is met with
a wide margin and does not constrain our models, whereas the requirement
$V\lesssim3\Lambda_s^{2}$ is the only one that fails over sizeable regions
of parameter space, and is in this sense the most restrictive of the
three at fixed compactification volume. Whether it remains so once the
volume dependence of $\Lambda_s$ is restored is, as discussed above, an
open question, and the natural continuation of this work.

\section{Conclusions and Outlook}
\label{sec:conclusions}

We have explored three swampland conjectures in the setup of type IIB
compactifications on one-parameter Calabi--Yau manifolds: the refined de
Sitter conjecture, the Trans-Planckian Censorship Conjecture and the
bounds associated with the species scale. Restricting to
$h^{2,1}=1$ has allowed us to work with the exact periods obtained from
the Picard--Fuchs equations rather than with their large-complex-structure
expansion, and therefore to explore the neighbourhood of the conifold on
the same footing as the asymptotic region. Evaluating the three conditions
on one and the same set of models has also made it possible to compare
their relative strength, which is difficult when each of them is studied
in a different setup.

We have made a numerical study of no-scale potentials arising from type
IIB string theory compactifications on the CY mirror quintic in
$\mathbb{P}^4$, a variety with Hodge numbers
$(h^{2,1},h^{1,1})=(1,101)$. We explored multiple flux configurations,
and studied the critical points of the resulting potential. A convergence
study of the extrema was carried out in terms of the series for the
periods of the compactification: we determine that $\sim50$ terms are
required to discard apparent vacua and to reach convergence of the
solutions. Since the K\"ahler directions remain flat at this stage, these
are critical points of the slice $\hat V(\tau,z)$ with $\hat V>0$ rather
than de Sitter minima of the full potential.

We find a statistical satisfaction of the RdSC: of the $\sim10^2$ models
in which a convergent critical point was identified, only two violated the
first condition.
%porcentaje
The conjecture is analyzed in two steps, first by stabilizing the
axio-dilaton and the complex structure with the flux-generated potential,
and then by considering a K\"ahler modulus that dominates the process and
determining its influence on the fulfilment of the conjecture. In fact,
the outcome of the second step is stronger than a statistical statement.
At any critical point of $\hat V$ the whole gradient of the full potential
comes from the K\"ahler directions, and the no-scale condition fixes its
logarithmic slope to the universal value
\begin{equation}
    \frac{\left|\nabla V\right|}{V}\bigg|_{z_0,\tau_0}=\sqrt{6}\,,
\end{equation}
independently of the number of K\"ahler moduli, of the detailed form of
the volume, and of the location of the extremum in complex-structure
moduli space, the only input being that the K\"ahler sector be exactly
no-scale. The first condition of the RdSC is therefore satisfied at every
extremum, for any $c\leq\sqrt6$, and the tension associated with a
positive minimum of $\hat V$ is resolved by the K\"ahler runaway.

This value is the familiar volume runaway rate of type IIB, obtained
explicitly for toroidal GKP models in \cite{CaboBizet:2019sgr} and
appearing as an asymptotic exponent in \cite{Casas:2024LightTowers},
and the contraction $K^{a\bar b}K_aK_{\bar b}=n$ was used in
\cite{CalderonInfante:2022AAE} to bound the de Sitter coefficient from
below by $\sqrt{2n}$, with $n=3$ at large volume. What we add is that
the bound is saturated at every critical point of $\hat V$, that $n$ is
twice the degree of homogeneity of the volume and hence equals $3$ for
any volume cubic in the two-cycle sizes, and that the resulting
$\sqrt6$, rather than obstructing acceleration, is what prevents these
critical points from being de Sitter vacua.

We also characterized the combination of fields defining the unstable
direction. Taken together, the eigenvector analysis and the geodesic
deviations suggest that the canonically normalized fields
$\phi_{\tau_1}$ and $\phi_{\tau_2}$ provide the dominant contributions to
the unstable direction, whereas the heaviest modes are dominated by the
complex-structure components. The two sectors thus separate statistically,
with the complex structure accounting for the heavy, stable directions and
the axio-dilaton for the light, unstable ones.

We check the Trans-Planckian Censorship Conjecture by considering
four-dimensional spacetime solutions with FRW metric. In those models we
have an evolving multifield scalar potential; we write the scalar field
equations and the Einstein equations and study the trajectories of the
multifield evolution. We employ the affine parameter to describe the
trajectories of the real fields $r$, $\theta$, $\tau_1$ and $\tau_2$;
trajectories parametrized by the affine parameter are equivalent to
trajectories in the canonically normalized directions. This conclusion
should hold when considering potential terms that aid in the stabilization
of the K\"ahler moduli, because the trajectories considered are possible
in the complete moduli space. Studying the asymptotic regions of moduli
space, we obtain that the TCC is satisfied in all the examples analyzed.
Two of these results are analytic and hold for the whole class of
one-parameter models. In the large-complex-structure limit the
logarithmic slope approaches the universal value
$\left|\nabla V\right|/V=2$, the triple intersection number cancelling
between numerator and denominator; near the conifold, which sits at
finite distance, the same quantity diverges, so that no bound of this
form can be violated there. In the axio-dilaton directions the leading
potential takes a single $SL(2)$-covariant form, for which the
axio-dilaton sector contributes exactly $2$ to
$\left|\nabla V\right|^2/V^2$ at every point of the $(\tau_1,\tau_2)$
half-plane, so that $\left|\nabla V\right|/V\geq\sqrt2$ throughout, which
is the quantity the conjecture bounds in its multifield form
\cite{Bedroya:2019snp}. The dilatonic direction saturates this value. We
note that along the purely axionic direction the {\it directional} slope
\eqref{eq:tcc-directional} instead decays; this does not affect the
above, since the bound \eqref{eq:tcc-both} applies to the norm of the
gradient, and in addition the potential grows in that direction, so that
no causal trajectory starting from the interior of moduli space is driven
along it.

The most restrictive consequence of the TCC in this setting is not a
constraint on the shape of the potential, which is satisfied throughout,
but a bound on the compactification volume. The number of e-folds that an
expanding phase can sustain is limited by the original form of the
conjecture, $N<\tfrac12\log(3M_{pl}^4/V_f)$, and since the K\"ahler
moduli enter the potential only through the overall factor
$e^{K_{\rm k}}$, this translates into
\begin{equation}
    \mathcal{V}\;>\;\mathcal{V}_0\equiv e^{N}\sqrt{\hat V_f/3}\,,
\end{equation}
in reduced Planck units. The longest inflationary stretch found in our
scan, with $N=28.79$, gives $\mathcal{V}>1.1\times10^{13}$, that is a
linear size $R\sim\mathcal{V}^{1/6}\sim1.5\times10^{2}$ in string units.
The constraint therefore runs opposite to the usual concern about
large-volume compactifications: once a given amount of accelerated
expansion has taken place, the internal volume is bounded from
{\it below}. Regarding the potential application of these multifield
models to inflation, it is important to notice that a single patch of the
complex-structure moduli space is not sufficient to achieve a large number
of e-folds.

Finally, in the trajectories studied we determine the bound on the scalar
field displacements. The displacements are below the Planck scale, in
agreement with the bounds identified from the species scale. We stress,
however, that this compatibility is obtained with a wide margin: the
largest excursion in our sample is $\Delta\phi_{\rm tot}\simeq0.89$, and
the trajectories are trapped by the critical points of the flux potential
and never travel far in field space, so that the field-range bound is not
tested sharply within this setup. What the analysis establishes is the
absence of any tension. We find that the overall bound
$V\lesssim\Lambda_s^2$ is a much more restrictive requirement on the
models. In the approximation used here $\Lambda_s$ is determined by the
complex-structure moduli and is independent of the fluxes and of the
axio-dilaton, and one needs potentials in a restricted regime in order to
accomplish this bound. This last statement, however, holds at fixed
compactification volume. Writing the condition as
$\hat V\lesssim3\,\mathcal{V}^2\Lambda_s^2$ and restoring the volume
dependence of the species scale, which for the Kaluza--Klein and string
towers of an isotropic compactification scales as
$\Lambda_s\sim\mathcal{V}^{-1/2}$, the condition becomes
$\hat V\lesssim3\,g_s^{3/2}\mathcal{V}$ and is satisfied by many orders of
magnitude at the volumes required by the TCC. More generally, for
$\Lambda_s\sim\mathcal{V}^{-p}$ the condition would bound the volume from
above only if $p>1$, which the species-scale slope bound itself excludes.
The two conditions are thus not in competition: the lower bound from the
TCC is the genuine constraint, and the species-scale requirement follows
from it.

As a next exploration step one has to account for the stabilization of the
K\"ahler moduli. One would need to explore the geometries considered here,
as well as others, in order to determine the extent to which these
conjectures are satisfied in type IIB CY orientifold compactifications.
Three directions seem to us particularly worth pursuing: a careful
treatment of the species scale in the orientifold, including the
anisotropic case, where the exponent $p$ may vary within the window
allowed by the slope bound; extending the analysis to the remaining
one-parameter models, for which the universality of the value $2$ at
large complex structure suggests that part of our conclusions carries
over unchanged; and following the K\"ahler directions along the
cosmological evolution rather than freezing them, which is the main
approximation made throughout this work.

\section{Acknowledgements}
We thank Alejandro Cabo Montes de Oca, Alon Faraggi, Shinji Hirano, Albrecht Klemm,
Oscar Loaiza Brito, Damián Mayorga Peña, Yessenia Olgu\'in, Annette Torres, Susha
Parameswaran, Cumrun Vafa, Gianmassimo Tasinato and Ivonne Zavala for useful discussions and
comments. We thank the Isaac Newton Institute for Mathematical Sciences,
Cambridge, for support and hospitality during the programme ``Black
holes: bridges between number theory and holographic quantum
information'', where work on this paper was undertaken. This work was
supported by EPSRC grant no.\ EP/R014604/1. We thank the support of the
University of Guanajuato grants 
CIIC 224/2023 ``Conjeturas de gravedad cu\'antica y el paisaje de la
teor\'ia de cuerdas'', CIIC 251/2024 ``Teorías efectivas de cuerdas y exploraciones de aprendizaje de máquina" and CIIC/2026 ``Implicaciones de gravedad
cu\'antica en las teor\'ias de campo 4-dimensionales efectivas'', and of
CONAHCyT grant A-1-S-37752 ``Teor\'ias efectivas de cuerdas y sus
aplicaciones a la f\'isica de part\'iculas y cosmolog\'ia''. NGCB would
like to acknowledge support from the ICTP through the Associates
Programme (2023--2028). JDJPL acknowledges financial support from the Secretar\'ia de Ciencia,
Humanidades, Tecnolog\'ia e Innovaci\'on (SECIHTI), Mexico, through a
doctoral fellowship.

\newpage

\begin{appendices}
\section{Table with vacua}
\label{app:vacua}
This appendix summarizes the numerical results for the critical points of the scalar potential $V$ arising from the compactification of type IIB string theory on the Mirror Quintic Calabi–Yau threefold. The data presented here provide the complete set of minima and saddle points considered in the analysis of the main text, for the different choices of three-form fluxes.

For each critical point, we report the value of the scalar potential $V$, together with the corresponding vacuum expectation values of the scalar fields $r$, $\theta$, $\tau_1$, and $\tau_2$. The local properties of each extremum are characterized by the smallest and largest eigenvalues of the mass matrix of the scalar potential, denoted by $\lambda_{\mathrm{min}}$ and $\lambda_{\mathrm{max}}$, respectively. These quantities provide a direct characterization of the local stability of the corresponding critical point.

We further report the flux contribution to the induced D3-brane charge, $N_{\mathrm{flux}}$, as well as the integer flux quanta specifying the RR and NS-NS three-form fluxes, $F_3$ and $H_3$, respectively. Altogether, the table contains more than two hundred critical points and provides the numerical data underlying the statistical and phenomenological analysis presented in the main text. It also allows for a direct comparison between the vacuum structure, stability properties, and flux configurations of the different solutions.

\begin{equation*}
	\resizebox{1\hsize}{!}{$
\begin{array}{|c|c|c|c|c|c|c|c|c|c|c|c|c|c|c|c|c|}
\hline
\text{case}&V&r&\theta&\tau_1&\tau_2&\lambda_{\text{min}}&\lambda_{\text{max}}&N_{\text{flux}}&F_{(3),1}&F_{(3),2}&F_{(3),3}&F_{(3),4}&H_{(3),1}&H_{(3),2}&H_{(3),3}&H_{(3),4}\\
    \hline
 1 & 6.31277 & 0.2355 & 6.42538 & 6.27781 & 7.51705 & -27.7098 & 3392.09 & 50 & 17 & -1 & 0 & -1 & -1 & -1 & 3 & 0 \\
 2 & 55.9521 & 0.242666 & 7.80969 & -5.97712 & 3.37539 & -127.502 & 15912.1 & 134 & -12 & 12 & -13 & 8 & 6 & -1 & -5 & -1 \\
 3 & 3.19638 & 0.71178 & 6.25008 & 5.14129 & 2.49383 & -131.725 & 2460.88 & 325 & -11 & 7 & 11 & -20 & -3 & 18 & 3 & -5 \\
 4 & 282.83 & 0.081815 & 8.47622 & 1.7234 & 1.86737 & -487.904 & 119988. & 94 & -1 & -11 & 17 & -10 & 10 & 20 & -9 & -5 \\
 5 & 297.88 & 0.325206 & 5.87467 & 0.237925 & 2.8392 & -392.264 & 37868.8 & 427 & -17 & -3 & -10 & 8 & 20 & 2 & -15 & 4 \\
 6 & 6.77929 & 0.706884 & 6.92315 & -1.50281 & 1.30448 & -30.9995 & 4267.07 & 249 & -8 & -6 & -10 & -1 & 16 & 7 & -11 & 1 \\
 7 & 14.1129 & 0.506075 & 6.13772 & -1.2219 & 2.54186 & -85.8616 & 10002.2 & 497 & -12 & -2 & -16 & -1 & 19 & 1 & -16 & 0 \\
 8 & 173.411 & 0.353161 & 6.46256 & 0.361938 & 2.45806 & -360.631 & 26977.5 & 500 & 18 & -6 & 10 & -2 & -18 & -11 & 18 & -3 \\
 9 & 16.0182 & 0.22344 & 6.82712 & -1.77601 & 4.1746 & -1305.69 & 4929.66 & 239 & -10 & 17 & -2 & -13 & 3 & 20 & 1 & -1 \\
 10 & 46.7283 & 0.21329 & 5.85485 & -0.630401 & 1.09436 & -152.315 & 18118.1 & 188 & -13 & 7 & 0 & 0 & 17 & 12 & -15 & -1 \\
 11 & 54.7732 & 0.00216921 & -3.77166 & 0.705654 & 2.12053 & 81.4352 & 3.624\times 10^7 & 490 & -18 & -8 & 19 & -3 & -13 & -3 & -14 & 0 \\
 12 & 461.225 & 0.275477 & 3.56973 & 0.24355 & 3.54336 & -605.658 & 72277.3 & 282 & 7 & -9 & 15 & -13 & -20 & -8 & 11 & -1 \\
 13 & 218.404 & 0.253658 & 6.78138 & 1.34134 & 4.6316 & -575.131 & 36331.9 & 294 & 18 & -17 & 10 & -7 & -10 & -11 & 16 & 1 \\
 14 & 279.691 & 0.385563 & 5.45807 & 0.53057 & 1.58466 & -168.161 & 16473. & 58 & -15 & -9 & 4 & 7 & -20 & -17 & 10 & -1 \\
 15 & 184.292 & 0.099282 & 9.73434 & 1.42819 & 1.09376 & -740.346 & 67059.3 & 130 & 8 & 18 & -19 & 8 & -6 & -13 & 4 & 6 \\
 16 & 453.285 & 0.489845 & 6.77969 & 0.603122 & 7.67629 & -326.484 & 17357.4 & 11 & -16 & 17 & -4 & 17 & -9 & -13 & 13 & 2 \\
 17 & 136.043 & 0.135022 & 4.78868 & -1.37603 & 2.79612 & -1261.53 & 33470.9 & 125 & -2 & -16 & -6 & 11 & -8 & -15 & 4 & -1 \\
 18 & 13.1898 & 0.320201 & 6.2692 & -4.43205 & 5.2769 & -89.7738 & 11051.4 & 306 & 19 & 0 & 19 & -1 & -9 & 2 & 7 & 0 \\
 19 & 33.4174 & 0.396537 & 0.81339 & 1.83054 & 1.80596 & -80.8339 & 12516.5 & 322 & -3 & 14 & -20 & 3 & 12 & -8 & -10 & 2 \\
 20 & 95.7262 & 0.241254 & 6.6859 & -0.954912 & 1.14066 & -306.094 & 29873. & 360 & -5 & 11 & -15 & 2 & 20 & -1 & -16 & -2 \\
 21 & 17.8495 & 0.73636 & 8.88524 & -4.88933 & 4.80503 & -55.1177 & 8411.74 & 294 & -17 & -19 & -18 & 0 & 9 & -14 & -10 & 2 \\
 22 & 162.502 & 0.298172 & 1.05065 & -3.97493 & 6.15271 & -1056.14 & 15098.1 & 110 & -14 & 13 & -4 & 14 & -4 & -12 & 3 & 0 \\
 23 & 3.93343 & 0.496995 & 4.02473 & -0.583307 & 3.10553 & -32.3001 & 5457.79 & 164 & -4 & -3 & 19 & 0 & -9 & 17 & 4 & -3 \\
 24 & 3.31181 & 0.300824 & 1.94376 & 0.0932002 & 1.22746 & -51.2619 & 16916.6 & 348 & -10 & 4 & 17 & -2 & -20 & 19 & 1 & -5 \\
 25 & 32.3516 & 0.214323 & 7.13452 & 1.70189 & 9.16615 & -100.295 & 11598.9 & 109 & 13 & 20 & 16 & -3 & -4 & 0 & 5 & -1 \\
 26 & 134.387 & 0.277193 & 6.80821 & -1.1608 & 2.53895 & -347.985 & 23474.4 & 253 & 7 & -12 & 14 & -4 & -14 & -6 & 15 & 2 \\
 27 & 9.11939 & 0.60661 & 2.45937 & 0.243092 & 1.19589 & -60.3658 & 7936.73 & 274 & 3 & -1 & -15 & 0 & 20 & -14 & -7 & 5 \\
 28 & 182.209 & 0.278947 & 12.1526 & -5.52687 & 1.32713 & -382.755 & 15746.1 & 28 & -19 & 1 & 1 & 19 & 7 & 4 & -6 & -3 \\
 29 & 169.099 & 0.404173 & 5.77308 & -2.01603 & 1.48906 & -130.735 & 9027.72 & 57 & 4 & -11 & 2 & -9 & 8 & 14 & -5 & 3 \\
 30 & 22.1837 & 0.0767054 & 3.71031 & 0.411498 & 1.59875 & -24.0028 & 31286.7 & 66 & 13 & 0 & -10 & 2 & 5 & 5 & 2 & -1 \\
 31 & 3.25855 & 0.719692 & 11.9302 & -5.27164 & 9.46379 & -102.308 & 1428.32 & 173 & 9 & -3 & -17 & -12 & -2 & 14 & 4 & -1 \\
 32 & 302.499 & 0.342656 & 5.97944 & 2.21698 & 1.90242 & -365.016 & 28092.6 & 251 & 20 & 6 & 9 & -10 & -9 & -4 & 12 & -5 \\
 33 & 99.6656 & 0.143549 & 6.33739 & -1.10884 & 1.38034 & -264.222 & 54146.1 & 301 & 6 & -2 & 17 & -3 & -17 & -10 & 7 & 0 \\
 34 & 66.1578 & 0.253498 & 6.45998 & 2.31609 & 5.62684 & -162.267 & 10269.5 & 79 & -16 & -13 & -7 & 4 & 4 & 4 & -5 & 1 \\
 35 & 530.913 & 0.304245 & 6.34128 & 1.4149 & 1.11273 & -1172.86 & 41129.1 & 111 & -18 & -16 & -11 & 10 & 9 & 18 & -16 & 6 \\
  \hline
		\end{array}
		$}
\end{equation*}
\newpage
\begin{equation*}
	\resizebox{1\hsize}{!}{$
\begin{array}{|c|c|c|c|c|c|c|c|c|c|c|c|c|c|c|c|c|}
\hline
\text{case}&V&r&\theta&\tau_1&\tau_2&\lambda_{\text{min}}&\lambda_{\text{max}}&N_{\text{flux}}&F_{(3),1}&F_{(3),2}&F_{(3),3}&F_{(3),4}&H_{(3),1}&H_{(3),2}&H_{(3),3}&H_{(3),4}\\
    \hline
 36 & 410.542 & 0.175268 & 1.42045 & 0.58309 & 2.16791 & -923.227 & 104205. & 369 & -18 & 2 & -17 & 9 & 18 & 19 & -13 & 0 \\
 37 & 28.8273 & 0.386426 & 6.02903 & 1.47651 & 3.61631 & -175.363 & 17710.5 & 640 & -18 & 3 & -20 & 0 & 14 & -4 & -20 & 0 \\
 38 & 25.1863 & 0.304207 & 7.04692 & 3.32209 & 3.71246 & -128.379 & 22307.6 & 481 & -19 & 12 & -20 & 0 & 8 & -9 & -15 & 3 \\
 39 & 98.4718 & 0.1257 & 5.15695 & -0.65766 & 1.08576 & -191.329 & 28793. & 41 & -9 & 14 & -1 & 1 & 13 & 16 & -8 & -2 \\
 40 & 215.593 & 0.561633 & 6.92411 & -2.51031 & 1.90922 & -232.137 & 9305.38 & 179 & 8 & -16 & -19 & 17 & -9 & -14 & 6 & -4 \\
 41 & 63.9282 & 0.300609 & 8.03988 & -1.54569 & 1.20065 & -231.738 & 26384.6 & 385 & -17 & -5 & -10 & -2 & 20 & -20 & -15 & 6 \\
 42 & 631.293 & 0.449388 & 4.33162 & -0.282836 & 5.8526 & 109.741 & 41814.1 & 57 & 16 & -2 & 3 & -20 & -17 & -7 & 9 & -1 \\
 43 & 94.0078 & 0.290615 & 5.8919 & 1.74551 & 1.62323 & -177.273 & 13611.6 & 134 & -12 & 10 & -6 & 3 & 6 & 10 & -9 & 2 \\
 44 & 33.968 & 0.418965 & 1.28586 & 1.10651 & 1.063 & -79.3664 & 18927.5 & 478 & 4 & 0 & 19 & -4 & -18 & 17 & 17 & -4 \\
 45 & 571.605 & 0.367825 & 7.03891 & 2.34107 & 1.16154 & -1534.07 & 35121.5 & 108 & 14 & -4 & 7 & -19 & 16 & 20 & -14 & -9 \\
 46 & 214.898 & 0.227586 & 6.19414 & -1.03412 & 2.21903 & -743.997 & 21678.1 & 15 & 15 & 14 & -17 & 8 & -12 & -18 & 5 & 0 \\
 47 & 10.8466 & 0.402729 & 7.21171 & -1.84161 & 2.20792 & -85.7773 & 11205.8 & 379 & -9 & 5 & -18 & -2 & 15 & 0 & -11 & 2 \\
 48 & 83.5095 & 0.458708 & 0.181928 & -0.101538 & 1.97842 & -166.979 & 11061.9 & 365 & -15 & -10 & 20 & 0 & -17 & -6 & -3 & 2 \\
 49 & 80.1771 & 0.090074 & 9.06535 & 2.28057 & 2.37509 & -2366.55 & 38279.7 & 175 & -5 & -16 & 11 & -9 & 4 & 15 & -4 & -4 \\
 50 & 46.9986 & 0.25729 & 6.79221 & 2.83904 & 3.48 & -146.268 & 12081.3 & 149 & 5 & 1 & 20 & -3 & -6 & -7 & 10 & 0 \\
 51 & 132.356 & 0.114605 & 5.34327 & 2.31734 & 1.52158 & -211.013 & 51402.7 & 102 & 12 & -6 & 12 & -6 & -4 & -11 & 9 & -2 \\
 52 & 62.9678 & 0.38167 & 2.43531 & 0.836713 & 2.35637 & -74.2644 & 16151.5 & 214 & 10 & -6 & 8 & -4 & -12 & 4 & 9 & -2 \\
 53 & 204.886 & 0.243954 & 5.34227 & 0.371112 & 3.76402 & -360.82 & 40164.3 & 285 & 15 & -8 & 11 & -7 & -14 & -5 & 10 & -2 \\
 54 & 45.5777 & 0.176403 & 8.45297 & -1.16755 & 3.93286 & -147.421 & 29986.7 & 210 & -16 & -15 & -9 & -2 & 9 & -9 & -12 & 3 \\
 55 & 31.5754 & 0.460218 & 8.74802 & 2.55669 & 3.04449 & -301.906 & 2723.53 & 38 & -10 & 8 & -1 & -7 & -3 & 15 & 4 & -3 \\
 56 & 238.051 & 0.0853619 & 8.11691 & 3.46103 & 4.13972 & -2866.27 & 108339. & 320 & 14 & -16 & -7 & -17 & 8 & 20 & -10 & -4 \\
 57 & 6.13748 & 0.0671229 & 6.12435 & -0.375221 & 2.7054 & -46.9578 & 107481. & 338 & 11 & -11 & 15 & 1 & -12 & 7 & 14 & -1 \\
 58 & 9.61257 & 0.258543 & 5.50512 & 0.956371 & 1.24449 & -55.9227 & 12210.1 & 218 & 3 & -17 & 13 & 2 & -12 & -7 & 16 & 0 \\
 59 & 41.1045 & 0.161212 & 7.82611 & 3.8419 & 4.70093 & -4605.09 & 14711.1 & 286 & -15 & -7 & -1 & -16 & -1 & 19 & 3 & -4 \\
 60 & 275.058 & 0.345037 & 12.57 & 4.06936 & 1.73244 & -1023.96 & 18071.4 & 144 & -15 & 9 & 3 & 18 & -9 & -12 & 9 & 4 \\
 61 & 446.653 & 0.205157 & 1.27544 & 0.387295 & 4.15597 & -1578.07 & 57294.2 & 26 & 17 & -16 & 11 & -12 & 12 & 19 & -6 & -2 \\
 62 & 237.237 & 0.154808 & 4.99586 & -2.01316 & 1.06073 & -597.298 & 41932.9 & 4 & -18 & 7 & -3 & 8 & -1 & -17 & 6 & -3 \\
 63 & 72.4837 & 0.239842 & 5.00416 & 0.175921 & 3.81532 & -133.68 & 41514.8 & 477 & -14 & 3 & -17 & 5 & 15 & -9 & -12 & 3 \\
 64 & 256.815 & 0.175204 & 1.81211 & 0.404303 & 5.51357 & -1181.91 & 43884.3 & 70 & 16 & 12 & 8 & -14 & 8 & 15 & -4 & -1 \\
 65 & 197.525 & 0.322356 & 1.46921 & -1.52338 & 1.41117 & -800.368 & 17230.3 & 86 & -9 & 4 & -2 & 9 & -7 & -18 & 6 & -2 \\
 66 & 218.653 & 0.27401 & 6.04033 & 0.219221 & 3.84329 & -741.11 & 17780. & 30 & 19 & 10 & 1 & -7 & 8 & 19 & -5 & 0 \\
 67 & 63.5228 & 0.242586 & 6.36658 & 1.29398 & 2.38444 & -175.224 & 9283. & 58 & 0 & -18 & 13 & -1 & -7 & -15 & 8 & 1 \\
 68 & 35.5621 & 0.352505 & 11.8672 & 0.0283119 & 1.7839 & -75.8203 & 5841.04 & 97 & -10 & 15 & 9 & -1 & 9 & 13 & -18 & -1 \\
 69 & 296.626 & 0.220354 & 3.78095 & -1.74338 & 1.15686 & -610.379 & 39751.9 & 9 & -8 & -12 & -7 & 12 & -9 & -7 & 9 & -5 \\
 70 & 288.248 & 0.0443085 & 9.85935 & 0.438554 & 3.41864 & 86.6358 & 250129. & 1 & -2 & 8 & -15 & 9 & 10 & 19 & -15 & -1 \\
 71 & 267.995 & 0.287181 & 3.50385 & 3.05847 & 2.65194 & -378.865 & 31570. & 49 & 10 & 15 & -18 & 13 & 12 & 9 & -8 & 2 \\
 72 & 587.758 & 0.361172 & 1.58824 & 0.289353 & 2.63149 & -1024. & 45762.6 & 73 & 17 & 13 & 8 & -14 & -19 & -19 & 11 & 0 \\
 73 & 310.088 & 0.213202 & 6.6147 & 2.48656 & 2.1367 & -1908.86 & 36239.5 & 85 & 15 & -17 & 5 & -10 & 11 & 19 & -9 & -5 \\
 74 & 36.6186 & 0.374624 & 8.90244 & -3.93024 & 3.80776 & -87.8898 & 4847.06 & 38 & -18 & -15 & 4 & -5 & 5 & -4 & -6 & 2 \\
 75 & 63.1351 & 0.0552323 & 0.97131 & -0.537288 & 1.65069 & -23.9547 & 55842.5 & 34 & 2 & -9 & -11 & 4 & 7 & 15 & 4 & -1 \\
 76 & 174.551 & 0.120026 & 1.59241 & 1.28924 & 1.16621 & -299.184 & 57059.7 & 78 & -9 & 13 & -10 & 5 & 7 & 19 & -10 & 1 \\
 77 & 271.259 & 0.408511 & 6.58637 & 0.388567 & 1.53714 & -299.556 & 15127.1 & 82 & 19 & 11 & -20 & 8 & 17 & 18 & -6 & 0 \\
 78 & 266.025 & 0.888496 & 13.5452 & 3.23091 & 1.85806 & -296542. & 300268. & 62 & -11 & 5 & 5 & 13 & -9 & -9 & 10 & 2 \\
 79 & 82.4549 & 0.248992 & 6.13396 & -4.31056 & 5.85211 & -748.661 & 8698.55 & 62 & -13 & 5 & -4 & 8 & -2 & -12 & 2 & 0 \\
 80 & 9.68651 & 0.262992 & 6.81884 & 3.31489 & 2.88637 & -23.6381 & 2102.15 & 23 & -17 & 18 & 20 & 0 & -2 & 5 & 1 & 0 \\
 81 & 107.692 & 0.00775718 & 9.48181 & -0.0706286 & 1.40464 & 154.61 & 1.81954\times 10^6 & 87 & 7 & -2 & 2 & -3 & -10 & -19 & 18 & 1 \\
 82 & 56.8537 & 0.300876 & 7.18832 & 0.614271 & 5.11699 & -296.641 & 4904.04 & 28 & 5 & -15 & -2 & -5 & 4 & 7 & -6 & -1 \\
 83 & 197.851 & 0.102309 & 7.72403 & 6.80612 & 6.82142 & -521.088 & 60302.9 & 39 & 12 & -17 & 11 & -15 & 5 & 8 & -5 & -2 \\
 84 & 463.959 & 0.200409 & 10.2736 & 2.57271 & 4.74432 & -1000.58 & 67891.3 & 19 & 0 & -6 & -18 & 19 & -12 & -13 & 16 & 2 \\
 85 & 354.551 & 0.146209 & 2.87255 & 1.23899 & 1.21779 & -610.897 & 88643.6 & 89 & -4 & -15 & -15 & 11 & 15 & 12 & -14 & 4 \\
 86 & 147.756 & 0.247919 & 5.96993 & -0.93166 & 1.36272 & -731.038 & 14845.6 & 77 & 17 & 0 & -12 & 8 & -14 & -20 & 5 & 0 \\
 87 & 130.276 & 0.253906 & 12.1042 & -11.6564 & 5.72131 & -281.947 & 14888.3 & 54 & -12 & 16 & -14 & 17 & 4 & 2 & -4 & -1 \\
 88 & 182.267 & 0.322753 & 10.3096 & 5.88558 & 5.36409 & -600.03 & 16339.2 & 10 & -20 & 7 & -15 & -18 & -6 & 16 & 8 & -4 \\
 89 & 136.64 & 0.182482 & 11.5597 & 4.66737 & 3.38205 & -1059.74 & 21806.3 & 87 & 18 & -5 & -17 & -12 & 7 & 10 & -9 & -2 \\
 90 & 108.681 & 0.0912678 & 5.36244 & 0.432182 & 1.12373 & -127.242 & 53720.8 & 73 & 9 & 7 & 3 & -3 & -9 & -13 & 11 & -2 \\
 91 & 161.878 & 0.310777 & 9.53483 & 2.49435 & 1.0005 & -535. & 17232.4 & 84 & 3 & -17 & -12 & 16 & -6 & -15 & 6 & 6 \\
 92 & 24.2565 & 0.308855 & 7.60631 & -2.18539 & 2.77198 & -383.191 & 2727.36 & 39 & 8 & 15 & -14 & 3 & -3 & -14 & 3 & 1 \\
 93 & 192.907 & 0.286244 & -0.390772 & 0.12295 & 3.03035 & -339.295 & 19469.6 & 93 & 15 & 8 & -5 & 9 & 12 & 8 & 7 & 0 \\
 94 & 469.646 & 0.314301 & 6.8081 & -1.04226 & 4.66856 & -1190.78 & 35277.6 & 78 & -20 & 6 & -3 & 13 & 12 & 20 & -16 & -3 \\
 95 & 339.588 & 0.256423 & 6.41587 & 0.109185 & 5.2514 & -1005.7 & 32175.8 & 57 & 20 & 18 & 13 & -10 & -9 & -20 & 7 & 0 \\
 96 & 310.052 & 0.179678 & 11.3202 & 3.19928 & 3.86421 & -1380.15 & 48671.9 & 95 & 20 & -20 & -17 & -12 & 11 & 14 & -15 & -2 \\
 97 & 240.27 & 0.0877578 & 6.76878 & -1.33617 & 2.6409 & -369.172 & 90992.4 & 39 & -4 & 13 & -17 & 6 & 10 & 20 & -7 & -3 \\
 98 & 78.0458 & 0.341163 & 9.07915 & 3.62228 & 3.02237 & -813.59 & 8388.09 & 68 & 19 & 2 & -5 & 12 & 5 & -14 & -7 & 4 \\
 99 & 386.292 & 0.242192 & 12.121 & 2.37792 & 4.97429 & -1394.57 & 37697.8 & 69 & -12 & -7 & 0 & 16 & -11 & -16 & 15 & 1 \\
 100 & 212.419 & 0.447286 & 1.48745 & 1.41749 & 1.77325 & -206.543 & 14593.5 & 66 & 3 & 5 & -12 & 8 & 14 & 10 & -9 & 1 \\
 101 & 581.177 & 0.367683 & 5.77389 & 2.3194 & 2.89174 & -484.039 & 37847.8 & 140 & -2 & 12 & -17 & 17 & 17 & 13 & -12 & 4 \\
 \hline

 \end{array}
		$}
\end{equation*}

\newpage
\begin{equation*}
	\resizebox{1\hsize}{!}{$
\begin{array}{|c|c|c|c|c|c|c|c|c|c|c|c|c|c|c|c|c|}
\hline
\text{case}&V&r&\theta&\tau_1&\tau_2&\lambda_{\text{min}}&\lambda_{\text{max}}&N_{\text{flux}}&F_{(3),1}&F_{(3),2}&F_{(3),3}&F_{(3),4}&H_{(3),1}&H_{(3),2}&H_{(3),3}&H_{(3),4}\\
    \hline
 102 & 82.7723 & 0.136556 & 2.77182 & 3.8585 & 3.02183 & -124.666 & 24549.6 & 34 & 10 & -9 & 9 & -7 & -3 & -6 & 4 & -1 \\
 103 & 160.946 & 0.456013 & 1.21642 & 1.38284 & 2.37331 & -158.702 & 9176.68 & 25 & -9 & -5 & -5 & 8 & 8 & 8 & -6 & 1 \\
 104 & 43.6099 & 0.297358 & 6.39572 & -1.01968 & 3.98789 & -109.696 & 5975.22 & 67 & 9 & -10 & 4 & -2 & -6 & -5 & 7 & 1 \\
 105 & 109.601 & 0.181361 & 1.81842 & -3.17635 & 2.53492 & -454.9 & 17802.7 & 23 & -5 & 2 & -9 & 9 & -4 & -9 & 4 & -1 \\
 106 & 11.2903 & 0.133131 & 6.02904 & -1.17046 & 2.986 & -31.164 & 11262.4 & 68 & -4 & -6 & -9 & 2 & 6 & -2 & -4 & 1 \\
 107 & 107.913 & 0.416261 & 6.3383 & 1.48114 & 1.30471 & -74.4157 & 6671.3 & 63 & -5 & -8 & -7 & 6 & 8 & 1 & -9 & 4 \\
 108 & 18.1961 & 0.11491 & 6.01367 & -1.62407 & 1.42859 & -59.6924 & 23995.7 & 125 & -9 & 9 & -8 & 0 & 10 & 1 & -5 & 0 \\
 109 & 14.9401 & 0.390954 & 6.55903 & 2.20935 & 2.20055 & -170.781 & 1192.12 & 32 & 2 & -1 & 1 & -4 & 2 & 9 & -2 & -2 \\
 110 & 2.61828 & 0.122243 & 10.1032 & -0.831402 & 1.35475 & -7.3317 & 3748.76 & 18 & -6 & 9 & 7 & -1 & 4 & -2 & -8 & 0 \\
 111 & 0.391487 & 0.458105 & 12.4724 & -3.9158 & 2.16178 & -65.5846 & 967.711 & 121 & -3 & 2 & -8 & -10 & 2 & 10 & -1 & 1 \\
 112 & 37.4021 & 0.560157 & 6.6532 & -0.272075 & 10.3862 & -94.4088 & 1937.46 & 76 & 0 & 6 & -5 & 10 & -2 & -8 & 3 & 1 \\
 113 & 18.21 & 0.0000412828 & 3.6971 & 0.0750008 & 1.33977 & 36.1341 & 3.7805\times 10^9 & 47 & 5 & -10 & 4 & 0 & -3 & -9 & 9 & 1 \\
 114 & 20.17 & 0.0343919 & 10.799 & -1.72141 & 1.86313 & -5.80814 & 104958. & 87 & -5 & -10 & -8 & 3 & 7 & 3 & -8 & 0 \\
 115 & 0. & 0.0742471 & 6.1767 & 0.252715 & 1.24866 & 40.3858 & 17677.8 & 89 & -10 & -7 & 9 & 1 & -1 & -10 & -7 & 0 \\
 116 & 7.75111 & 0.17792 & 6.29368 & -1.27893 & 1.16395 & -42.8202 & 9612.83 & 106 & -6 & 4 & -7 & 0 & 10 & -1 & -6 & 0 \\
 117 & 37.0509 & 0.216986 & 12.4218 & 3.89944 & 1.25205 & -412.408 & 4689.68 & 30 & -10 & 2 & 6 & 7 & -4 & -6 & 4 & 2 \\
 118 & 7.55577 & 0.222068 & 6.00905 & 7.31911 & 9.65284 & -571.987 & 1912.43 & 66 & 9 & 2 & -5 & -9 & 1 & 7 & 0 & -1 \\
 119 & 59.6659 & 0.150971 & 4.29953 & -0.00709421 & 1.94432 & -96.8825 & 15659.6 & 28 & -6 & 9 & -2 & 2 & 8 & 9 & -5 & 0 \\
 120 & 11.881 & 0.251915 & 6.98591 & 2.71719 & 1.11323 & -716.872 & 2605.92 & 95 & 7 & 4 & 2 & 9 & 1 & -9 & 0 & 4 \\
 121 & 26.7164 & 0.252835 & 6.7925 & -2.34868 & 2.75249 & -74.0472 & 6989.25 & 81 & 9 & 9 & 8 & -2 & -7 & -1 & 4 & -1 \\
 122 & 92.851 & 0.26737 & 12.5841 & -3.65624 & 2.68215 & -242.084 & 8515.94 & 19 & 9 & 6 & 1 & -7 & -6 & -8 & 7 & 1 \\
 123 & 85.752 & 0.364748 & 6.43144 & 1.34841 & 1.14834 & -137.263 & 5967.81 & 40 & 10 & 6 & 3 & -3 & -3 & -9 & 7 & -2 \\
 124 & 65.4174 & 0.22952 & 3.21949 & 0.940675 & 1.66595 & -104.754 & 16506.7 & 84 & 3 & 10 & 8 & -5 & -10 & 0 & 8 & -2 \\
 125 & 19.8751 & 0.0897351 & 7.2878 & -1.51344 & 1.00898 & -40.8142 & 24857.6 & 77 & 5 & 1 & 8 & -2 & -8 & -4 & 4 & 1 \\
 126 & 3.46233 & 0.538098 & 5.61001 & -0.258827 & 1.68202 & -14.3456 & 2412.21 & 106 & -10 & 9 & 0 & -2 & 10 & 3 & -10 & 0 \\
 127 & 2.17462 & 0.305313 & -0.260276 & -2.54022 & 1.91259 & -11.6418 & 1265.57 & 30 & 7 & -1 & -9 & 0 & 1 & -1 & 3 & 0 \\
 128 & 57.6647 & 0.205638 & 9.5123 & -2.80617 & 1.19186 & -93.8977 & 18165.1 & 92 & 8 & 10 & 6 & -8 & -8 & -4 & 7 & 2 \\
 129 & 117.333 & 0.0970207 & 7.88275 & -1.62891 & 2.43308 & -161.846 & 43836.8 & 28 & -5 & 10 & -6 & 5 & 8 & 9 & -9 & -2 \\
 130 & 31.1305 & 0.289601 & 6.57585 & 1.39826 & 1.76239 & -86.2186 & 7976.6 & 125 & 9 & 0 & 6 & -1 & -6 & -1 & 10 & -2 \\
 131 & 33.7568 & 0.349 & 6.60123 & -3.63437 & 3.36068 & -71.5716 & 3843.51 & 54 & -7 & 8 & -8 & 3 & 5 & 2 & -4 & -1 \\
 132 & 5.30679 & 0.28058 & 7.40816 & -1.05812 & 1.4609 & -18.5819 & 2136.1 & 32 & -8 & -6 & 5 & 1 & 6 & 2 & -8 & 0 \\
 133 & 14.1411 & 0.103549 & 7.16252 & 0.104556 & 4.41339 & -37.6314 & 19153.8 & 80 & -5 & 10 & -9 & 1 & 5 & 0 & -7 & 0 \\
 134 & 102.247 & 0.158537 & 2.34253 & 1.36293 & 2.30606 & -188.361 & 26378.9 & 51 & -7 & -5 & -8 & 6 & 7 & 7 & -6 & 1 \\
 135 & 49.4988 & 0.173692 & 0.869493 & -0.174315 & 1.01931 & -119.619 & 10661. & 32 & -7 & 10 & -1 & 1 & 9 & 9 & -6 & -1 \\
 136 & 0. & 0.392527 & 3.55955 & -1.45839 & 1.77204 & 31.4902 & 1695.34 & 54 & 9 & 2 & 3 & 3 & -5 & -7 & 2 & 0 \\
 137 & 73.5732 & 0.166883 & 5.76741 & 2.49436 & 1.0945 & -211.032 & 11704.8 & 6 & 6 & -1 & 8 & -6 & -2 & -7 & 5 & -2 \\
 138 & 6.3344 & 0.20298 & 6.38996 & -2.73195 & 2.72905 & -33.8153 & 5983.13 & 78 & -9 & 3 & -7 & 0 & 6 & 1 & -4 & 0 \\
 139 & 56.0625 & 0.293927 & 6.5332 & 0.0114528 & 1.5872 & -129.334 & 4843.66 & 17 & -2 & -6 & -5 & 3 & 8 & 9 & -5 & 1 \\
 140 & 4.05707 & 0.0817872 & 6.89945 & 0.470889 & 4.24254 & -11.8519 & 16747.3 & 60 & 6 & 5 & 9 & 1 & -4 & 7 & 6 & -1 \\
 141 & 10.5842 & 0.358783 & 6.08975 & -4.13092 & 4.26286 & -275.225 & 1410.42 & 69 & 9 & -6 & -6 & -6 & 0 & 10 & 1 & 0 \\
 142 & 59.6001 & 0.358635 & 6.54137 & 1.36743 & 1.82232 & -107.472 & 6903.35 & 104 & -10 & -3 & -5 & 2 & 6 & 5 & -9 & 2 \\
 143 & 20.9356 & 0.348008 & 6.13034 & -2.50908 & 5.29953 & -36.8586 & 4596.47 & 93 & -10 & 7 & -7 & 2 & 6 & -2 & -4 & 1 \\
 144 & 23.7936 & 0.176433 & 6.07227 & -3.24109 & 3.61019 & -734.3 & 4607.38 & 45 & -9 & -9 & 2 & 6 & 0 & -9 & 1 & 0 \\
 145 & 97.8418 & 0.312133 & 12.3764 & 3.09927 & 2.28831 & -407.126 & 7346.03 & 52 & -5 & 3 & 0 & 9 & -6 & -9 & 7 & 2 \\
 146 & 28.1598 & 0.507796 & 7.12963 & 0.317332 & 4.28189 & -57.4181 & 1417.16 & 21 & 1 & -5 & -6 & 5 & -3 & -8 & 4 & 1 \\
 147 & 46.2581 & 0.389346 & 6.54385 & 0.283234 & 3.38211 & -74.5755 & 3744.5 & 48 & 4 & -4 & 8 & -2 & -6 & -10 & 5 & 0 \\
 148 & 46.6555 & 0.489082 & 6.66721 & 2.49686 & 2.50773 & -58.6511 & 3933.82 & 105 & -8 & -7 & -9 & 4 & 5 & 5 & -10 & 0 \\
 149 & 3.08835 & 0.437325 & 6.42051 & 0.0906062 & 4.31071 & -17.2695 & 3151.46 & 125 & 8 & -9 & 5 & -1 & -7 & 9 & 9 & -1 \\
 150 & 24.1197 & 0.422148 & 7.8399 & -1.19966 & 1.404 & -60.8892 & 4269. & 78 & -9 & -6 & -2 & -1 & 10 & -5 & -9 & 3 \\
 151 & 97.845 & 0.142312 & 8.57434 & 2.3688 & 2.51091 & -347.444 & 21821.8 & 27 & 8 & -4 & -4 & -7 & 7 & 9 & -9 & -2 \\
 152 & 195.699 & 0.240175 & 11.8934 & 2.17016 & 1.55545 & -551.916 & 19300.9 & 15 & -1 & 2 & -9 & 10 & -9 & -10 & 10 & 3 \\
 153 & 40.9523 & 0.164492 & 5.19451 & 1.15471 & 1.64035 & -74.8257 & 13674.7 & 61 & -6 & -4 & -5 & 3 & 6 & 1 & -7 & 2 \\
 154 & 13.3844 & 0.405063 & 6.93228 & 2.38497 & 1.25681 & -262.964 & 1528.23 & 72 & 0 & 10 & 3 & 6 & -2 & -6 & 4 & 3 \\
 155 & 9.51216 & 0.759828 & 7.48509 & -1.98571 & 2.17441 & -1.72268 & 349.668 & 3 & 8 & 7 & -6 & 2 & -2 & 1 & 3 & -1 \\
 156 & 18.485 & 0.337452 & 6.26886 & -0.702824 & 1.21161 & -48.7766 & 3101.55 & 58 & -10 & -4 & 4 & 1 & 9 & 10 & -10 & -1 \\
 157 & 78.9587 & 0.186863 & 1.11459 & 1.00186 & 1.05515 & -189.824 & 22877.6 & 125 & 10 & -7 & 8 & -3 & -6 & -10 & 10 & -1 \\
 158 & 21.9227 & 0.0671488 & 9.62091 & -2.21575 & 2.2559 & -31.2265 & 59947.8 & 128 & 8 & 2 & 7 & -3 & -8 & 0 & 9 & 0 \\
 159 & 94.1646 & 0.253438 & 3.58193 & -0.154061 & 2.91661 & -249.107 & 11515.8 & 22 & 4 & 1 & 2 & -6 & 8 & 9 & -4 & 0 \\
 160 & 61.7015 & 0.229135 & 5.13217 & 0.106031 & 2.01335 & -100.489 & 11716.6 & 62 & -10 & -7 & -1 & 3 & 9 & 1 & -7 & 2 \\
 161 & 8.39879 & 0.222076 & 1.05819 & 0.719368 & 1.08515 & -38.5878 & 7722.86 & 100 & -7 & -1 & -7 & 1 & 6 & -3 & -8 & 1 \\
 162 & 15.168 & 0.447068 & 3.04808 & 0.929361 & 2.12557 & -28.4045 & 1364.35 & 7 & 9 & -5 & -5 & 3 & 2 & 7 & 2 & 0 \\
 163 & 3.47829 & 0.766315 & 1.7411 & 3.30597 & 1.55642 & -164.278 & 922.497 & 84 & -1 & 1 & 4 & 8 & 0 & -10 & -1 & 3 \\
 164 & 22.1594 & 0.40724 & 6.86529 & 2.38808 & 2.36814 & -40.1197 & 2491.05 & 47 & -10 & -8 & -3 & 1 & 2 & 1 & -5 & 1 \\
 165 & 13.8788 & 0.55352 & 5.9007 & -1.57506 & 1.52752 & -30.3406 & 2303.5 & 115 & 5 & -8 & 8 & 0 & -10 & 9 & 7 & 0 \\
 166 & 36.0127 & 0.132143 & 5.5728 & 0.468141 & 1.3939 & -79.7703 & 21800.6 & 102 & 6 & -5 & 5 & -1 & -9 & -8 & 10 & -1 \\
 167 & 29.6627 & 0.159784 & 6.3613 & 0.425091 & 1.98292 & -85.9975 & 14006. & 91 & -4 & 4 & -8 & 1 & 8 & 9 & -9 & 0 \\
 \hline
 \end{array}
		$}
\end{equation*}

\newpage
\begin{equation*}
	\resizebox{1\hsize}{!}{$
\begin{array}{|c|c|c|c|c|c|c|c|c|c|c|c|c|c|c|c|c|}
\hline
\text{case}&V&r&\theta&\tau_1&\tau_2&\lambda_{\text{min}}&\lambda_{\text{max}}&N_{\text{flux}}&F_{(3),1}&F_{(3),2}&F_{(3),3}&F_{(3),4}&H_{(3),1}&H_{(3),2}&H_{(3),3}&H_{(3),4}\\
    \hline
 168 & 1.88073 & 0.275895 & 8.62633 & 0.133866 & 1.53647 & -2.06307 & 1902.47 & 25 & 4 & 4 & -9 & 0 & 5 & -7 & -6 & 1 \\
 169 & 90.3239 & 0.328965 & 6.07412 & -1.93296 & 1.38156 & -460.558 & 6908.29 & 89 & -4 & -7 & -2 & 9 & -5 & -10 & 3 & -3 \\
 170 & 30.6035 & 0.408284 & 6.29048 & 0.130304 & 1.59432 & -38.6519 & 3001.16 & 56 & -3 & 6 & -4 & 1 & 9 & 7 & -7 & 1 \\
 171 & 0. & 0.0106376 & 6.82038 & -1.81925 & 1.24712 & 9.42028 & 404312. & 87 & 9 & -7 & 5 & 0 & -7 & 10 & 5 & -1 \\
 172 & 139.333 & 0.17706 & 4.15126 & 0.921444 & 3.3515 & -227.894 & 30449.2 & 61 & -8 & -1 & -7 & 7 & 9 & 7 & -6 & 1 \\
 173 & 4.45581 & 0.0626256 & 8.3152 & 2.55666 & 1.76186 & -13.1296 & 30736.4 & 73 & -10 & 10 & -5 & -3 & 1 & -4 & -8 & 0 \\
 174 & 22.6069 & 0.62945 & 0.846483 & -0.729405 & 1.39278 & -16.5067 & 1498.04 & 43 & -10 & -10 & 5 & -1 & -2 & -7 & -5 & 1 \\
 175 & 22.1993 & 0.3314 & 6.75961 & 1.0327 & 2.55815 & -83.7447 & 1490.35 & 5 & 4 & 10 & 1 & -3 & 3 & 10 & -3 & -1 \\
 176 & 97.1986 & 0.0985965 & 5.57159 & 1.93828 & 4.78516 & -224.503 & 29348.8 & 14 & 5 & -9 & 9 & -6 & 5 & 10 & -2 & -1 \\
 177 & 3.38498 & 0.565017 & 7.43976 & -2.23126 & 1.45538 & -17.0845 & 1679.84 & 73 & -6 & -1 & -7 & -1 & 7 & 1 & -4 & 1 \\
 178 & 184.098 & 0.309876 & 7.62638 & -2.20859 & 1.22672 & -409.319 & 16206.2 & 25 & -6 & 10 & -5 & 9 & 10 & 5 & -10 & -4 \\
 179 & 35.2298 & 0.524467 & 0.95395 & 2.31195 & 2.71692 & -33.4308 & 3909.5 & 95 & 9 & 5 & 10 & -5 & -5 & 1 & 5 & -1 \\
 180 & 0. & 0.154723 & 8.4877 & -1.41444 & 3.16663 & 117.332 & 5649.2 & 109 & 3 & -8 & 10 & 6 & -4 & -10 & 3 & 0 \\
 181 & 48.4597 & 0.340632 & 6.75922 & 4.17973 & 1.73645 & -474.196 & 4014.95 & 63 & -5 & 5 & -4 & 10 & -3 & -7 & 2 & 3 \\
 182 & 0. & 0.0000195051 & 9.75971 & -0.0337254 & 2.64354 & 57.1681 & 1.53488\times 10^{10} & 99 & 9 & -7 & -6 & 4 & -1 & -10 & 8 & 1 \\
 183 & 105.151 & 0.246702 & 6.06861 & -2.04071 & 2.07287 & -271.31 & 12113.1 & 43 & 9 & -7 & 7 & -4 & -9 & -10 & 3 & 1 \\
 184 & 3.70487 & 0.43947 & 6.22502 & -4.63756 & 6.97796 & -17.7312 & 1614.63 & 68 & 9 & -4 & 10 & 1 & -4 & -1 & 3 & 0 \\
 185 & 7.34647 & 0.209608 & 1.58629 & 0.652333 & 2.14323 & -31.6 & 9471.94 & 107 & -10 & 1 & -7 & 1 & 6 & -4 & -6 & 1 \\
 186 & 93.557 & 0.292056 & 12.4676 & 2.14386 & 2.2499 & -239.857 & 6712.93 & 1 & 7 & 0 & -2 & -5 & 7 & 10 & -9 & -1 \\
 187 & 21.4564 & 0.0889381 & 6.1477 & 0.460011 & 1.28972 & -36.0229 & 18928.2 & 50 & 2 & 6 & -9 & 1 & 8 & 8 & -7 & 0 \\
 188 & 0. & 0.0560053 & 4.55866 & 1.03802 & 1.984 & 29.9903 & 12646.2 & 46 & -3 & -9 & 7 & -1 & -4 & 9 & 3 & -2 \\
 189 & 23.1649 & 0.00089899 & 0.19875 & -0.517303 & 1.20469 & 42.988 & 1.79853\times 10^7 & 39 & -1 & 2 & 10 & -2 & -5 & -9 & -5 & 1 \\
 190 & 3.54245 & 0.263853 & 6.14248 & 5.26993 & 5.91362 & -565.764 & 1695.21 & 108 & 1 & -9 & -2 & -9 & 0 & 10 & 0 & -2 \\
 191 & 36.6276 & 0.0956382 & 6.06525 & -5.34218 & 7.92051 & -1029.47 & 14196.8 & 63 & 1 & 3 & -9 & 10 & -2 & -8 & 1 & 0 \\
 192 & 9.15858 & 0.256176 & -0.189145 & -0.679235 & 1.52037 & -44.3382 & 5376.48 & 94 & -6 & -1 & 10 & 0 & -7 & 5 & -4 & 0 \\
 193 & 4.22984 & 0.350746 & 6.24368 & 3.80054 & 1.35551 & -310.087 & 1130.59 & 76 & -1 & -4 & -2 & -9 & 0 & 7 & -1 & -3 \\
 194 & 1.17967 & 0.0871412 & 10.5656 & -2.45474 & 2.19294 & -11.9816 & 29190.3 & 124 & -7 & 10 & -10 & -2 & 7 & -6 & -8 & 1 \\
 195 & 44.4678 & 0.402449 & 6.3668 & 0.73327 & 2.59023 & -69.117 & 3127.72 & 38 & -7 & 7 & -1 & 2 & 5 & 8 & -8 & -1 \\
 196 & 28.9084 & 0.341661 & 6.41368 & 6.79379 & 3.63737 & -582.689 & 3031.17 & 88 & -9 & 8 & -4 & 10 & -2 & -8 & 0 & 2 \\
 197 & 9.33116 & 0.291534 & 5.63587 & -0.229953 & 2.0475 & -19.329 & 5458.95 & 98 & -8 & 8 & -2 & 0 & 9 & -6 & -8 & 2 \\
 198 & 44.5641 & 0.292604 & 11.9496 & 3.50248 & 2.75796 & -97.7096 & 3482.75 & 7 & 3 & 9 & 1 & -7 & 4 & 5 & -5 & -1 \\
 199 & 24.8725 & 0.0377419 & 5.90196 & 0.857256 & 1.32556 & -12.7013 & 103132. & 99 & 10 & 3 & 2 & -1 & -3 & -4 & 10 & -1 \\
 200 & 97.1755 & 0.415535 & 7.25366 & 1.36573 & 3.65636 & -247.747 & 5784.16 & 40 & 0 & -6 & 7 & -7 & 6 & 10 & -8 & -2 \\
 201 & 5.65876 & 0.323473 & 5.13602 & 0.454429 & 1.13473 & -11.8393 & 2412.45 & 42 & 4 & -4 & 3 & 2 & -6 & 8 & 9 & -1 \\
 202 & 123.351 & 0.197875 & 1.97744 & -1.4521 & 2.11252 & -407.224 & 18039.8 & 12 & 7 & 7 & 5 & -7 & 6 & 10 & -5 & 1 \\
 203 & 56.506 & 0.181503 & 4.08041 & 2.58802 & 2.10714 & -107.888 & 9557.73 & 4 & -7 & -2 & 10 & -6 & -6 & -5 & 4 & -1 \\
 204 & 43.5707 & 0.273782 & 6.7574 & 0.269774 & 1.63782 & -111.635 & 7910.24 & 86 & 9 & 8 & 2 & -1 & -8 & -4 & 10 & -2 \\
 205 & 108.826 & 0.155983 & 2.71037 & 1.07268 & 1.15029 & -189.067 & 26936. & 39 & -5 & -2 & -6 & 5 & 8 & 9 & -8 & 2 \\
 206 & 107.853 & 0.095206 & 7.9142 & -1.55366 & 1.08706 & -145.099 & 39894.8 & 20 & 8 & -5 & 0 & -5 & -10 & -9 & 10 & 3 \\
 207 & 12.1944 & 0.228251 & 2.29969 & -2.54864 & 1.5003 & -708.89 & 2768.42 & 50 & 7 & -4 & 2 & 7 & -2 & -9 & -3 & -1 \\
 208 & 103.814 & 0.374823 & 1.68348 & -0.166531 & 3.17867 & -182.565 & 7419.77 & 8 & 10 & 1 & 0 & -6 & 7 & 8 & -4 & 0 \\
 209 & 30.577 & 0.27156 & 5.99303 & 0.113748 & 1.41486 & -55.3045 & 4591.34 & 39 & 3 & -3 & -7 & 3 & 9 & 2 & -5 & 1 \\
 210 & 1.71285 & 0.572969 & 10.0981 & -3.49949 & 2.16817 & -8.65437 & 3753.6 & 116 & 10 & -2 & 8 & 3 & -7 & 4 & 7 & -1 \\
 211 & 150.175 & 0.287756 & 5.40498 & 0.702923 & 3.68023 & -230.496 & 15749. & 63 & 4 & -8 & 8 & -7 & -10 & -7 & 6 & -1 \\
 212 & 13.6365 & 0.241873 & 0.468693 & -3.61283 & 1.84259 & -634.358 & 2487.67 & 64 & 5 & 0 & -3 & -8 & 0 & 8 & 0 & 1 \\
 213 & 1.1948 & 0.69907 & 6.66654 & -1.18582 & 3.87357 & -3.41255 & 441.058 & 28 & -4 & -4 & -3 & 0 & 4 & 2 & -4 & 0 \\
 214 & 18.3933 & 0.415927 & 0.755909 & -1.83846 & 2.25914 & -33.9326 & 2039.75 & 37 & 9 & 6 & -10 & 1 & 2 & 4 & 3 & -1 \\
 215 & 0. & 0.311901 & 9.19936 & -2.60153 & 9.66258 & 91.0307 & 2114.33 & 83 & 8 & -7 & -8 & -6 & -2 & 10 & 4 & -1 \\
 216 & 24.3711 & 0.239997 & 7.30907 & 3.91615 & 5.33458 & -169.604 & 2768.43 & 11 & 1 & 0 & 6 & -5 & 2 & 5 & -2 & -1 \\
 217 & 25.3366 & 0.174634 & 6.5688 & -1.51454 & 2.4083 & -89.0291 & 14823.6 & 127 & -10 & -1 & -7 & 1 & 9 & 6 & -7 & 0 \\
 218 & 4.02174 & 0.124185 & 4.82299 & 0.851983 & 1.92484 & -12.5975 & 9876.86 & 58 & 9 & 3 & 2 & 1 & -3 & 8 & 7 & -1 \\
 219 & 8.07429 & 0.386887 & 0.939611 & 1.50111 & 1.08491 & -22.6184 & 2538.95 & 60 & 2 & 3 & 9 & -2 & -5 & 3 & 6 & -1 \\
 220 & 147.98 & 0.413995 & 6.14108 & 2.12793 & 2.36565 & -92.1672 & 9832.41 & 104 & -10 & 1 & -7 & 7 & 8 & 5 & -8 & 3 \\
 221 & 8.9441 & 0.119585 & 8.90027 & -1.76052 & 1.08839 & -30.3881 & 23665.4 & 123 & 9 & 5 & 4 & -2 & -10 & 1 & 9 & 0 \\
 222 & 111.861 & 0.207227 & 4.04838 & 2.1524 & 2.61045 & -179.186 & 20888.6 & 50 & -6 & -9 & -8 & 8 & 7 & 3 & -6 & 2 \\
 223 & 3.45496 & 0.491668 & 6.11536 & -0.506443 & 3.17967 & -22.8364 & 2845.55 & 136 & 8 & -1 & 7 & 1 & -9 & -1 & 9 & 0 \\
 224 & 46.6011 & 0.311108 & 6.53189 & -3.25167 & 3.35277 & -652.408 & 4626.11 & 94 & -1 & -7 & -4 & 10 & -3 & -10 & 1 & -1 \\
 225 & 7.16221 & 0.310337 & 6.53199 & 0.0583652 & 2.43364 & -40.6144 & 4513.21 & 112 & 8 & 1 & 4 & 0 & -8 & -3 & 10 & 0 \\
 226 & 21.987 & 0.287017 & 8.18977 & -1.06644 & 4.21919 & -422.141 & 3046.99 & 50 & 6 & 6 & -10 & 5 & -3 & -10 & 4 & 1 \\
 227 & 158.174 & 0.407412 & 6.46957 & 0.0970163 & 4.19542 & -173.414 & 8213.1 & 31 & 9 & 8 & 8 & -7 & -8 & -10 & 5 & -1 \\
 228 & 196.322 & 0.378153 & 5.83806 & 1.58068 & 4.19372 & -133.271 & 11952.5 & 41 & -10 & -7 & -5 & 10 & 9 & 5 & -6 & 2 \\
 229 & 35.2118 & 0.0925053 & 5.08202 & -4.67604 & 9.50995 & -342.954 & 13673.9 & 40 & 2 & -1 & 6 & -9 & 2 & 6 & -1 & 0 \\
 230 & 16.7994 & 0.305247 & 12.4275 & -3.77601 & 1.60866 & -37.549 & 4217.2 & 71 & -6 & 7 & -7 & 3 & 5 & -2 & -5 & 0 \\
 231 & 27.6927 & 0.883695 & 8.22673 & -0.335981 & 2.7456 & -181141. & 183535. & 62 & -10 & 9 & 3 & -1 & 6 & 9 & -8 & -1 \\
 232 & 3.91638 & 0.58918 & 11.6834 & 1.18657 & 1.85655 & -86.8981 & 505.462 & 36 & 1 & -10 & -5 & -2 & 0 & 9 & -2 & -2 \\
 233 & 80.8275 & 0.199532 & 5.92187 & 0.901818 & 3.47069 & -246.542 & 9845.6 & 3 & 4 & 9 & -10 & 5 & 6 & 9 & -3 & 0 \\
 \hline
 \end{array}
		$}
\end{equation*}

\newpage
\begin{equation*}
	\resizebox{1\hsize}{!}{$
\begin{array}{|c|c|c|c|c|c|c|c|c|c|c|c|c|c|c|c|c|}
\hline
\text{case}&V&r&\theta&\tau_1&\tau_2&\lambda_{\text{min}}&\lambda_{\text{max}}&N_{\text{flux}}&F_{(3),1}&F_{(3),2}&F_{(3),3}&F_{(3),4}&H_{(3),1}&H_{(3),2}&H_{(3),3}&H_{(3),4}\\
    \hline
 234 & 3.73803 & 0.389371 & 6.14838 & 0.333372 & 3.43349 & -25.2148 & 2687.29 & 102 & 4 & 1 & 10 & 0 & -7 & 1 & 8 & 0 \\
 235 & 149.864 & 0.146761 & 3.68873 & 1.43168 & 1.02315 & -241.07 & 35686.6 & 29 & 2 & 0 & 9 & -7 & -9 & -10 & 9 & -3 \\
 236 & 22.6232 & 0.418495 & 6.58863 & 2.104 & 1.45268 & -47.7197 & 4819.96 & 140 & -8 & 8 & -9 & 1 & 5 & 1 & -10 & 2 \\
 237 & 2.94981 & 0.285997 & 3.54461 & 1.25096 & 1.37444 & -8.88426 & 3921.25 & 56 & -10 & -6 & -2 & 0 & 1 & -8 & -6 & 1 \\
 238 & 15.0115 & 0.42737 & 6.53918 & -0.200973 & 3.12689 & -33.8559 & 4142.8 & 133 & -4 & 8 & -9 & 1 & 9 & -4 & -8 & 2 \\
 239 & 34.6545 & 0.60498 & 6.62346 & 2.16342 & 2.98232 & -16.0429 & 2288.67 & 89 & -9 & -9 & -4 & 4 & 5 & 3 & -9 & 0 \\
 240 & 18.6196 & 0.258323 & 7.27626 & -0.849507 & 4.22561 & -78.2254 & 11002.5 & 167 & -9 & -10 & -10 & 3 & 8 & 1 & -10 & 0 \\
 241 & 46.7434 & 0.188137 & 6.07305 & -1.26795 & 2.03869 & -140.309 & 7193.56 & 14 & -5 & 5 & -4 & 2 & 6 & 10 & -3 & -1 \\
 242 & 93.5953 & 0.36972 & 11.8821 & -2.83688 & 1.01416 & -63.4957 & 8700.1 & 82 & 9 & -4 & 0 & -9 & -9 & 0 & 10 & 2 \\
 243 & 21.9146 & 0.331745 & 6.7967 & -0.11409 & 2.406 & -67.0854 & 4872.21 & 95 & -7 & -5 & -5 & 1 & 8 & 8 & -9 & 0 \\
 244 & 4.98426 & 0.790281 & 6.0022 & -2.54981 & 4.04907 & -10.6194 & 2104.02 & 151 & 10 & -4 & 9 & 2 & -9 & 0 & 7 & 0 \\
 245 & 10.1141 & 0.0961037 & 6.56938 & -0.934076 & 2.26003 & -322.741 & 3950.03 & 18 & 0 & 5 & 3 & -3 & 2 & 8 & -1 & 0 \\
 246 & 3.07592 & 0.124821 & 9.59485 & 1.87976 & 1.99889 & -13.7207 & 9513.3 & 54 & 5 & -9 & 5 & 2 & -3 & 3 & 9 & 0 \\
 247 & 40.1462 & 0.217437 & 2.41288 & 1.4099 & 1.20003 & -71.2983 & 10908.4 & 57 & 4 & 7 & -10 & 3 & 9 & 4 & -7 & 1 \\
 248 & 128.076 & 0.381898 & 6.00521 & 0.0862243 & 5.50823 & -113.589 & 9221.28 & 77 & 10 & -6 & 7 & -7 & -8 & -5 & 5 & -1 \\

 \hline
 \end{array}
		$}
\end{equation*}

\section{Stabilizing first the axio-dilaton}
\label{axiodilatonHeavier}

In Sec.~\ref{sec:vacua} the critical points of $\hat V$ were obtained by
solving the four real equations \eqref{extre} simultaneously for $z$ and
$\tau$. An alternative, and often more economical, procedure consists in
first solving $\partial_\tau\hat V=0$ for the axio-dilaton at fixed
complex structure, which yields $\tau^\ast(z)$, and then extremizing the
resulting one-parameter potential $\hat V(z,\tau^\ast(z))$ with respect
to $z$ alone. This two-step procedure is justified provided that, for
every $z$, the potential does develop a minimum along the axio-dilaton
directions, so that $\tau$ can be integrated out. In this appendix we
show that this is indeed the case: the dependence of $\hat V$ on $\tau$
is constrained enough by the structure of the flux superpotential that
the extremum in $\tau$ can be found in closed form, its existence
follows from a simple inequality, and the value of the potential at the
extremum takes a compact expression. The result also bears on the
question of which sector is stabilized first, discussed in
\cite{CaboBizet:2024conifold}: the mass hierarchy between the
axio-dilaton and the complex-structure modulus is not automatic, and the
explicit form of $\tau^\ast(z)$ obtained below is the starting point for
addressing it.

The argument can be made precise as follows. As a function of the real
components of $\tau$, the potential takes the polynomial form
\begin{equation}
     \hat V(\tau,z)\simeq
     \alpha_{1}(z)
     +\frac{\alpha_2(z)}{\tau_2}
     +\alpha_3(z)\frac{\tau_1}{\tau_2}
     +\alpha_4(z)\frac{\tau_1^2}{\tau_2}
     +\alpha_5(z)\,\tau_2
     =\frac{P(\tau_1)}{\tau_2}+\alpha_1(z)+\alpha_5(z)\,\tau_2\,,
     \label{eq:Vtau}
\end{equation}
with $P(\tau_1)=\alpha_2+\alpha_3\tau_1+\alpha_4\tau_1^2$ and
$\alpha_i\in\mathbb{R}$. The five coefficients are not independent. Using
$W=A+B\tau$ from \eqref{eq:AB} one finds that both the axio-dilaton and
the complex-structure contributions to \eqref{nsV} depend on $\tau$ only
through $|A+B\bar\tau|^2$ and $|A+B\tau|^2$, each of which contains
$\tau_1$ and $\tau_2$ in the combination $\tau_1^2+\tau_2^2$ with no
cross term. Hence
\begin{equation}
    \alpha_4=\alpha_5\,,
    \label{eq:alpha45}
\end{equation}
which is the statement that $\hat V$ is $SL(2,\mathbb{Z})$ covariant in
the axio-dilaton sector. Extremizing \eqref{eq:Vtau} then gives, at fixed
$z$,
\begin{equation}
    \tau_1^\ast=-\frac{\alpha_3}{2\alpha_4}\,,
    \qquad
    \left(\tau_2^\ast\right)^2
    =\frac{\alpha_2}{\alpha_4}-\frac{\alpha_3^2}{4\alpha_4^2}
    =\frac{4\alpha_2\alpha_4-\alpha_3^2}{4\alpha_4^2}\,,
    \qquad
    \hat V^\ast=2\alpha_4\,\tau_2^\ast+\alpha_1\,.
    \label{eq:taumin}
\end{equation}
A minimum with $\tau_2^\ast>0$ therefore exists whenever
$4\alpha_2\alpha_4>\alpha_3^2$, which follows from the Cauchy--Schwarz
inequality applied to $A$ and $B$, with equality precisely in the
degenerate imaginary-self-dual case.

The analysis above shows that the axio-dilaton can always be integrated
out at fixed $z$. Whether it {\it should} be is a different matter. The
two-step procedure implicitly assumes that $\tau$ is the heavier field,
so that it settles to $\tau^\ast(z)$ before the complex-structure modulus
has moved appreciably. Our numerical results point in the opposite
direction. The eigenvector analysis of Sec.~\ref{sec:vacua},
Figs.~\ref{fig:unstable} and \ref{fig:heaviest}, shows that the lightest
modes of the mass matrix are dominated by $\phi_{\tau_1}$ and
$\phi_{\tau_2}$, while the heaviest ones are dominated by the
complex-structure components $r$ and $\theta$: statistically, over the
$\sim10^{2}$ convergent extrema, the axio-dilaton is the {\it lighter}
sector. In that regime the hierarchy assumed by the two-step procedure is
inverted, and it is $z$, not $\tau$, that should be integrated out first.
This is consistent with the conclusions of \cite{CaboBizet:2024conifold},
where the same inversion was found near the conifold for generic
one-parameter geometries. The closed-form solution \eqref{eq:taumin}
therefore remains useful as an analytic characterization of the
axio-dilaton extremum, and as a check of the numerical solutions, but not
as a justification for stabilizing the axio-dilaton first: the reliable
procedure, and the one adopted in the main text, is to solve the four
equations \eqref{extre} simultaneously.

\end{appendices}
% Bibliography
%-----------------------------------------------------------------
\bibliographystyle{utphys}
\bibliography{biblioCP26}

\end{document}